\documentclass[11pt]{article}
\usepackage{txfonts}
\usepackage{amssymb}

\renewcommand{\arraystretch}{1.4}

\def\appendix{
    \setcounter{section}{0}
    \renewcommand{\thesection}{Appendix \Alph{section}:}
    \renewcommand{\thesubsection}{\Alph{section}.\arabic{subsection}.}
    \renewcommand{\theequation}{\Alph{section}.\arabic{equation}}}
\def\section#1{
    \addtocounter{section}{1}
    \setcounter{subsection}{0}
    \setcounter{subsubsection}{0}
    \setcounter{equation}{0}
    \vskip8mm\begin{center}{\bf\large\thesection~#1}\end{center}}
\def\subsection#1{
    \addtocounter{subsection}{1}
    \setcounter{subsubsection}{0}
    \vskip6mm\noindent{\bf\large\thesubsection~#1}\vskip4mm}
\def\subsubsection#1{
    \addtocounter{subsubsection}{1}
    \vskip6mm\noindent{\bf #1}\par\vskip3mm}
\def\paragraph#1{
    \vskip6mm\noindent\underline{\sf #1}\par\vskip3mm}
\def\thebibliography#1{\list
 {[\arabic{enumi}]}{\settowidth\labelwidth{[#1]}\leftmargin\labelwidth
  \advance\leftmargin\labelsep
  \usecounter{enumi}}
  \def\newblock{\hskip .11em plus .33em minus .07em}
  \sloppy\clubpenalty4000\widowpenalty4000
  \sfcode`\.=1000\relax}

\newcommand{\nn}{\nonumber}
\newcommand{\vev}[1]{\langle{#1}\rangle}
\newcommand{\bra}[1]{\langle{#1}|}
\newcommand{\ket}[1]{|{#1}\rangle}
\newcommand{\bbra}[1]{{\!\!\langle\!\langle{#1}|}}
\newcommand{\bket}[1]{{|{#1}\rangle\!\rangle}}

\newcommand{\tprod}{{\textstyle\prod}}
\newcommand{\tfrac}[2]{{\textstyle\frac{#1}{#2}}}

\newcommand{\bG}{{\bf G}}
\newcommand{\bS}{{\bf S}}
\newcommand{\bs}{{\bf s}}
\newcommand{\bc}{{\bf c}}

\newcommand{\chat}{{\hat{c}}}

\newcommand{\gsix}[6]{{
  \def\arraystretch{0.4}
  \left[\!\!\begin{array}{lll}
  #1 \!\!&\!\! #2 \!\!&\!\! #3 \\
  #4 \!\!&\!\! #5 \!\!&\!\! #6 \end{array}\!\!\right] }}

\DeclareFontFamily{U}{rsf}{}
\DeclareFontShape{U}{rsf}{m}{n}{
  <5> <6> rsfs5 <7> <8> <9> rsfs7 <10-> rsfs10}{}
\DeclareMathAlphabet\Scr{U}{rsf}{m}{n}

\newcommand{\mZ}{{\mathbb Z}}
\newcommand{\mR}{{\mathbb R}}

\begin{document}

%%%%%%%%%%%%%%%%%%%%%%%%%%%%%%%%%%%%%%%%%%%%%%%%%%
%%%%% T I T L E   P A G E %%%%%%%%%%%%%%%%%%%%%%%%
%%%%%%%%%%%%%%%%%%%%%%%%%%%%%%%%%%%%%%%%%%%%%%%%%%

\renewcommand{\thefootnote}{\fnsymbol{footnote}}
\setcounter{footnote}{0}
\setcounter{section}{0}
\baselineskip = 0.6cm
\pagestyle{empty}

%%%%% title %%%%%%%%%%%%%%%%%%%%%%%%%%%%%%%%%%%%%%

%%%%% preprint number %%%%%%%%%%%%%%%%%%%%%%%%%%%%
\baselineskip 5mm
\hfill\vbox{\hbox{hep-th/0408172} }

\baselineskip0.8cm\vskip2cm

\vskip10mm

\begin{center}
 {\large\bf $\bf N=2$ Liouville Theory with Boundary}
\end{center}

\vskip10mm

%%%%% author %%%%%%%%%%%%%%%%%%%%%%%%%%%%%%%%%%%%%
\baselineskip0.6cm
\begin{center}
    Kazuo Hosomichi
 \\ \vskip2mm
{\it Service de Physique Th\'eorique, CEA Saclay\\
     F-91191 Gif sur Yvette, France} \vskip3mm
\end{center}

%%%%% abstract %%%%%%%%%%%%%%%%%%%%%%%%%%%%%%%%%%%
\vskip8mm\baselineskip=3.5ex
\begin{center}{\bf Abstract}\end{center}\par\smallskip

We study $N=2$ Liouville theory with arbitrary
central charge in the presence of boundaries.
After reviewing the theory on the sphere and deriving some important
structure constants, we investigate the boundary states of the
theory from two approaches, one using the modular transformation
property of annulus amplitudes and the other using the bootstrap
of disc two-point functions containing degenerate bulk
operators.
The boundary interactions describing the boundary states are
also proposed, based on which the precise correspondence
between boundary states and boundary interactions is obtained.
The open string spectrum between D-branes is studied from
the modular bootstrap approach and also from the reflection
relation of boundary operators, providing a consistency
check for the proposal.

\vspace*{\fill}
\noindent Second version~:~May~~2006
\setcounter{page}{0}
\newpage

%%%%%%%%%%%%%%%%%%%%%%%%%%%%%%%%%%%%%%%%%%%%%%%%%%
%%%%%%%%%%%%%%%% T I T L E   P A G E   E N D %%%%%
%%%%%%%%%%%%%%%%%%%%%%%%%%%%%%%%%%%%%%%%%%%%%%%%%%

\setcounter{footnote}{0}
\setcounter{section}{0}
\pagestyle{plain}
\renewcommand{\thesection}{\arabic{section}.}
\renewcommand{\thesubsection}{\arabic{section}.\arabic{subsection}.}
\renewcommand{\theequation}{\arabic{section}.\arabic{equation}}
\renewcommand{\thefootnote}{\arabic{footnote}}
\setcounter{footnote}{0}

%%%%%%%%%%%%%%%%%%%%%%%%%%%%%%%%%%%%%%%%%%%%%%%%%%
%%%%% M A I N   T E X T %%%%%%%%%%%%%%%%%%%%%%%%%%
%%%%%%%%%%%%%%%%%%%%%%%%%%%%%%%%%%%%%%%%%%%%%%%%%%

%%%%%%%%%%%%%%%%%%%%%%%%%%%%%%%%%%%%%%%%%%%%%%%%%%
\section{Introduction}%%%%%%%%%%%%%%%%%%%%%%%%%%%%
%%%%%%%%%%%%%%%%%%%%%%%%%%%%%%%%%%%%%%%%%%%%%%%%%%

$N=2$ Liouville theory has a wide variety of
applications in string theory.
Although the theory is interacting, the $N=2$ superconformal
symmetry will allow one to compute various structure constants
and correlation functions exactly.
In the last decade there has been a great progress in the understanding of
non-compact, interacting CFTs such as Liouville theories with
various supersymmetry\cite{Goulian-L}-\cite{Ahn-RS}.
These recent works have studied the theories by combining the
knowledge of the representations of the symmetry algebra
together with the Lagrangian description as perturbed
free conformal field theories.

A particularly important progress has been made in $N=0$ and $N=1$
Liouville theories in the understanding of boundary states or D-branes,
where Cardy's construction of boundary states has been
successfully applied and we have found various boundary
states in consistency with the representation theory
of Virasoro or super-Virasoro algebras\cite{Fateev-ZZ,
Zamolodchikov-Z2, Fukuda-H2, Ahn-RS}.
For some boundary states the field theory descriptions in terms
of boundary interactions have also been provided,
whereas the others are interpreted as the theories being
realized on the pseudosphere.
This big breakthrough was followed by the determination of
various exact structure constants on
disc\cite{Hosomichi,Ponsot-T}.

In this paper we try to extend this success to $N=2$ Liouville
theory with boundary.
There have been quite a few works \cite{Eguchi-S}-\cite{Ahn-SY2}
on this theory and also some related works on the dual coset model
or the $H_3^+$ WZW model\cite{Giveon-KS}-\cite{Fotopoulos-NP}
along the path explained above.
However, there still remain certain confusing issues which
we attempt to resolve in the present paper.
One source of confusion is the additional periodic
direction $\theta$.
As we will see, the open and closed string states carry
momentum and winding number along $\theta$ obeying
a certain quantization law,
and one has to take a proper account of the quantization
law in analyzing the theory.
For example, the boundary states in $N=2$ Liouville theory
are classified as A-branes or B-branes according to the choice
of boundary conditions on supercurrents, and the momentum/winding number
quantization law makes these two families qualitatively very
different.
In this paper, we are able to take the correct account of
the quantization law.

We also study some other difficult problems in $N=2$ Liouville theory
in detail.
One of them is related to the property of operators
belonging to degenerate representations.
The $N=2$ Liouville theory actually has few properties
in common with the less supersymmetric theories.
For example, unlike the Liouville theories with less supersymmetry,
$N=2$ Liouville theory does not have a simple strong-weak coupling duality.
It instead has as the dual theory the $N=2$ supersymmetric
$SL(2,\mR)/U(1)$ coset model\cite{Hori-K}.
One important difference between $N=2$ theory and $N=0,1$ theories
is the spectrum of degenerate representations.
The degenerate representations of $N=2$ superconformal algebra
are generated by two fundamental degenerate representations
with Liouville momentum $j=1/2$ and $j=k/2$.
These two representations are totally different in quality since
the former is chiral and the latter is non-chiral, so that
they behave very differently under fusion.
Another is related to the boundary fermions we are lead to introduce
in describing D-branes in terms of boundary interactions.
They introduce the Chan-Paton degree of freedom on the boundary
and make the properties of boundary operators quite complicated.

~

This paper is organized as follows.
In section 2 we give a rather thorough review of the theory on
the sphere, where some OPE coefficients and the reflection coefficients
of bulk operators are presented.
Section 3 starts the analysis of the theory with boundary,
where we find the wave functions for A-branes
by analyzing annulus partition functions.
We also argue that the similar analysis for B-branes does not
work as long as there is a continuous spectrum of closed string
states obeying reflection relation.
Section 4 gives another derivation of the wave functions
which makes use of the Ward identity of disc two-point functions
containing degenerate bulk operators.
In section 5 we first propose the boundary interactions
preserving B-type supersymmetry using the construction well-known
in $N=2$ Landau-Ginzburg models, and then attempt to extend it to A-branes.
Using them we calculate some structure constants on the disc
and find the relation between boundary couplings and the
labels of boundary states.
Section 6 analyzes the reflection property of boundary operators,
where we find the open string spectrum
from the phase of reflection coefficients and check the
consistency with the result of modular bootstrap analysis.
In section 7 we give some brief concluding remarks.
Some useful formulae are recorded in the appendix.

%%%%%%%%%%%%%%%%%%%%%%%%%%%%%%%%%%%%%%%%%%%%%%%%%%
\section{N=2 Super-Liouville Theory}%%%%%%%%%%%%%%
%%%%%%%%%%%%%%%%%%%%%%%%%%%%%%%%%%%%%%%%%%%%%%%%%%

%%%%%%%%%%%%%%%%%%%%%%%%%%%%%%%%%%%%%%%%%%%%%%%%%%
\subsection{Action}%%%%%%%%%%%%%%%%%%%%%%%%%%%%%%%
%%%%%%%%%%%%%%%%%%%%%%%%%%%%%%%%%%%%%%%%%%%%%%%%%%

The $N=(2,2)$ superspace has four anti-commuting coordinates
$\theta^\pm$ and $\bar{\theta}^\pm$, and they are
related by hermitian conjugation as $(\theta^\pm)^\dag=\bar{\theta}^\mp$.
The action of $N=2$ Liouville theory on a flat Euclidean
worldsheet is given by
\begin{equation}
  I =\frac{1}{8\pi}\int d^2z
   d\theta^+d\theta^-d\bar{\theta}^+d\bar{\theta}^-\Phi\bar\Phi
  +\frac{\mu}{2\pi}      \int d^2z d\theta^+d\bar{\theta}^+
   e^{-\sqrt{\frac{k}{2}}\Phi}
  +\frac{\bar{\mu}}{2\pi}\int d^2z d\theta^-d\bar{\theta}^-
   e^{-\sqrt{\frac{k}{2}}\bar{\Phi}},
\end{equation}
where $\Phi$ is a chiral superfield satisfying
\begin{eqnarray}
  \left(\frac{\partial}{\partial\theta^-}
       -i\theta^+\partial\right)\Phi
  ~=~
  \left(\frac{\partial}{\partial\bar{\theta}^-}
       -i\bar{\theta}^+\bar{\partial}\right)\Phi
  &=&
  0,
 \nn \\
  \left(\frac{\partial}{\partial\theta^+}
       -i\theta^-\partial\right)\bar{\Phi}
  ~=~
  \left(\frac{\partial}{\partial\bar{\theta}^+}
       -i\bar{\theta}^-\bar{\partial}\right)\bar{\Phi}
  &=&
  0,
\end{eqnarray}
and obeying the $\theta$-expansion:
\begin{eqnarray}
 \Phi &=&
  \phi+i\sqrt{2}\theta^+\psi_+ +i\sqrt{2}\bar{\theta}^+\bar{\psi}_+
  +2\theta^+\bar{\theta}^+F+\cdots
 \nn \\
 \bar{\Phi} &=&
   \bar{\phi}+i\sqrt{2}\theta^-\psi_- +i\sqrt{2}\bar{\theta}^-\bar{\psi}_-
  +2\theta^-\bar{\theta}^-\bar{F}+\cdots.
\end{eqnarray}
Writing in components the action becomes, up to total derivatives,
\begin{eqnarray}
  I &=& \frac{1}{2\pi}\int d^2z
  \left[
   \tfrac{1}{2}\partial\phi\bar{\partial}\bar{\phi}
  +\tfrac{1}{2}\bar{\partial}\phi\partial\bar{\phi}
  +i\psi_+\bar{\partial}\psi_-
  +i\bar{\psi}_+\partial\bar{\psi}_-
  -F\bar{F}
  \right]
 \nn \\
 &&
 - \frac{\mu}{2\pi}\int d^2z
   (k\psi_+\bar{\psi}_+-\sqrt{2k}F)
   e^{-\sqrt{\frac{k}{2}}\phi}
 - \frac{\bar{\mu}}{2\pi}\int d^2z
   (k\psi_-\bar{\psi}_--\sqrt{2k}\bar{F})
   e^{-\sqrt{\frac{k}{2}}\bar{\phi}}.
\end{eqnarray}

In analyzing supersymmetric field theories, we usually integrate
out the auxiliary fields such as $F$ and $\bar{F}$ here to obtain
the action written in terms of dynamical fields only.
After eliminating the auxiliary fields we obtain the exponential
potential for the real part of $\phi$, so the theory describes
the dynamics of strings in the presence of Liouville-like potential wall.
However, to make use of the calculational techniques in CFT we would
rather not integrate over $F,\bar{F}$ first.
Generic vertex operators are therefore local functionals of dynamical
fields as well as $F,\bar{F}$.
Since the auxiliary fields give contact interaction, under some
restriction on the momenta of vertex operators we may simply neglect
their contributions to the correlation functions.
We can also see that by simply putting $F=\bar{F}=0$ the action
reduces to the system of free fields with $N=2$ superconformal symmetry
perturbed by exponential operators
\begin{eqnarray}
  I &=& \frac{1}{2\pi}\int d^2z
  \left[
   \partial\rho\bar{\partial}\rho -\sqrt{\tfrac{2}{k}}\tfrac{R\rho}{4}
  +\partial\theta\bar{\partial}\theta
  +\psi_+\bar{\partial}\psi_-
  +\bar{\psi}_+\partial\bar{\psi}_-
  \right]
 \nn \\
 &&
 + \frac{ik\mu}{2\pi}\int d^2z
   \psi_+\bar{\psi}_+e^{-\sqrt{\frac{k}{2}}\phi}
 + \frac{ik\bar{\mu}}{2\pi}\int d^2z
   \psi_-\bar{\psi}_-e^{-\sqrt{\frac{k}{2}}\bar{\phi}}.
\label{SLTac}
\end{eqnarray}
Here we introduced the real bosons $\rho, \theta$ by
$\phi= \rho+i\theta$, and redefined the fermions so as to
satisfy the canonical OPEs
\begin{equation}
  \rho(z)\rho(w)     \sim  \theta(z)\theta(w) \sim -\ln|z-w|^2, ~~~
  \psi_+(z)\psi_-(w) \sim  \frac{2}{z-w},
\end{equation}
The system of free fields defined by the first line of the action
(\ref{SLTac}) represents a $N=2$ superconformal algebra with the
central charge $\hat{c}=\frac{c}{3}=1+\frac{2}{k}$:
\begin{eqnarray}
  T&=& -\tfrac{1}{2}(\partial\rho\partial\rho
           +\sqrt{\tfrac{2}{k}}\partial^2\rho
           +\partial\theta\partial\theta)
       -\tfrac{1}{4}(\psi_+\partial\psi_-+\psi_-\partial\psi_+),
 \nn \\
  \sqrt{2}T_F^+
  &=& i\psi_+\partial\bar{\phi} +i\sqrt{\tfrac{2}{k}}\partial\psi_+,
 \nn \\
  \sqrt{2}T_F^-
  &=& i\psi_-\partial\phi +i\sqrt{\tfrac{2}{k}}\partial\psi_-,
 \nn \\
  J &=& \tfrac{1}{2}\psi_+\psi_- +i\sqrt{\tfrac{2}{k}}\partial\theta,
\end{eqnarray}
and the interaction terms in the second line commute with
these currents.
From this we see that by simply dropping the auxiliary fields
we still have a superconformal symmetry.
As in the $N=0$ and $N=1$ Liouville theories, the interaction
terms screen the momentum along $\rho$ or $\theta$ directions and
therefore the momentum along these directions does not conserve.
For this reason we sometimes refer to these interaction terms
as {\it screening operators}.
The easiest way to calculate various quantities
is therefore to restrict first the momenta of vertex operators
so that the contact terms do not contribute, and then make
the analytic continuation in the momenta.
Some quantities are expressed as correlators of free fields
with some screening operators inserted.
As we will illustrate later in a few examples, the role of
auxiliary fields is to cancel some of the divergences that arise
in naive screening integral expressions.
For more detailed discussions on these matters, see
\cite{Green-S, Dine-S, DiFrancesco-K}.

From the viewpoint of free CFT perturbed by screening operators,
there is another screening operator which can be written as a
D-term invariant
\begin{eqnarray}
\lefteqn{
  \int d^2z d\theta^+d\theta^-d\bar{\theta}^+d\bar{\theta}^-
   \exp\left(-\tfrac{1}{\sqrt{2k}}(\Phi+\bar{\Phi})\right)
}
 \nn \\
 &=&
   \sqrt{\tfrac{2}{k}}
    \partial\bar{\partial}\rho
    e^{-\sqrt{\frac{2}{k}}\rho}
 +  \tfrac{i}{k}\left\{
    \psi_+\bar{\partial}\psi_-
   -\bar{\partial}\psi_+\psi_-
   +\bar{\psi}_+\partial\bar{\psi}_-
   -\partial\bar{\psi}_+\bar{\psi}_-
  \right\}
    e^{-\sqrt{\frac{2}{k}}\rho}
 \nn \\
 &&
 +  \tfrac{1}{k^2}\left\{
   -2kF\bar{F}
   +\sqrt{2k}\psi_+\bar{\psi}_+\bar{F}
   +\sqrt{2k}\psi_-\bar{\psi}_-F
  \right\}
    e^{-\sqrt{\frac{2}{k}}\rho}
 \nn \\
 &&
 +\tfrac{1}{k^2}
   (\psi_+\psi_- -\sqrt{2k}\partial\theta)
   (\bar{\psi}_+\bar{\psi}_- -\sqrt{2k}\bar{\partial}\theta)
    e^{-\sqrt{\frac{2}{k}}\rho}.
\label{Dscr}
\end{eqnarray}
The first two lines are neglected since they only give
contact interactions, and the last line gives,
after canonical normalization of fermions, the following screening
operator:
\begin{equation}
   (\psi_+\psi_- -i\sqrt{2k}\partial\theta)
   (\bar{\psi}_+\bar{\psi}_- -i\sqrt{2k}\bar{\partial}\theta)
    e^{-\sqrt{\frac{2}{k}}\rho}.
\end{equation}

It is useful to bosonize the fermions in terms of a compact boson $H$:
\begin{equation}
        \psi_\pm = \sqrt{2}e^{\pm iH_L},~~~
  \bar{\psi}_\pm = \sqrt{2}e^{\pm iH_R}.
\end{equation}
where the suffices $L,R$ indicate the holomorphic and anti-holomorphic parts.
The two screening operators that were in the original action
are rewritten as follows:
\begin{equation}
  \mu S + \bar\mu\bar S ~=~
 - \frac{k\mu}{\pi}      \int d^2z e^{-\sqrt{\frac{k}{2}}\phi +iH}
 - \frac{k\bar{\mu}}{\pi}\int d^2z e^{-\sqrt{\frac{k}{2}}\bar{\phi} -iH}.
\end{equation}

%%%%%%%%%%%%%%%%%%%%%%%%%%%%%%%%%%%%%%%%%%%%%%%%%%
\subsection{Vertex Operators}%%%%%%%%%%%%%%%%%%%%%
%%%%%%%%%%%%%%%%%%%%%%%%%%%%%%%%%%%%%%%%%%%%%%%%%%

As bulk operators which are inserted in the interior of
the worldsheet, we mainly consider those of the form
\begin{equation}
  V^{j(s,\bar{s})}_{m,\bar{m}} \equiv
 \exp\left[\sqrt{\tfrac{2}{k}}\left\{j\rho
          +i(m+s)\theta_L+i(\bar{m}+\bar{s})\theta_R\right\}
          +isH_L         +i\bar{s}H_R\right].
\end{equation}
The labels $(s,\bar s)$ determines the monodromy of fermions
$(\psi_\pm,\bar\psi_\pm)$ around this operator.
In particular, NS sector corresponds to $s,\bar{s}\in\mZ$ and the
R sector to $s,\bar{s}\in\mZ+\tfrac{1}{2}$.
These labels are also regarded as the amounts of {\it spectral flow}
explained later.

For the correlators of these vertex operators to be
calculable perturbatively, the interaction terms must be
single-valued around them.
This gives the constraint $m-\bar{m}\in\mZ$.
Hereafter we shall restrict our attention to such operators
and call them {\it perturbatively well-defined}.
Note that this is not enough to ensure the mutual locality
of these vertex operators.

The $\theta$ corresponds to the phase of the chiral field
$\exp(-\sqrt{k/2}\Phi)$, and it has the periodicity
$2\pi\sqrt{2/k}$.
The periodicity can also be read off from the behavior
of $\theta$ around a perturbatively well-defined operator,
\begin{equation}
  \theta(z)V^j_{m\bar{m}}(0)\sim
  -i\sqrt{\tfrac{2}{k}}(m\ln z+\bar{m}\ln \bar{z})V^j_{m\bar{m}}(0),
\end{equation}
and $m-\bar m$ therefore corresponds to the winding number along
$\theta$-direction.
The $\theta$-momentum $\frac{1}{\sqrt{2k}}(m+\bar{m})$ should
also be quantized in unit of $\sqrt{\frac{k}{2}}$.
We thus have the quantization law
\begin{equation}
  m,\bar{m} = \frac{kn\pm w}{2}.~~(n,w\in \mZ)
\end{equation}
Physical spectrum of {\it closed string states} i.e. the states on
a circle, should obey this condition. But we sometimes consider bulk
operators not satisfying this in the calculation of correlators on
the sphere or the disc.
The same argument for operators with nonzero $s,\bar s$
leads to the quantization law
\begin{equation}
  m-\bar{m}\in\mZ,~~~
  m+s+\bar{m}+\bar{s}\in k\mZ.
\label{quant}
\end{equation}

%%%%%%%%%%%%%%%%%%%%%%%%%%%%%%%%%%%%%%%%%%%%%%%%%%%%%%%%%%%%%%%%%%%%
\subsection{Representations of $\bf N=2$ Superconformal Algebra}%%%%
%%%%%%%%%%%%%%%%%%%%%%%%%%%%%%%%%%%%%%%%%%%%%%%%%%%%%%%%%%%%%%%%%%%%

The $N=2$ superconformal algebra is generated by the currents
$T,T_F^\pm$ and $J$ satisfying the OPEs
\begin{equation}
\begin{array}{rcl}
 T(z)T(0) &\sim& \frac{3\chat}{2z^4}+\frac{2T(0)}{z^2}+\frac{\partial T(0)}{z},\\
 T(z)T_F^\pm(0) &\sim& \frac{3T_F(0)}{2z^2}+\frac{\partial T_F(0)}{z},\\
 T_F^+(z)T_F^-(0)&\sim& \frac{2\chat}{z^3}+\frac{2J(0)}{z^2}
    +\frac{2T(0)}{z}+\frac{\partial J(0)}{z},
\end{array}
\begin{array}{rcl}
  T(z)J(0) &\sim& \frac{J(0)}{z^2}+\frac{\partial J(0)}{z},\\
  J(0)T_F^\pm(0)&\sim& \pm\frac{T_F^\pm(0)}{z}, \\
  J(z)J(0)&\sim& \frac{\chat}{z^2}.
\end{array}
\end{equation}
If we define their modes as follows
\begin{equation}
  T(z)=\sum_{n\in\mZ} L_nz^{-n-2},~~~
  T_F^\pm(z)=\sum_{r\in\mZ\pm\alpha+1/2} G_r^\pm z^{-n-3/2},~~~
  J(z)=\sum_{n\in\mZ} J_nz^{-n-1},
\end{equation}
they obey the (anti-)commutation relations
\begin{equation}
\begin{array}{rcl}
  \left[L_m,L_n\right] &=& (m-n)L_{m+n}+\frac{\chat}{4}(m^3-m)\delta_{m+n,0},\\
  \left[L_m,G_n^\pm\right] &=& \left(\tfrac{m}{2}-n\right)G_{m+n}^\pm,\\
  \left\{G_m^+,G_n^-\right\}
   &=& 2L_{m+n}+(m-n)J_{m+n}+\chat(m^2-\tfrac{1}{4})\delta_{m+n,0},
\end{array}
\begin{array}{rcl}
  \left[L_m,J_n\right] &=& -nJ_{m+n}, \\
  \left[J_m,G_n^\pm\right] &=& \pm G_{m+n}^\pm, \\
  \left[J_m,J_n\right] &=& \chat m\delta_{m+n,0}.
\end{array}
\label{SCA}
\end{equation}
NS and R algebras are labelled by $\alpha=0$ and $\alpha=\frac12$,
respectively.
For {\it open string states}, i.e. states on a strip, we will have
to consider other algebras labelled by arbitrary real $\alpha$.
Such algebras are related to one another by spectral flow:
\begin{eqnarray}
  U^{-\alpha} L_n U^\alpha &=&
  L_n +\alpha J_n +\tfrac{\alpha^2\chat}{2}\delta_{m+n,0},
 \nn \\
  U^{-\alpha} G_n^\pm U^\alpha &=& G_{n\pm\alpha}^\pm,
 \nn \\
  U^{-\alpha} J_n U^\alpha &=& J_n+\alpha\chat\delta_{n,0}.
\end{eqnarray}
The spectral flows labelled by $\alpha\in\mZ$ are automorphisms
of the $N=2$ superconformal algebra.

To any vertex operator there corresponds a representation of 
superconformal algebra.
Let us take as an example the bulk operator introduced
in the previous subsection and focus on its left-moving part:
\begin{equation}
  V^{j(s)}_{m}(z) =
   \exp\left[\sqrt{\tfrac{2}{k}}\left\{
         j\rho_L +i(m+s)\theta_L\right\} +isH_L  \right].
\end{equation}
It corresponds to the state $\ket{j,m,s}$ which has
the $L_0$ and $J_0$ eigenvalues $(h,Q)$:
\begin{equation}
  h = \frac{(m+s)^2-j(j+1)}{k} +\frac{s^2}{2},~~~
  Q = \frac{2(m+s)}{k}+s,
\end{equation}
and is annihilated by $G^+_{r\ge\frac{1}{2}-s}$ and
$G^-_{r\ge\frac{1}{2}+s}$.
The operators with $s=0$ correspond to NS sector primary states,
and $s$ represents the amount of spectral flow:
\begin{equation}
  \ket{j,m,s}=U^s\ket{j,m,0}.
\end{equation}
The action of supercurrents on them reads
\begin{eqnarray}
 T_F^\pm(z)V^{j(s)}_m(0)
 &\sim& -i\sqrt{\tfrac{2}{k}}(j\pm m)z^{\pm s-1}
 V^{j(s\pm1)}_{m\mp1}(0).
\label{GVope}
\end{eqnarray}
The states with $j=\mp m$ are annihilated by $G^\pm_{\pm s-\frac{1}{2}}$.
They are (anti-)chiral primary states spectral flowed by $s$ units.
The above formula also shows that the two highest weight representations
are related by an integer spectral flow when their $m$ labels
differ by an integer.

%%%%%%%%%%%%%%%%%%%%%%%%%%%%%%%%%%%%%%%%%%%%%%%%%%
\subsubsection{Degenerate representations}%%%%%%%%
%%%%%%%%%%%%%%%%%%%%%%%%%%%%%%%%%%%%%%%%%%%%%%%%%%

As can be read off from the determinant formula of \cite{Boucher-FK,Nam},
the Verma module of NS algebra labelled by conformal weight $h$
and R-charge $Q$ contains a null vector when
\begin{equation}
 f_{r,s}(h,Q)\equiv
 2(\chat-1)h-Q^2-\tfrac{1}{4}(\chat-1)^2+\tfrac{1}{4}\{(\chat-1)r+2s\}^2=0
 ~~~(r,s\in\mZ_{>0}),
\label{deg1}
\end{equation}
and the null vector appears at level $rs$.
It follows from this that NS vertex operator $V^j_m$
corresponds to a degenerate representation when
\begin{equation}
  2j+1 = \pm\left(r+ks\right). ~~~(r,s\in\mZ_{>0})
\end{equation}
The most fundamental degenerate operators within this category are
those with $(r,s)=(1,1)$ or $j=\frac{k}{2}$.
According to \cite{Boucher-FK} the determinant also vanishes when
\begin{equation}
 g_p(h,Q)\equiv
  2h-2pQ+(\chat-1)(p^2-\tfrac{1}{4})=0  ~~~~ (p\in\mZ+\tfrac{1}{2}).
\label{deg2}
\end{equation}
In the simplest case $p=\pm\frac{1}{2}$ we have $2h=\pm Q$.
For generic $p$ we can easily find a null vector $\chi$ of the form
\begin{eqnarray}
  (p>0) && \ket{\chi}~=~  G^+_{-p}\cdots G^+_{-3/2}G^+_{-1/2}\ket{h,Q},
  \nn \\
  (p<0) && \ket{\chi}~=~\;G^-_{ p}\cdots G^-_{-3/2}G^-_{-1/2}\ket{h,Q},
\end{eqnarray}
so they are chiral representations spectral flowed by $p-\frac12$
units, or anti-chiral representations spectral flowed by $p+\frac12$ units.
In particular, the representation has two null vectors\footnote{
 Note that the relevant Verma module contains more than two null vectors.
 For example, the Verma module over the NS ground state has three
 null vectors, $G_{-1/2}^\pm\ket{0}$ and $L_{-1}\ket{0}$.
}
if $(h,Q)$ are such that there are two different
values of $p$ satisfying (\ref{deg2}).
In terms of the labels $(j,m)$ the condition is simply
\begin{equation}
 (  j  \pm m\in \mZ_{\ge 0})~~{\rm or}~~
 ( -j-1\pm m\in \mZ_{\ge 0}).
\end{equation}
Though we will not explain in detail, classification of representations
of $SL(2,\mR)$ current algebra at level $k+2$ is also known, and
degenerate representations appear precisely in the same manner
as those of $N=2$ superconformal algebra.
For example, The representations with two null vectors correspond
to finite dimensional representations of $SL(2,\mR)$.

%%%%%%%%%%%%%%%%%%%%%%%%%%%%%%%%%%%%%%%%%%%%%%%%%%
\subsubsection{Unitary representations}%%%%%%%%%%%
%%%%%%%%%%%%%%%%%%%%%%%%%%%%%%%%%%%%%%%%%%%%%%%%%%

In \cite{Boucher-FK} the conditions for unitary representations were
also given.
For $k>0$ there are two classes of unitary representations of NS algebra.
The first ones satisfy 
\begin{equation}
  g_p(h,Q)\ge 0 \mbox{ ~~ or~  }
 (j+\tfrac{1}{2})^2\le (m-p)^2
  \mbox{ ~~ for all~ }p\in\mZ+\tfrac{1}{2}.
\end{equation}
The representations with $j\in-\frac{1}{2}+i\mR$ are therefore
all unitary irrespective of the value of $m$.
There are also some unitary representations with $-1<j<0$
in this class, depending on the value of $m$.
The second ones satisfy
\begin{equation}
  g_p(h,Q)=0,~~
  g_{p+{\rm sgn}(p)}(h,Q)<0~~\mbox{and}~
  f_{1,1}(h,Q)\ge 0.
\end{equation}
In terms of $j,m$ this condition becomes, up to
$j\leftrightarrow -j-1$ equivalence,
\begin{equation}
  -\tfrac{k}{2}-1<j<-\tfrac{1}{2}~~{\rm and}~~
  \pm m\in j-\mZ_{\ge 0}
\end{equation}
Hereafter we will use the term discrete/continuous
representations for these two classes of representations.
The bound for $j$ is called the {\it unitarity bound},
and we actually expect a little more stringent bound for
discrete series from recent works.
This can also be understood from the reflection relation
for chiral operators sending $j$ to $-j-1-\frac k2$,
which we will explain later.

%%%%%%%%%%%%%%%%%%%%%%%%%%%%%%%%%%%%%%%%%%%%%%%%%%
\subsection{Perturbed Linear Dilaton CFT}%%%%%%%%%
%%%%%%%%%%%%%%%%%%%%%%%%%%%%%%%%%%%%%%%%%%%%%%%%%%

The $N=2$ Liouville theory can be analyzed as a linear dilaton
theory (free CFT) with exponential type perturbations.
As was found in \cite{Goulian-L}, the correlators of such
theories are calculable simply as Wick contractions of free CFT
with a certain number of screening operators inserted.

The simplest example of such theories is the bosonic Liouville theory,
defined by the action
\begin{equation}
  I = \frac{1}{8\pi}\int d^2x\sqrt{g}
  \left[g^{mn}\partial_m\phi\partial_n\phi +\sqrt{2}QR\phi
       +8\pi\mu e^{\sqrt{2}b\phi}\right],~~~
  Q=b+b^{-1}.
\end{equation}
We are interested in the correlators typically of the form
\begin{equation}
 \vev{\prod_ie^{\sqrt{2}\alpha_i\phi(z_i)}}
 = \int {\cal D}\phi e^{-I}\prod_ie^{\sqrt{2}\alpha_a\phi(z_i)}.
\end{equation}
Extracting the dependence on the zero-mode of $\phi$ we find,
on a worldsheet with genus $g$, the following integral:
\begin{eqnarray}
\lefteqn{
  \int d\phi_0
  \exp\left[\sqrt{2}\{\sum_i\alpha-Q(1-g)\}\phi_0
      -e^{\sqrt{2}b\phi_0}\mu\int d^2x e^{\sqrt{2}b(\phi-\phi_0)}\right]
}
 \nn \\
 &=&
  \frac{\Gamma(-N)}{\sqrt{2}b}
  \left\{\mu\int d^2x e^{\sqrt{2}b(\phi-\phi_0)}\right\}^N.
  ~~~(bN=Q(1-g)-\sum\alpha_i)
\end{eqnarray}
$N$ is the number of screening operators necessary to
cancel the momentum carried by vertices and also by the background
with genus $g$.
Since the integration over non-zero mode of $\phi$ is equivalent to
taking the Wick contraction using the free correlator, we obtain
the following formal expression
\begin{equation}
 \vev{\prod_ie^{\sqrt{2}\alpha_i\phi(z_i)}}
 =\frac{\Gamma(-N)}{\sqrt{2}b}
 \vev{\prod_ie^{\sqrt{2}\alpha_i\phi(z_i)}(\mu S)^N}_{\rm free},
 ~~~\mu S=\mu\int d^2x e^{\sqrt{2}b\phi}.
\end{equation}
This shows that the correlator diverges when the total momentum
of vertices and the background can be cancelled by a non-negative
integer insertion of screening operators,
\begin{equation}
  Q(1-g)-\sum\alpha_i \in b\mZ_{\ge 0},
\end{equation}
and the residue of such divergences is given by the Wick contraction
of free fields.
More explicitly, by rewriting the Gamma function as a sum of simple
poles we obtain an expression
\begin{equation}
 \vev{\prod_ie^{\sqrt{2}\alpha_i\phi(z_i)}}
 \simeq
 \sum_{n\ge 0}\frac{1}{\sqrt{2}}\frac{1}{nb+\Sigma\alpha_i-Q(1-g)}
 \vev{\prod_ie^{\sqrt{2}\alpha_i\phi(z_i)}
      \frac{(-\mu S)^n}{n!}}_{\rm free},
\end{equation}
which well approximates the behavior of correlators near the poles
$Q(1-g)-\sum\alpha_i\in b\mZ_{\ge 0}$.
 
The above argument applies also to the $N=2$ Liouville theory.
Let us consider the correlator
\begin{equation}
  \vev{\prod_iV^{j_i(s_i,\bar{s}_i)}_{m_i,\bar{m}_i}(z_i)}
 =\int{\cal D}\rho{\cal D}\theta{\cal D}\psi
  e^{-I}\prod_iV^{j_i(s_i,\bar{s}_i)}_{m_i,\bar{m}_i}(z_i).
\end{equation}
In this theory we have two screening operators (with couplings $\mu$
and $\bar{\mu}$) in the defining action, so let us expand into power
series in $\mu$ and then integrate over the zero mode of $\rho$.
Then we finally obtain
\begin{equation}
  \vev{\prod_iV^{j_i(s_i,\bar{s}_i)}_{m_i,\bar{m}_i}(z_i)}
 \simeq \sqrt{\frac{k}{2}}
  \sum_{n,\bar{n}\ge0}
  \frac{1}{n!\bar{n}!}
  \frac{
   \vev{\prod_iV^{j_i(s_i,\bar{s}_i)}_{m_i,\bar{m}_i}(z_i)
        (-\mu S)^n(-\bar\mu\bar S)^{\bar{n}} }_{\rm free} } 
       {-\sum_ij_i-1+g+\frac{k}{2}(n+\bar{n})},
\label{cor1}
\end{equation}
which well approximates the behavior of correlators near the
poles $\sum j_i +1-g\in\frac{k}{2}\mZ_{\ge 0}$.

%%%%%%%%%%%%%%%%%%%%%%%%%%%%%%%%%%%%%%%%%%%%%%%%%%
\subsubsection{Three-point function}%%%%%%%%%%%%%%
%%%%%%%%%%%%%%%%%%%%%%%%%%%%%%%%%%%%%%%%%%%%%%%%%%

Using this formula we calculate the three-point function
of operators $V^{j(s,\bar{s})}_{m\bar{m}}$.
We first evaluate the residues of the poles corresponding to
integer insertions of screening operators, and then obtain
the correlator by some kind of {\it extrapolation}.
Similar calculations were performed for bosonic Liouville theory
in \cite{Dorn-O,Zamolodchikov-Z1} and $N=1$ Liouville theory
in \cite{Rashkov-S,Poghosian}.

The residues of the poles in (\ref{cor1}) are given by the
Wick contraction:
\begin{equation}
   \frac{1}{n!\bar{n}!}
   \vev{\prod_{i=1}^3V^{j_i(s_i,\bar{s}_i)}_{m_i,\bar{m}_i}(z_i)
       (-\mu S)^n(-\bar\mu\bar S)^{\bar{n}}}_{\rm free}.
\end{equation}
In order to account for the anti-commutativity of Grassmann odd
quantities, we need to include cocycle factors in doing
the contraction.
We therefore assign the factor
\begin{equation}
 \exp\left(
 \frac{i\pi}{k}((m_1+s_1)(\bar{m}_2+\bar{s}_2)-(m_2+s_2)(\bar{m}_1+\bar{s}_1))
+\frac{i\pi}{2}(s_1\bar{s}_2-s_2\bar{s}_1)
 \right)
\label{cocycle}
\end{equation}
in contracting the product
 $V^{j_1(s_1\bar{s}_1)}_{m_1\bar{m}_1}
  V^{j_2(s_2\bar{s}_2)}_{m_2\bar{m}_2}$.
This ensures that the operators are simply commuting or anti-commuting
according to their Grassmann parity if they satisfy (\ref{quant})
as well as $s,\bar s\in\mZ$.
In particular, the positions of screening operators
do not matter in calculating correlation functions.
Note also that the momentum conservation of
linear dilaton theory requires
\begin{equation}
  \Sigma j_a +1     =\tfrac{k}{2}(n+\bar{n}),~~~
  \Sigma m_a        =
  \Sigma \bar{m}_a  =(1+\tfrac{k}{2})(n-\bar{n}),~~~
  \Sigma s_a        =
  \Sigma \bar{s}_a  =-n+\bar{n}.
\end{equation}
After the Wick contraction we encounter the integral
\begin{eqnarray}
 I &=&
    \frac{(-)^{(n-\bar{n})(m_1-\bar{m}_1)}}{n!\bar{n}!}
  \int
  \prod_{i=1}^nd^2z_i
    z_i^{j_1-m_1}\bar{z}_i^{j_1-\bar{m}_1}
   (1-z_i)^{j_2-m_2}(1-\bar{z}_i)^{j_2-\bar{m}_2}
 \nn \\ && \times
  \prod_{\hat{i}=1}^{\bar{n}}d^2z_{\hat{i}}
    z_{\hat{i}}^{j_1+m_1}\bar{z}_{\hat{i}}^{j_1+\bar{m}_1}
    (1-z_{\hat{i}})^{j_2+m_2}(1-\bar{z}_{\hat{i}})^{j_2+\bar{m}_2}
  \prod_{i<j}|z_{ij}|^2
  \prod_{\hat{i}<\hat{j}}|z_{\hat{i}\hat{j}}|^2
  \prod_{i,\hat{j}}|z_{i\hat{j}}|^{-2k-2},
\end{eqnarray}
which is calculated using the formula\footnote{
 Some typos by unnecessary sign factors there are corrected here.}
in \cite{Fukuda-H1}.
It is non-vanishing only when $n-\bar{n}=\pm1$ or $0$.
\begin{eqnarray}
 && (n=\bar{n}+1)
  \nn \\
  I &=&
 \pi^{2n-1}
 \left[\prod_{r=1}^{n-1}
       \gamma(-rk)\gamma(1+2j_1-rk)\gamma(1+2j_2-rk)\gamma(1+2j_3-rk)\right]
 F_-(j_a,m_a,\bar{m}_a),
 \nn \\
\lefteqn{
 F_\pm(j_a,m_a,\bar{m}_a)
 ~=~
  (-)^{m_2-\bar{m}_2}
\frac{\Gamma(1+j_1\pm m_1)\Gamma(1+j_2\pm m_2)\Gamma(1+j_3\pm m_3)}
    {\Gamma(-j_1\mp\bar{m}_1)\Gamma(-j_2\mp\bar{m}_2)\Gamma(-j_3\mp\bar{m}_3)},
} \nn \\
 && (n=\bar{n})
  \nn \\
  I &=&
 \pi^{2n}
 \left[
 \prod_{r=1}^{n-1}
 \gamma(-rk)\gamma(1+2j_1-rk)\gamma(1+2j_2-rk)\gamma(1+2j_3-rk)
 \right]
 F(j_a,m_a,\bar{m}_a),
 \nn \\
\lefteqn{
 F(j_a,m_a,\bar{m}_a) = \pi^{-2}
 \int d^2z d^2w
 z^{j_1-m_1}\bar{z}^{j_1-\bar{m}_1}(1-z)^{j_2-m_2}(1-\bar{z})^{j_2-\bar{m}_2}
}
 \nn \\ &&
 \hskip14mm\times
 w^{j_1+m_1}\bar{w}^{j_1+\bar{m}_1}(1-w)^{j_2+m_2}(1-\bar{w})^{j_2+\bar{m}_2}
 |z-w|^{-4-2(j_1+j_2+j_3)}.
\label{FH}
\end{eqnarray}
where $\gamma(x)=\frac{\Gamma(x)}{\Gamma(1-x)}$.
Using $k=b^{-2}$ and the special function $\Upsilon$ introduced
in \cite{Zamolodchikov-Z1} (see the appendix for the definition) we can rewrite
the products of $\gamma$ functions as follows:
\begin{eqnarray}
\lefteqn{ n=\bar{n}+1=b^2(j_{1+2+3}+1)+\tfrac{1}{2}: }
  \nn \\
 I &=&
  F_-(j_a,m_a,\bar{m}_a)
  \frac{\pi^{2n-1}b^{2(1+k)(n-1)}\Upsilon'(0)
        \Upsilon(b(2j_1+1))\Upsilon(b(2j_2+1))\Upsilon(b(2j_3+1))}
       {\Upsilon'(\frac{1}{b}(1-n))
        \Upsilon(\frac{1}{2b}+bj_{1-2-3})
        \Upsilon(\frac{1}{2b}+bj_{2-3-1})
        \Upsilon(\frac{1}{2b}+bj_{3-1-2})},
 \nn \\
\lefteqn{ n=\bar{n}= b^2(\Sigma j_a+1): }
  \nn \\
  I &=&
  F(j_a,m_a,\bar{m}_a)
  \frac{\pi^{2n}b^{2(n-1)}\Upsilon'(0)
        \Upsilon(b(2j_1+1))\Upsilon(b(2j_2+1))\Upsilon(b(2j_3+1))}
       {\Upsilon'(\frac{1}{b}(1-n))
        \Upsilon(\frac{1}{b}+bj_{1-2-3})
        \Upsilon(\frac{1}{b}+bj_{2-3-1})
        \Upsilon(\frac{1}{b}+bj_{3-1-2})}.
\end{eqnarray}
The derivative of $\Upsilon$ function appears because we re-wrote
a product of $\gamma$ functions by the ratio of $\Upsilon$
functions both are vanishing.
The function $\Upsilon(\frac{1}{b}(1-n))$ vanishes precisely
at the poles of the three-point function appearing in (\ref{cor1}).
So we combine the sum over poles into a single function using
the {\it extrapolation}
\begin{equation}
  \sum_{a\in\mbox{\scriptsize \{simple zeroes of }F(x)\}}
  \frac{1}{(x-a)F'(a)}\simeq \frac{1}{F(x)}.
\end{equation}
Combining with other factors we obtain
\begin{eqnarray}
\lefteqn{
 \vev{\tprod_{a=1}^3V^{j_a(s_{a}\bar{s_a})}_{m_a\bar{m}_a}(z_a)}
}
 \nn \\
 &=& \prod_{\{abc\}=\{321\},\{312\},\{213\}}
   z_{ab}^{h_c-h_a-h_b}
   \bar{z}_{ab}^{\bar{h}_c-\bar{h}_a-\bar{h}_b}
    e^{\frac{i\pi}{k}
     \{(m_a+s_a)(\bar{m}_b+\bar{s}_b)
      -(m_b+s_b)(\bar{m}_a+\bar{s}_a)\}
      +\frac{i\pi}{2}
     (s_a\bar{s}_b-s_b\bar{s}_a)}
 \nn \\ && \times
 \left\{
 \delta^2(\Sigma_i m_i-\tfrac{k\chat}{2})\delta^2(\Sigma_i s_i+1)
  F_-(j_a,m_a,\bar{m}_a)D_-
  \right. \nn \\ && \left. ~
 +\delta^2(\Sigma_i m_i+\tfrac{k\chat}{2})\delta^2(\Sigma_i s_i-1)
  F_+(j_a,m_a,\bar{m}_a)D_+
 +\delta^2(\Sigma_i m_i)\delta^2(\Sigma_i s_i)
  F(j_a,m_a,\bar{m}_a)D_0
 \right\},
 \nn
\end{eqnarray}
\begin{eqnarray}
 D_\pm &=&
  \frac{(\nu b^{2-2b^2})^{j_{1+2+3}+1}\Upsilon'(0)
        \Upsilon(b(2j_1+1))\Upsilon(b(2j_2+1))\Upsilon(b(2j_3+1))}
       {\sqrt{2}b^{1+k}\Upsilon(\frac{1}{2b}-b(j_{1+2+3}+1))
        \Upsilon(\frac{1}{2b}+bj_{1-2-3})
        \Upsilon(\frac{1}{2b}+bj_{2-3-1})
        \Upsilon(\frac{1}{2b}+bj_{3-1-2})},
 \nn \\
  D_0 &=&
   \frac{(\nu b^{-2b^2})^{j_{1+2+3}+1}\Upsilon'(0)
         \Upsilon(b(2j_1+1))\Upsilon(b(2j_2+1))\Upsilon(b(2j_3+1))}
       {\sqrt{2}b^2\Upsilon(\frac{1}{b}-b(j_{1+2+3}+1))
        \Upsilon(\frac{1}{b}+bj_{1-2-3})
        \Upsilon(\frac{1}{b}+bj_{2-3-1})
        \Upsilon(\frac{1}{b}+bj_{3-1-2})},
\end{eqnarray}
where we set $\mu=\bar\mu=\nu^{\frac k2}$ using the translation
along $\theta$-direction.

The three-point structure constants have much more poles than the
path-integral formula predicts.
Some of the poles can be accounted for by incorporating
another screening operator of D-term type given in (\ref{Dscr}).
The poles in the $\rho$-momentum space at
\begin{equation}
  j_1+j_2+j_3+1=\tfrac{k}{2}(n+\bar{n})+m~~~~
  (n,\bar n,m\in\mZ_{\ge0})
\end{equation}
can be explained in this way.
Note that in bosonic Liouville theory we meet a similar situation,
where we have two screening operators although only one of them
is present in the defining action.
The two screening operators there are transformed to each other
by $b\leftrightarrow \frac{1}{b}$ flip, so it follows that the theory
has the strong/weak coupling ($b\leftrightarrow \frac{1}{b}$) duality.

All the other poles are related to the ones explained above
by reflection relations which are explained later.

%%%%%%%%%%%%%%%%%%%%%%%%%%%%%%%%%%%%%%%%%%%%%%%%%%
\subsubsection{Contact interactions}%%%%%%%%%%%%%%
%%%%%%%%%%%%%%%%%%%%%%%%%%%%%%%%%%%%%%%%%%%%%%%%%%

Let us make some comments here on the role of the auxiliary fields
which we have neglected in the calculations above.
Before doing this, we would like to note that at the stage
of evaluating the screening integral using the formula (\ref{FH})
we have already assumed the analyticity.
Otherwise the validity of the formula for correlators would be
very restricted in the momentum space,
because any integrals which look like
\begin{equation}
  \int d^2x x^{\alpha}\bar{x}^{\bar{\alpha}}\cdots~~~
  (\alpha-\bar{\alpha}\in\mZ)
\label{div}
\end{equation}
diverge around $x\sim 0$ when ${\rm max}(\alpha,\bar{\alpha})\le -1$.
After the analytic continuation, one can see the divergences
as poles of the functions like $\Gamma(1+\alpha)$ or
$\Gamma(1+\bar{\alpha})$.
Assuming analyticity, away from such poles in the momentum space
we are allowed to do all the naive operation such as partial
integration.
From this viewpoint, it would not be so bad to simply discard the
contact interactions, calculate correlators in a region of momentum
space where contact terms are negligible and then analytically extend.
But of course one can treat the contact terms honestly and find
that they play the role of {\it ensuring the analyticity}.

As will be illustrated below, the auxiliary fields {\it cancel} some
of the divergences in correlators.
We begin by recalling that, when a correlator contains a primary
operator $e^{-(p\phi+\bar{p}\bar{\phi})}$, the screening integral
behaves around it as (neglecting coefficients)
\begin{equation}
  \mu S\cdot e^{-(p\phi+\bar{p}\bar{\phi})}(0)
 \sim \int d^2z |z|^{-2\sqrt{2k}\bar p},~~~~~
  \bar\mu\bar S\cdot e^{-(p\phi+\bar{p}\bar{\phi})}(0)
 \sim \int d^2z |z|^{-2\sqrt{2k}p}.
\end{equation}
So there are no divergences from integrals over origin
as long as $p,\bar{p}<\sqrt{1/2k}$.
From the superconformal symmetry, it is natural to expect
that the screening integrals containing $e^{-(p\phi+\bar{p}\bar{\phi})}$
or {\it any of its descendants} are finite for those
values of $p,\bar{p}$.
Now, let us take a descendant:
\begin{equation}
  \int d\theta^+ d\bar{\theta}^+ e^{-(p\Phi+\bar{p}\bar{\Phi})}
 = 2p(F+ip\psi_+\bar{\psi}_+)e^{-(p\phi+\bar{p}\bar{\phi})},
\end{equation}
put at $z=0$ in a certain correlator.
We should then consider the divergences of screening integrals around
this operator.
One finds the singular behavior of $\mu S$ becomes milder,
whereas that of $\bar\mu\bar S$ gets stronger, since
the latter contains the fermions with opposite charge:
\begin{equation}
  \bar\mu\bar S
 = \frac{\bar\mu}{2\pi}\int d^2z
   \left(ik\psi_-\bar{\psi}_-+\sqrt{2k}\bar F\right)
   e^{-\sqrt{\frac{k}{2}}\bar{\phi}}.
\end{equation}
So the $\bar\mu\bar S$-integral is apparently finite only for negative $p$.

Let us show that, through suitable regularization, the
$\bar\mu\bar S$-integral is actually finite for $p$ slightly above zero
due to the cancellation between contact and non-contact interactions.
The screening integral is the sum of a non-contact and a contact terms,
both of which are divergent for positive $\bar{p}$.
In order to compare the two divergences, let us introduce
the following regularization.
First, reguralize the non-contact term by cutting off
the integration domain by a hole of radius $\epsilon$.
The non-contact interaction is then evaluated as follows:
\begin{eqnarray}
\lefteqn{
    \frac{ik\bar{\mu}}{2\pi}\int d^2z
    \psi_-\bar{\psi}_-
    e^{-\sqrt{\frac{k}{2}}\bar{\phi}}(z)
   \times
   (2ip^2)\psi_+\bar{\psi}_+e^{-(p\phi+\bar{p}\bar{\phi})}(0)
} \nn \\
   &\sim&
  \frac{4kp^2\bar{\mu}}{\pi}\int_{|z|\ge \epsilon} d^2z|z|^{-2-2\sqrt{2k}p}
 = 2\sqrt{2k}p\bar{\mu}\epsilon^{-2\sqrt{2k}p}.
\end{eqnarray}
Second, separate $F$ and $e^{-(p\phi+\bar{p}\bar{\phi})}$ by spreading
$F$ along the boundary of the same hole.
The contact interaction is also regularized, and it precisely
cancels with the non-contact interaction
\begin{equation}
   \frac{\bar{\mu}}{2\pi}\int d^2z
   \sqrt{2k}\bar{F}e^{-\sqrt{\frac{k}{2}}\bar{\phi}}(z)
   \times 2p\oint_\epsilon\frac{dx}{2\pi ix}
   F(x)e^{-(p\phi+\bar{p}\bar{\phi})}(0)
  = -2\sqrt{2k}p\bar{\mu}\epsilon^{-2\sqrt{2k}p}.
\end{equation}
It is expected that the auxiliary fields play similar roles of cancelling
the unwanted divergences in other correlators,
though we will not analyze it in a systematic way.

%%%%%%%%%%%%%%%%%%%%%%%%%%%%%%%%%%%%%%%%%%%%%%%%%%
\subsubsection{Reflection relations}%%%%%%%%%%%%%%
%%%%%%%%%%%%%%%%%%%%%%%%%%%%%%%%%%%%%%%%%%%%%%%%%%

In previous subsection, we introduced the primary states
$\ket{j,m,s}$ as highest weight states of $N=2$ superconformal
algebra labelled by $s$.
From a purely representation theoretical viewpoint, there are
the following equivalence relations between them:
\begin{equation}
  \ket{j,m,s}\sim\ket{-j-1,m,s},
  ~~~
  \ket{j,\pm j,s}\sim\ket{\tilde{j},\mp\tilde{j},s\mp1}.
  ~~~
  (\tilde{j}=-j-1-\tfrac{k}{2})
\end{equation}
Therefore, as in $N=0$ and $1$ Liouville theories, we expect the
following equivalence relations between operators
\begin{eqnarray}
  V^{j(s,\bar{s})}_{m,\bar{m}} &=&
  R(j,m,\bar{m}) V^{-j-1(s,\bar{s})}_{m,\bar{m}},
\nn \\
  V^{j(s,\bar{s})}_{\pm j,\pm j} &=&
  R_\mp(j)\times
  V^{\tilde{j}(s\mp1,\bar{s}\mp1)}_{\mp\tilde{j},\mp\tilde{j}},~~~
  \tilde{j}\equiv -j-\tfrac{k}{2}-1.
\end{eqnarray}
We will refer to these relations as {\it reflection relations}
and the coefficients $R,R_\mp$ as {\it reflection coefficients}.
They should be independent of the labels $s$ and $\bar{s}$, because
they are coordinates along the $S^1$ corresponding to R-rotation
which is an exact symmetry of the theory.
$R(j,m,\bar{m})$ is easily obtained from the
three-point structure constants $D_\pm$ and $F_\pm$:
\begin{equation}
  R(j,m,\bar{m}) = -
  \nu^{2j+1}
  \frac{\Gamma(1+j+m)\Gamma(1+j-\bar{m})}
       {\Gamma(-j+m)\Gamma(-j-\bar{m})}
  \frac{\Gamma(-b^2(2j+1))\Gamma(-2j-1)}
       {\Gamma(b^2(2j+1))\Gamma(2j+1)}.
\end{equation}
This can also be obtained from $D_0$ and $F$ using the equality
\begin{equation}
 \frac{F(j_a,m_a,\bar{m}_a)}
      {F(j_a,m_a,\bar{m}_a)|_{j_1\to -j_1-1}}
 =
 \frac{\Gamma(1+j_1+m_1)\Gamma(1+j_1-\bar{m}_1)}
      {\Gamma(-j_1+m_1)\Gamma(-j_1-\bar{m}_1)}
 \gamma(j_{2-3-1})\gamma(j_{3-1-2}).
\end{equation}
$R_\pm(j)$ are obtained by using
\begin{equation}
 F(j_a,m_a,\bar{m}_a)|_{m_1=\bar{m}_1=j_1}
 =   \frac{(-)^{m_2-\bar{m}_2}\gamma(2j_1+1)
           \Gamma(1+j_2-m_2)\Gamma(1+j_3-m_3)}
          {\gamma(1+j_{1+2-3})\gamma(1+j_{1-2+3})\gamma(2+j_{1+2+3})
           \Gamma(\bar{m}_2-j_2)\Gamma(\bar{m}_3-j_3)},
\end{equation}
and taking the ratios of $FD_0$ and $F_\pm D_\pm$:
\begin{equation}
  R_\pm(j) = \nu^{2j+1+\frac{k}{2}}
  \frac{\Gamma(-b^2(2j+1))}
       {\Gamma(-b^2(2\tilde{j}+1))}.
\end{equation}
Finally, the two-point function for operators belonging to
continuous representations can be written as
\begin{eqnarray}
  \vev{V^{j_1(s_1\bar{s}_1)}_{m_1\bar{m}_1}(z_1)
       V^{j_2(s_2\bar{s}_2)}_{m_2\bar{m}_2}(z_2)}
  &=&
  z_{12}^{-2h_1}\bar{z}_{12}^{-2\bar{h}_1}
    \delta^2(m_1+m_2)
    \delta^2(s_1+s_2)
 \nn \\ && \times
    \left\{
    \delta\left(i(j_1+j_2+1)\right)
   +\delta\left(i(j_1-j_2)\right)R(j_1,m_1,\bar{m}_1)
    \right\}.
\end{eqnarray}

%%%%%%%%%%%%%%%%%%%%%%%%%%%%%%%%%%%%%%%%%%%%%%%%%%
\subsection{OPE Involving Degenerate Fields}%%%%%%
%%%%%%%%%%%%%%%%%%%%%%%%%%%%%%%%%%%%%%%%%%%%%%%%%%

The three-point structure constant and the reflection
coefficients can also be obtained from the property of
degenerate operators.
The technology was first invented in bosonic Liouville theory
in \cite{Teschner} (see also \cite{Teschner2}).

We first study the OPEs involving degenerate operators.
The operators $V^{1/2}_{M\bar{M}}$ and $V^{k/2}_{M\bar{M}}$
are the most important, because any other degenerate operators
descend from their products.
When multiplied on a generic operator $V^j_{m\bar m}$,
they should satisfy the OPE formulae
\begin{eqnarray}
  V^{\frac{1}{2}}_{M\bar{M}}(z_1)V^j_{m\bar{m}}(z_2)
 &\sim& e^{\frac{i\pi}{k}(M\bar{m}-\bar{M}m)}\sum_\pm
   z_{12}^{\frac{4Mm+1\mp(2j+1)}{2k}}
   \bar{z}_{12}^{\frac{4\bar{M}\bar{m}+1\mp(2j+1)}{2k}}
   C_\pm
   V^{j\pm\frac{1}{2}}_{m+M,\bar{m}+\bar{M}}(z_2),
 \nn \\
  V^{\frac{k}{2}}_{M\bar{M}}(z_1)V^j_{m\bar{m}}(z_2)
 &\sim& e^{\frac{i\pi}{k}(M\bar{m}-\bar{M}m)}\sum_\pm
 z_{12}^{\frac{2Mm}{k}+\frac{1\mp(2j+1)}{2}}
 \bar{z}_{12}^{\frac{2\bar{M}\bar{m}}{k}+\frac{1\mp(2j+1)}{2}}
 \tilde{C}_\pm
 V^{j\pm\frac{k}{2}}_{m+\mu,\bar{m}+\bar{\mu}}(z_2)
 \nn \\ && \hskip-30mm
    +e^{\frac{i\pi}{k}(M\bar{m}-\bar{M}m)}\sum_{\updownarrow}
 z_{12}^{\frac{2Mm}{k}+\frac{k\chat}{2}\mp(m+M)}
 \bar{z}_{12}^{\frac{2\bar{M}\bar{m}}{k}
              +\frac{k\chat}{2}\mp(\bar{m}+\bar{M})}
 \tilde{C}_{\updownarrow}
 V^{j(\pm1,\pm1)}
  _{m+M\mp\frac{k\chat}{2},\bar{m}+\bar{M}\mp\frac{k\chat}{2}}(z_2).
\end{eqnarray}
where the coefficients $C_\pm,\tilde{C}_{\pm,\updownarrow}$
are functions of $(M,\bar{M};j,m,\bar{m})$.
Remember that $j=1/2$ operator must have $M,\bar{M}=\pm1/2$
in order to belong to a degenerate representation.

Throughout the paper we use the un-tilded or tilded letters
(like  $C$ or $\tilde C$) for OPE coefficients involving
$j=1/2$ or $j=k/2$ operators.
The suffix $\pm$ indicates the channels in which $j$ quantum number
changes by $\pm1/2$ or $\pm k/2$, and $\updownarrow$ indicates
the channels where the $s$ quantum number changes by $\pm1$.

The OPE coefficients can be calculated by the standard
perturbative argument.
The idea is that these finite number of terms in OPEs are the
contributions from poles in three-point correlators, so are
calculable as ordinary Wick contractions with some insertions
of screening operators.
We should use the two screening operators contained in the
original action as perturbation terms
\begin{equation}
 \mu S +\bar\mu\bar S =
 - \frac{k\mu}{\pi}      \int d^2z e^{-\sqrt{\frac{k}{2}}\phi +iH}
 - \frac{k\bar{\mu}}{\pi}\int d^2z e^{-\sqrt{\frac{k}{2}}\bar{\phi} -iH},
\end{equation}
as well as the other (D-type) one,
\begin{equation}
 \tilde\mu\tilde S = \frac{\tilde{\mu}}{4\pi}\int d^2z
   (\psi_+\psi_- -i\sqrt{2k}\partial\theta)
   (\bar{\psi}_+\bar{\psi}_- -i\sqrt{2k}\bar{\partial}\theta)
    e^{-\sqrt{\frac{2}{k}}\rho}.
\end{equation}
The OPE coefficients are calculated as the ratios of three-point
functions and two-point functions both of which are diverging,
so that we only have to take the ratios of the residues.
For example,
\begin{eqnarray}
 C_+(M\bar{M};jm\bar{m}) &=&
  e^{-\frac{i\pi}{k}(M\bar{m}-\bar{M}m)}
 \lim\frac
 {\vev{V^{-j-3/2}_{-m-1/2,-\bar{m}+1/2}V^{1/2}_{1/2,-1/2}V^j_{m,\bar{m}}}}
 {\vev{V^{-j-3/2}_{-m-1/2,-\bar{m}+1/2}V^{j+1/2}_{m+1/2,\bar{m}-1/2}}}
 ~=~ 1.
\end{eqnarray}
In the same way, $C_-$ is calculated as a Wick contraction with one
screening operator $\tilde\mu\tilde S$ inserted:
\begin{eqnarray}
 C_-(M\bar{M};jm\bar{m}) &=&
  e^{-\frac{i\pi}{k}(M\bar{m}-\bar{M}m)}
 \lim\frac
 {\vev{(-\tilde\mu\tilde S)
   V^{-j-1/2}_{-m-1/2,-\bar{m}+1/2}
   V^{1/2}_{1/2,-1/2}V^j_{m,\bar{m}}}_{\rm free}}
 {\vev{V^{-j-1/2}_{-m-1/2,-\bar{m}+1/2}
   V^{j-1/2}_{m+1/2,\bar{m}-1/2}}_{\rm free}}
 \nn \\
 &=&
 -\tfrac{\tilde{\mu}}{k^2}
 \gamma(-\tfrac{2j+1}{k})\gamma(\tfrac{2j}{k})\gamma(\tfrac{1}{k})
 (m-2jM)(\bar{m}-2j\bar{M}).
\label{OPE1}
\end{eqnarray}
The coefficients $\tilde{C}_{\pm,\updownarrow}$ are also
calculated as Wick contractions with some
$\mu S, \bar\mu\bar S$ inserted.
\begin{eqnarray}
 \tilde{C}_+(M\bar{M};jm\bar{m}) &=& 1,
 \nn \\
 \tilde{C}_\uparrow(M\bar{M};jm\bar{m}) &=&
 k\mu e^{\frac{i\pi}{2}(\bar{m}-m+M-\bar{M})}
  \frac{\Gamma(1+j-m)\Gamma(1+\frac{k}{2}-M)
        \Gamma(-j-\frac{k}{2}+\bar{m}+\bar{M}-1)}
       {\Gamma(-j+\bar{m})\Gamma(-\frac{k}{2}+\bar{M})
        \Gamma(2+j+\frac{k}{2}-m-M)},
 \nn \\
 \tilde{C}_\downarrow(M\bar{M};jm\bar{m}) &=&
 k\bar{\mu}e^{-\frac{i\pi}{2}(\bar{m}-m+M-\bar{M})}
  \frac{\Gamma(1+j+m)\Gamma(1+\frac{k}{2}+M)
        \Gamma(-j-\frac{k}{2}-\bar{m}-\bar{M}-1)}
       {\Gamma(-j-\bar{m})\Gamma(-\frac{k}{2}-\bar{M})
        \Gamma(2+j+\frac{k}{2}+m+M)},
 \nn \\
 \tilde{C}_-(M\bar{M};jm\bar{m}) &=&
 k^2\mu\bar{\mu}\gamma(-2j-1)\gamma(1+2j-k)
 \frac{\Gamma(1+j+m)\Gamma(1+j-\bar{m})}
      {\Gamma(-j+m)\Gamma(-j-\bar{m})}
 \nn \\ && \times
 \frac{\Gamma(-j+\frac{k}{2}+m+M)\Gamma(-j+\frac{k}{2}-\bar{m}-\bar{M})}
      {\Gamma(1+j-\frac{k}{2}+m+M)\Gamma(1+j-\frac{k}{2}-\bar{m}-\bar{M})}.
\label{OPE2}
\end{eqnarray}
Here we restricted to operators which are not spectral flowed,
but the OPE coefficients depend on the labels $s,\bar{s}$ at most
through cocycle factors.
These coefficients are all obtained by the repeated use of the formula
\begin{equation}
  \int d^2z z^\alpha\bar{z}^{\bar{\alpha}}(1-z)^\beta(1-\bar{z})^{\bar{\beta}}
 =\pi\frac{\Gamma(1+\alpha)\Gamma(1+\beta)\Gamma(-\bar{\alpha}-\bar{\beta}-1)}
          {\Gamma(-\bar{\alpha})\Gamma(-\bar{\beta})\Gamma(\alpha+\beta+2)}.
\end{equation}

Setting $\mu=\bar{\mu}=\nu^{\frac k2}$ and combining
the OPE formulae with the reflection symmetry we find that the
reflection coefficients are given by
\begin{eqnarray}
 R(j,m,\bar{m}) &=&
 -\nu^{2j+1}
   \frac{\Gamma(1+j-m)\Gamma(1+j+\bar{m})}
        {\Gamma(-j-m)\Gamma(-j+\bar{m})}
   \frac{\Gamma(-2j-1)\Gamma(-\frac{2j+1}{k})}
        {\Gamma(2j+1)\Gamma(\frac{2j+1}{k})},
 \nn \\
  R_\pm(j) &=& \nu^{2j+1+\frac{k}{2}}\gamma(-\tfrac{2j+1}{k}).
\label{refcoef}
\end{eqnarray}
consistently with the expression obtained from three-point
structure constants.
At the same time, we also find the relation between coupling
constants $\nu$ and $\tilde{\mu}$:
\begin{equation}
  \nu = \tilde{\mu}\gamma(\tfrac{1}{k}).
\end{equation}

%%%%%%%%%%%%%%%%%%%%%%%%%%%%%%%%%%%%%%%%%%%%%%%%%%
\section{N=2 Liouville Theory with Boundary}%%%%%%
%%%%%%%%%%%%%%%%%%%%%%%%%%%%%%%%%%%%%%%%%%%%%%%%%%

Now we turn to the analysis of the theory in the presence of boundary.
As boundary conditions we only consider those preserving a half
of superconformal symmetry of the theory without boundary.
The simplest worldsheet with boundary is the upper half-plane
or the disc, through the analysis of which one can classify
all the possible boundary states.

Following the recent works on the boundary Liouville \cite{Zamolodchikov-Z2}
and $N=1$ super-Liouville theories \cite{Fukuda-H2, Ahn-RS},
we first analyze the annulus amplitudes using the modular
transformation property of characters of $N=2$ superconformal algebra.
They have been studied in some recent works
\cite{Eguchi-S, Ahn-SY1, Ahn-SY2} and \cite{Israel-PT1, Israel-PT2},
but let us analyze them carefully, taking the proper account of
the quantization law of $\theta$-momentum/winding number.

On the theory on the upper half-plane, there are two classes
of boundary conditions on the real line:
\begin{eqnarray}
 \mbox{\bf A-type} &:&
  T(z)=\bar{T}(\bar{z}),~~~
  T_F^\pm(z)=e^{\pm2\pi i\alpha}\bar{T}^\mp_F(\bar{z}),~~~
  J(z)=-\bar{J}(\bar{z}),
 \nn \\
 \mbox{\bf B-type} &:&
  T(z)=\bar{T}(\bar{z}),~~~
  T_F^\pm(z)=e^{\pm 2\pi i\alpha}\bar{T}^\pm_F(\bar{z}),~~~
  J(z)=\bar{J}(\bar{z}).
\end{eqnarray}
where $\alpha$ denotes the angle of R-rotation by $J_0\pm\bar J_0$.
Both of them preserve a copy of $N=2$ superconformal algebra.
By a conformal map that transforms the upper half-plane to the
unit disc, they are transformed to the condition on {\it boundary states}.
For A-type boundary states it becomes,
\begin{equation}
\begin{array}{rcl}
 0 &=& \bra{A^\alpha}(L_n-\bar{L}_{-n}), \\
 0 &=& \bra{A^\alpha}(G_r^\pm+ie^{\pm2\pi i\alpha}\bar{G}^\mp_{-r}), \\
 0 &=& \bra{A^\alpha}(J_n-\bar{J}_{-n}),
\end{array}
~~~
\begin{array}{rcl}
  (L_n-\bar{L}_{-n})\ket{A^\alpha} &=& 0, \\
  (G_r^\pm-ie^{\pm2\pi i\alpha}\bar{G}^\mp_{-r})\ket{A^\alpha}  &=& 0, \\
  (J_n-\bar{J}_{-n})\ket{A^\alpha} &=& 0,
\end{array}
\label{bsteqa}
\end{equation}
while the condition on B-types is
\begin{equation}
\begin{array}{rcl}
 0 &=& \bra{B^\alpha}(L_n-\bar{L}_{-n}), \\
 0 &=& \bra{B^\alpha}(G_r^\pm+ie^{\pm2\pi i\alpha}\bar{G}^\pm_{-r}), \\
 0 &=& \bra{B^\alpha}(J_n+\bar{J}_{-n}),
\end{array}
~~~
\begin{array}{rcl}
  (L_n-\bar{L}_{-n})\ket{B^\alpha} &=& 0, \\
  (G_r^\pm-ie^{\pm2\pi i\alpha}\bar{G}^\pm_{-r})\ket{B^\alpha}  &=& 0, \\
  (J_n+\bar{J}_{-n})\ket{B^\alpha} &=& 0.
\end{array}
\label{bsteqb}
\end{equation}

D-branes are described as boundary states, or the solutions
to (\ref{bsteqa}) or (\ref{bsteqb}) supporting a well-defined
spectrum of open string states.
Ishibashi states form the basis of solutions to the boundary condition,
and are constructed by summing up all the descendants of
a single primary state. We define A-type Ishibashi states by 
(here we use $h_0=\frac{m^2-j(j+1)}{k},q_0=\frac{2m}{k}$)
\begin{eqnarray}
\lefteqn{
 \bbra{A^\alpha_{j,m,\beta}}~=~
}\nn \\
  &&
 e^{-2\pi i\alpha(\frac{2m}{k}+\beta\chat)}
 \bra{V^{j(\beta,\beta)}_{m,m}}
 \left(1+\frac{ie^{-2\pi i\alpha}}{2h_0-q_0}
          G^+_{\frac{1}{2}+\beta}\bar{G}^+_{\frac{1}{2}+\beta}
        +\frac{ie^{2\pi i\alpha}}{2h_0+q_0}
          G^-_{\frac{1}{2}-\beta}\bar{G}^-_{\frac{1}{2}-\beta}
        +\cdots\right),
 \nn \\
\lefteqn{
 \bket{A^\alpha_{j,m,\beta}} ~=~
}\nn \\
  &&
 \left(\cdots
        +\frac{ie^{-2\pi i\alpha}}{2h_0-q_0}
          G^+_{-\frac{1}{2}-\beta}\bar{G}^+_{-\frac{1}{2}-\beta}
        +\frac{ie^{2\pi i\alpha}}{2h_0+q_0}
          G^-_{-\frac{1}{2}+\beta}\bar{G}^-_{-\frac{1}{2}+\beta}
        +1\right)
    \ket{V^{j(\beta,\beta)}_{m,m}}
 e^{-2\pi i\alpha(\frac{2m}{k}+\beta\chat)},
 \nn \\
 && (m+\beta\in\tfrac{k}{2}\mZ)
\end{eqnarray}
 and B-type Ishibashi states by
\begin{eqnarray}
\lefteqn{
 \bbra{B^\alpha_{j,m,\beta}}~=~
}\nn \\
  &&
 e^{-2\pi i\alpha(\frac{2m}{k}+\beta\chat)}
 \bra{V^{j(\beta,-\beta)}_{m,-m}}
 \left(1+\frac{ie^{-2\pi i\alpha}}{2h_0-q_0}
          G^+_{\frac{1}{2}+\beta}\bar{G}^-_{\frac{1}{2}+\beta}
        +\frac{ie^{2\pi i\alpha}}{2h_0+q_0}
          G^-_{\frac{1}{2}-\beta}\bar{G}^+_{\frac{1}{2}-\beta}
        +\cdots\right),
 \nn \\
\lefteqn{
 \bket{B^\alpha_{j,m,\beta}}~=~
}\nn \\
  &&
 \left(\cdots
        +\frac{ie^{-2\pi i\alpha}}{2h_0-q_0}
          G^+_{-\frac{1}{2}-\beta}\bar{G}^-_{-\frac{1}{2}-\beta}
        +\frac{ie^{2\pi i\alpha}}{2h_0+q_0}
          G^-_{-\frac{1}{2}+\beta}\bar{G}^+_{-\frac{1}{2}+\beta}
        +1\right)
    \ket{V^{j(\beta,-\beta)}_{m,-m}}
 e^{-2\pi i\alpha(\frac{2m}{k}+\beta\chat)},
 \nn \\
 && (m\in\tfrac{1}{2}\mZ)
\end{eqnarray}
Note the restriction on $m$ arising from the $\theta$-momentum/winding
number quantization law.
Note that, since $V^{j(\beta,\beta)}_{m,m}$ can be transformed to
$V^{j(\beta+n,\beta+n)}_{m-n,m-n}$ for any integer $n$ by a
multiplication of supercharges, there are proportionality
relations between Ishibashi states
\begin{equation}
  \bbra{A^\alpha_{j,m,\beta}}~\sim~
  \bbra{A^\alpha_{j,m+n,\beta-n}},~~~
  \bket{A^\alpha_{j,m,\beta}}~\sim~
  \bket{A^\alpha_{j,m+n,\beta-n}}.
\label{eqishi}
\end{equation}
The same holds also for B-type Ishibashi states.
In this paper we only consider the Ishibashi states lying in
continuous representations ($j\in-\frac{1}{2}+i\mR$), and set
their normalization by the formula
\begin{eqnarray}
\lefteqn{
  \bbra{A^\alpha_{j,m,\beta}}
  e^{i\pi\tau_c(L_0+\bar{L}_0-\frac{c}{12})}
  \bket{A^{\alpha'}_{j',m',-\beta}}
} \nn \\
 &=& 2\pi\delta_{m+m',0}
     \left\{\delta(i(j+j'+1))
           +\delta(i(j-j'))R(j,m,m)\right\}
     \chi_{j,m+\beta,\beta}(\tau_c,\alpha'-\alpha),
 \nn \\
\lefteqn{
  \bbra{B^\alpha_{j,m,\beta}}
  e^{i\pi\tau_c(L_0+\bar{L}_0-\frac{c}{12})}
  \bket{B^{\alpha'}_{j',m',-\beta}}
} \nn \\
 &=& 2\pi\delta_{m+m',0}
     \left\{\delta(i(j+j'+1))
           +\delta(i(j-j'))R(j,m,-m)\right\}
     \chi_{j,m+\beta,\beta}(\tau_c,\alpha'-\alpha),
\end{eqnarray}
where $R(j,m,\bar{m})$ is the reflection coefficient for
bulk operators, and $\chi_{j,m,\beta}(\tau,\alpha)$ is
the $N=2$ character for continuous representation
\begin{equation}
  \chi_{j,m,\beta}(\tau,\alpha)
 ~\equiv~
  q^{\frac{m^2}{k}-\frac{(2j+1)^2}{4k}+\frac{\beta^2}{2}}
  z^{\frac{2m}{k}+\beta}
 \vartheta(\alpha+\beta\tau,\tau)\eta(\tau)^{-3},
\end{equation}
with $q = e^{2\pi i\tau},~z = e^{2\pi i\alpha}$.
$\theta(\nu,\tau)$ is Jacobi theta function and $\eta(\tau)$
is Dedekind eta function; see the appendix for their definition
and modular transformation property.
It follows from this that the character is periodic in $\beta$
with period $1$, corresponding to the equivalence of Ishibashi states
(\ref{eqishi}).

D-branes are expressed as suitable superpositions of the Ishibashi states.
We call them as A-branes or B-branes, depending on the choice of
boundary conditions.
A-branes are point-like along $\theta$-direction in the sense that
they source closed string states without winding number
along $\theta$-direction.
Similarly, B-branes are winding around $\theta$-direction.

~

We will also consider the spectrum of open string states or
corresponding boundary operators between two arbitrary D-branes.
We will restrict our discussion mainly to those between the
same type of branes.
We will later consider the boundary primary operators
labelled by $(l,m,s)$, and denote them as $[B^{l(s)}_m]^X_{~X'}$
making explicit the dependence on two D-branes $X$ and $X'$
they are ending on.
Similarly to the bulk operators, the boundary operators also obey
certain quantization law of momentum or winding number along
$\theta$-direction.

Let us first consider the open string states between two A-branes.
Since A-branes are not wrapping around $S^1$, they do not carry
$\theta$-momentum.
Therefore the open string states with both ends on the same A-brane
only have quantized winding numbers, $m\in\mZ$.
(Later we will see a mild modification to this.)
For a generic pair of A-branes it will be shifted as
$m\in\mZ+\delta$, but it should still be integer-spaced.
The index $s$ should also be quantized.
Generic A-brane satisfy the boundary condition twisted by $\alpha$, 
and the open strings stretched between two A-branes labelled by
$\alpha$ and $\alpha'$ are in the $(\alpha\!-\!\alpha')$-th
spectral flowed sector.
This can be understood in the following way.
If we put a boundary operator at the origin and A-branes with labels
($\alpha,\alpha'$) on the negative and positive real axis,
then we obtain
\begin{equation}
  T_F^\pm(ze^{2\pi i}) = e^{\pm 2\pi i(\alpha-\alpha')}T_F^\pm(z),
\end{equation}
indicating that the boundary operator should belong to the
spectral flowed sector.
One can argue in a similar way for B-branes, so the quantization
laws are summarized as follows:
\begin{equation}
[B^{l(s)}_m]^X_{X'}:~~
\left\{
\begin{array}{lll}
X,X'~\mbox{\bf A-branes}~\Rightarrow
 & m\in\mZ+\delta,& s\in\mZ+\alpha-\alpha',
\\
X,X'~\mbox{\bf B-branes}~\Rightarrow
 & m+s\in k\mZ+\delta, & s\in\mZ+\alpha-\alpha'.
\end{array}
\right.
\end{equation}
One can immediately check the compatibility with the superconformal
symmetry: if a primary operator connects two D-branes, so does any of
its descendants.

%%%%%%%%%%%%%%%%%%%%%%%%%%%%%%%%%%%%%%%%%%%
\subsection{Modular Bootstrap for A-branes}
%%%%%%%%%%%%%%%%%%%%%%%%%%%%%%%%%%%%%%%%%%%

From the previous discussion we expect that the open string spectrum
between two A-branes involves summing over $m$ quantum number
with unit periodicity.
Based on this, we propose that
\begin{quote}
 {\it the open string spectrum between
A-branes is a sum over integer spectral flow}.
\end{quote}
This means that the presence of $[B^{l(s)}_m]^X_{X'}$ implies
the presence of $[B^{l(s+n)}_m]^X_{X'}$ for any integer $n$.
This is proved in the following way.
Consider an open string state between two A-branes
labelled by $\alpha,\alpha'$.
Its $s$ label has to satisfy $s+\alpha'-\alpha\in\mZ$.
Now rotate one A-brane adiabatically so that $\alpha$ increases by one.
The A-brane labelled by $\alpha$ should come back to itself up to
an overall phase, so the open string spectrum in particular will not
change by the unit shift of $\alpha$.
On the other hand, during the adiabatic process the $s$ label of
each open string state increases, and is shifted by one in the end.
This means the invariance of open string spectrum under
integer spectral flow.
So the open string amplitudes between A-branes involve characters of
a {\it large} $N=2$ superconformal algebra which includes integer spectral
flows.
Note that the sum of characters over integer spectral flows is
equivalent to the sum over integer shifts of $m$ quantum number, due to
the periodicity of the character $\chi_{j,m,\beta}(\tau,\alpha)$
explained before.

Let us start with presenting several useful formulae
for later calculations.
First, we will frequently consider the sum of characters spectral
flowed by integer amounts.
So let us work out the modular transformation law for such quantity here.
Using the Gauss integral and Poisson resummation formula one finds
\begin{eqnarray}
\lefteqn{
  e^{-2\pi i\alpha\beta}
  \sum_{n\in\mZ+\alpha}
  e^{-\frac{4\pi i(M+n)\beta}{k}}
  \chi_{J,M+n,n}(\tau_o,\beta)
}
 \nn \\
 &=&
  -i\sum_{m\in\frac{k}{2}\mZ}\int_{{\cal C}_0} dj
  e^{-\frac{4\pi iMm}{k}+\frac{i\pi}{k}(2j+1)(2J+1)}
  \chi_{j,m,\beta}(\tau_c,-\alpha).
 ~~~~({\cal C}_0\equiv \{-\tfrac{1}{2}+i\mR\})
\end{eqnarray}
Next, let us present a formula involving characters for chiral
representations.
\begin{eqnarray}
  \frac{\chi_{J,M,\alpha}(\tau,\beta)}
       {1+e^{2\pi i\beta}q^{M\pm(J+\frac{1}{2})}}
 &=&
  \sum_{l\in\mZ_{\ge 0}}
  (-)^l
  \chi_{J\mp\frac{kl}{2},M+\frac{kl}{2},\alpha}(\tau,\beta),
 \nn \\
  \frac{\chi_{J,M,\alpha}(\tau,\beta)}
       {1+e^{-2\pi i\beta}q^{-M\pm(J+\frac{1}{2})}}
 &=&
  \sum_{l\in\mZ_{\ge 0}}
  (-)^l
  \chi_{J\mp\frac{kl}{2},M-\frac{kl}{2},\alpha}(\tau,\beta).
\end{eqnarray}

%%%%%%%%%%%%%%%%%%%%%%%%%%%%%%%%%%%%%%%%%%%%%%%%%%%%%%%
\subsubsection{Identity representation}%%%%%%%%%%%%%%%%
%%%%%%%%%%%%%%%%%%%%%%%%%%%%%%%%%%%%%%%%%%%%%%%%%%%%%%%

Of all the boundary states satisfying the A-type boundary condition,
the most important is the one corresponding to the identity
representation, $\bra{A_{[1]}}$ and $\ket{A_{[1]}}$.
We start from the fact that the annulus amplitude with both ends
on it is given by the sum of the character for identity representation
over integer spectral flows.
\begin{eqnarray}
Z &=&
  \bra{A^{\alpha,\beta}_{[1]}}
  e^{i\pi\tau_c(L_0+\bar{L}_0-\frac{c}{12})}
  \ket{A^{\alpha',-\beta}_{[1]}}
 \nn \\
  &=&
 \sum_{n\in\mZ+\alpha-\alpha'}
 y^{\alpha'-\alpha-\frac{2n}{k}}
 \frac{\chi_{0,n,n}(\tau_o,\beta)(1-q_o)}
      {(1+y     q_o^{\frac{1}{2}+n})
       (1+y^{-1}q_o^{\frac{1}{2}-n})}
 ~~~~(y = e^{2\pi i\beta})
 \nn \\
  &=&
 \sum_{l\in\mZ_{\ge 0}}(-)^l
 \sum_{n\in\mZ+\alpha-\alpha'}
 y^{\alpha'-\alpha-\frac{2n}{k}}
 \left\{
  \chi_{-\frac{kl}{2},n+\frac{kl}{2},n}(\tau_o,\beta)
 -\chi_{ \frac{kl}{2},n+\frac{kl}{2},n}(\tau_o,\beta)
 \right\}.
\end{eqnarray}
The open strings are in the $(\alpha-\alpha')$-th spectral flowed sector.
In the second line, the powers of $y$ in the sum is chosen in accordance
with the periodicity of the label $\alpha\sim \alpha+1$ of boundary
states.
It therefore follows that, when the closed string states are chosen from
$(\beta,\beta)$-spectral flowed sector, the trace over open string
states should be taken with the phase $e^{2\pi i\beta F}$, where
$F$ is defined by
\begin{equation}
 F[B^{0(n)}_0]
 ~\equiv~ n-\alpha+\alpha'
 ~=~ J_0[B^{0(n)}_0]+\alpha'-\alpha-\tfrac{2n}{k},
~~~~
 F[G_r^\pm]=\pm1.
\end{equation}
After the modular S transformation, the annulus amplitude is expressed
as a sum over closed string exchanges,
\begin{eqnarray}
Z &=&
 -i\sum_{l\in\mZ_{\ge 0}}(-)^l
   \int_{{\cal C}_0} dj
 \!\!\!
 \sum_{m+\beta\in\frac{k}{2}\mZ}
 e^{\frac{i\pi}{k}(2j+1)}
 \left\{
 e^{-i\pi l(2m+2j+1)}
-e^{-i\pi l(2m-2j-1)}
 \right\}
 \chi_{j,m+\beta,\beta}(\tau_c,\alpha'-\alpha)
 \nn \\ &\simeq&
  \int_{{\cal C}_0} dj
 \sum_{m+\beta\in\frac{k}{2}\mZ}
 \frac{i\sin\pi(2j+1)\sin\frac{\pi}{k}(2j+1)}
      {2\sin(j+m)\pi\sin(j-m)\pi}
 \chi_{j,m+\beta,\beta}(\tau_c,\alpha'-\alpha),
\end{eqnarray}
where $\simeq$ means the equality up to possible emergence of
discrete series states from changing the order of $l$-sum
and $j$-integration.
On the other hand, A-branes are written as superpositions
of A-type Ishibashi states
\begin{eqnarray}
  \bra{A^{\alpha,\beta}_{[1]}} &=&
  \sum_{m+\beta\in{\frac{kZ}{2}}}\int_{{\cal C}_0}
  \frac{dj}{2\pi i}U_{[1]}(-j-1,-m,-\beta)\times
  \bbra{A^\alpha_{j,m,\beta}} ~+~\mbox{(discrete reps.)},
  \nn \\
  \ket{A^{\alpha,\beta}_{[1]}} &=&
  \sum_{m+\beta\in{\frac{kZ}{2}}}\int_{{\cal C}_0}
  \frac{dj}{2\pi i}
  \bket{A^\alpha_{j,m,\beta}}\times
   U_{[1]}(-j-1,-m,-\beta) ~+~\mbox{(discrete reps.)},
\end{eqnarray}
the wave function for the identity A-brane $U_{[1]}(j,m,\beta)$
has to satisfy
\begin{eqnarray}
  U_{[1]}(j,m,\beta) &=& R(j,m,m)U_{[1]}(-j-1,m,\beta),
 \nn \\
  U_{[1]}(j,m,\beta)U_{[1]}(-j-1,-m,-\beta)
  &=& -\frac{\pi\sin\pi(2j+1)\sin\tfrac{\pi(2j+1)}{k}}
            {2  \sin\pi(j+m)\sin\pi(j-m)}.
\end{eqnarray}
So we obtain, up to $\pm$ sign,
\begin{equation}
  U_{[1]}(j,m,\beta) ~=~
  (\tfrac{k\pi}{2})^{\frac{1}{2}}\nu^{j+\frac{1}{2}}
  \frac{\Gamma(1+j+m)\Gamma(1+j-m)}{\Gamma(2j+2)\Gamma(\frac{2j+1}{k})}.
\end{equation}
 
The wave functions for other A-branes are obtained by
considering annulus amplitudes bounded by one identity
and one generic A-branes.
In the following we consider five classes of them, and we label them
by the highest weights $\ket{J,M}$ of $N=2$ superconformal algebra.
In the following we will simply neglect the contribution from
closed string states in discrete representations, because the wave
function $U(j,m,\beta)$ for $j\in-\frac12+i\mR$ is enough to determine
the disc one-point function of bulk operators completely
under the assumption of analyticity.

%%%%%%%%%%%%%%%%%%%%%%%%%%%%%%%%%%%%%%%%%%%%%%%%%%%%%%%%%%%
\subsubsection{Non-chiral non-degenerate representations}%%
%%%%%%%%%%%%%%%%%%%%%%%%%%%%%%%%%%%%%%%%%%%%%%%%%%%%%%%%%%%

The first example we consider is the A-brane
$\ket{A_{[J,M]}}$ corresponding to the Verma module
over highest weight state $\ket{J,M}$.
The annulus amplitude between this and an identity A-branes
is given by a sum of characters over integer spectral flows,
\begin{eqnarray}
\lefteqn{
  \bra{A^{\alpha,\beta}_{[1]}}
  e^{i\pi\tau_c(L_0+\bar{L}_0-\frac{c}{12})}
  \ket{A^{\alpha',-\beta}_{[J,M]}}
 ~=~
 \sum_{n\in\mZ+\alpha-\alpha'}
 y^{\alpha'-\alpha-\frac{2}{k}(M+n)}\chi_{J,M+n,n}(\tau_o,\beta)
}
 \nn \\
 &=&
 -i\sum_{m+\beta\in\frac{k}{2}\mZ}\int_{{\cal C}_0}dj
 \chi_{j,m+\beta,\beta}(\tau_c,\alpha'-\alpha)
 e^{-\frac{4\pi iM}{k}(m+\beta)}\cos\{\tfrac{\pi}{k}(2j+1)(2J+1)\}.
\end{eqnarray}
From this we obtain
\begin{equation}
 U_{[1]}(j,m,\beta)U_{[J,M]}(-j-1,-m,-\beta)
 = \pi e^{-\frac{4\pi iM}{k}(m+\beta)}\cos\{\tfrac{\pi}{k}(2j+1)(2J+1)\}.
\end{equation}
The wave function for this A-brane thus becomes
\begin{equation}
  U_{[J,M]}(j,m,\beta)
 =(\tfrac{2\pi}{k})^{\frac{1}{2}}\nu^{j+\frac{1}{2}}
  \frac{\Gamma(-2j)\Gamma(-\frac{2j+1}{k})}{\Gamma(-j+m)\Gamma(-j-m)}
  e^{\frac{4\pi iM}{k}(m+\beta)}\cos\{\tfrac{\pi}{k}(2j+1)(2J+1)\}.
\end{equation}

%%%%%%%%%%%%%%%%%%%%%%%%%%%%%%%%%%%%%%%%%%%%%%%%%%%%%%%%
\subsubsection{Non-chiral degenerate representations}%%%
%%%%%%%%%%%%%%%%%%%%%%%%%%%%%%%%%%%%%%%%%%%%%%%%%%%%%%%%

When $J=J_{r,s}=\frac{1}{2}(r-1+ks)~(r,s\in\mZ_{>0})$
the Verma module over $\ket{J,M}$ has a null vector at the level $rs$,
and an irreducible representation is defined by the subtraction
of the null submodule.
Denoting the corresponding A-brane by $\ket{A_{[J_{r,s},M]}}$,
the annulus amplitude between this and the identity A-branes becomes
\begin{eqnarray}
\lefteqn{
  \bra{A^{\alpha,\beta}_{[1]}}
  e^{i\pi\tau_c(L_0+\bar{L}_0-\frac{c}{12})}
  \ket{A^{\alpha',-\beta}_{[J_{r,s},M]}}
}
 \nn \\
 &=&
  \sum_{n\in\mZ+\alpha-\alpha'}
  y^{\alpha'-\alpha-\frac{2}{k}(M+n)}
  \left\{
  \chi_{J_{ r,s},M+n,n}(\tau_o,\beta)
 -\chi_{J_{-r,s},M+n,n}(\tau_o,\beta)
  \right\},
\end{eqnarray}
from which we obtain, in the same way as before,
\begin{equation}
  U_{[J_{r,s},M]}(j,m,\beta)
 =-2(\tfrac{2\pi}{k})^{\frac{1}{2}}\nu^{j+\frac{1}{2}}
  \frac{\Gamma(-2j)\Gamma(-\frac{2j+1}{k})}{\Gamma(-j+m)\Gamma(-j-m)}
  e^{\frac{4\pi iM}{k}(m+\beta)}
  \sin\tfrac{(2j+1)r\pi}{k}\sin(2j+1)s\pi.
\end{equation}

From the momentum quantization for bulk operators
$m+\beta\in \frac{k}{2}\mZ$ it follows that the label $M$ has
period $1$ for these two classes of branes.

%%%%%%%%%%%%%%%%%%%%%%%%%%%%%%%%%%%%%%%%%%%%%
\subsubsection{Anti-chiral representations}%%
%%%%%%%%%%%%%%%%%%%%%%%%%%%%%%%%%%%%%%%%%%%%%

When the highest weight $\ket{J,M}$ satisfies $J-M\in\mZ_{\ge 0}$,
then an irreducible representation is obtained by putting
\begin{equation}
 0= G^-_{-J+M-\frac{1}{2}}\cdots
    G^-_{-\frac{3}{2}}G^-_{-\frac{1}{2}}\ket{J,M}.
\label{cnv1}
\end{equation}
The case $J=M$ gives an anti-chiral representation, and other
cases are its spectral flow.
Denoting the corresponding A-branes by $\ket{A_{[J,M]^-}}$,
the annulus amplitude between this and the identity A-branes is
calculated as follows:
\begin{equation}
  \bra{A^{\alpha,\beta}_{[1]}}
  e^{i\pi\tau_c(L_0+\bar{L}_0-\frac{c}{12})}
  \ket{A^{\alpha',-\beta}_{[J,M]^-}}
 =
 \sum_{n\in\mZ+\alpha-\alpha'}
 y^{\alpha'-\alpha-\frac{2}{k}(M+n)}
 \frac{\chi_{J,M+n,n}(\tau_o,\beta)}
      {1+y^{-1}q_o^{\frac{1}{2}+J-M-n}}.
\end{equation}
From this we obtain
\begin{eqnarray}
\lefteqn{
 U_{[J,M]^-}(j,m,\beta) ~=~
 i(8k\pi)^{-\frac{1}{2}}\nu^{j+\frac{1}{2}}e^{\frac{4\pi iM}{k}(m+\beta)}
  \Gamma(-2j)\Gamma(-\tfrac{2j+1}{k})
}
 \nn \\ &&\times
 \left\{
 e^{i\pi(m+j)+\frac{i\pi}{k}(2j+1)(2J+1)}\frac{\Gamma(1+j+m)}{\Gamma(-j+m)}
-e^{i\pi(m-j)-\frac{i\pi}{k}(2j+1)(2J+1)}\frac{\Gamma(1+j-m)}{\Gamma(-j-m)}
 \right\}.
\end{eqnarray}

%%%%%%%%%%%%%%%%%%%%%%%%%%%%%%%%%%%%%%%
\subsubsection{Chiral representations}%
%%%%%%%%%%%%%%%%%%%%%%%%%%%%%%%%%%%%%%%

Similarly to the above, when the highest weight satisfies
$J+M\in\mZ_{\ge 0}$ the irreducible representations are
defined by the null vector equation
\begin{equation}
0 = G^+_{-J-M-\frac{1}{2}}\cdots
    G^+_{-\frac{3}{2}}G^+_{-\frac{1}{2}}\ket{J,M}.
\label{cnv2}
\end{equation}
They are chiral representations or their spectral flows.
The corresponding A-branes are denoted as $\ket{A_{[J,M]^+}}$,
and from the analysis of annulus amplitude we obtain
\begin{eqnarray}
\lefteqn{
 U_{[J,M]^+}(j,m,\beta) ~=~
  i(8k\pi)^{-\frac{1}{2}}\nu^{j+\frac{1}{2}}e^{\frac{4\pi iM}{k}(m+\beta)}
  \Gamma(-2j)\Gamma(-\tfrac{2j+1}{k})
}
 \nn \\ &&\times
 \left\{
 e^{i\pi(-m+j)+\frac{i\pi}{k}(2j+1)(2J+1)}\frac{\Gamma(1+j-m)}{\Gamma(-j-m)}
-e^{i\pi(-m-j)-\frac{i\pi}{k}(2j+1)(2J+1)}\frac{\Gamma(1+j+m)}{\Gamma(-j+m)}
 \right\}.
\end{eqnarray}

~

Taking the quantization conditions on $M$ and $m+\beta$ into account,
one finds that the wave functions for (anti-)chiral A-branes are
independent of $M$.
We can also check the following equivalence
as required from representation theory:
\begin{equation}
 \ket{A_{[J,\pm J]^\mp}} ~=~{\rm const.}\times
 \ket{A_{[\tilde{J},\mp\tilde{J}]^\pm}}
 ~~~(\tilde{J}=-J-1-\tfrac{k}{2}).
\end{equation}

%%%%%%%%%%%%%%%%%%%%%%%%%%%%%%%%%%%%%%%%%%%%%%%%%%%
\subsubsection{Degenerate chiral representations}%%
%%%%%%%%%%%%%%%%%%%%%%%%%%%%%%%%%%%%%%%%%%%%%%%%%%%

When $J\pm M$ are both nonnegative integers, the Verma module
has two independent null vectors defined by (\ref{cnv1}) and
(\ref{cnv2}).
By setting them to zero we obtain an irreducible representation
which we call as degenerate chiral.
The corresponding A-branes will be denoted as
$\ket{A_{[J,M]^{\rm dc}}}$.
The calculation of annulus amplitudes involving them is
a little complicated.
By a little thought one finds that the representation space is
spanned by the following vectors
\begin{eqnarray}
 &&
 \bigoplus_{p=1}^{J+M}
 \left\{\mbox{polynomial of }
        L_{n\le -2},~G^\pm_{r\le\mp p-3/2},~J_{n\le -1}\right\}
 G^+_{-p+\frac{1}{2}}\cdots G^+_{-\frac{3}{2}}G^+_{-\frac{1}{2}}\ket{J,M}
 \nn \\ &&
 ~~~\oplus
 \left\{\mbox{polynomial of }
        L_{n\le -2},~G^\pm_{r\le-3/2},~J_{n\le -1}\right\}\ket{J,M}
 \nn \\ &&
 \oplus
 \bigoplus_{p=1}^{J-M}
 \left\{\mbox{polynomial of }
        L_{n\le -2},~G^\pm_{r\le\pm p-3/2},~J_{n\le -1}\right\}
 G^-_{-p+\frac{1}{2}}\cdots G^-_{-\frac{3}{2}}G^-_{-\frac{1}{2}}\ket{J,M}.
\end{eqnarray}
So the character for this representation
spectral flowed by $\alpha$ units is given by
\begin{eqnarray}
 {\rm Tr}[y^Fq^{L_0-\frac{c}{24}}]
 &=&
 \sum_{p=M-J}^{M+J}
 q^{\frac{(M+\alpha)^2-J(J+1)}{k}+\frac{(p+\alpha)^2}{2}-\frac{c}{24}}
 y^p
    \prod_{n\ge 1}\frac{(1+yq^{n+\frac{1}{2}+p+\alpha})
                        (1+y^{-1}q^{n+\frac{1}{2}-p-\alpha})}
                       {(1-q^n)(1-q^{n+1})}
\nn \\
 &=&
 \sum_{p=M-J}^{M+J}
 \frac{y^{-\frac{2M}{k}-\alpha\chat}
       \chi_{J,M+\alpha,p+\alpha}(\tau,\beta)(1-q)}
      {(1+yq^{\frac{1}{2}+p+\alpha})(1+y^{-1}q^{\frac{1}{2}-p-\alpha})}.
\end{eqnarray}
After summing over spectral flow we obtain the annulus amplitude
bounded by $\ket{A_{[J,M]^{\rm dc}}}$ and the identity A-branes:
\begin{eqnarray}
 Z &=&
  \bra{A^{\alpha,\beta}_{[1]}}
  e^{i\pi\tau_c(L_0+\bar{L}_0-\frac{c}{12})}
  \ket{A^{\alpha',-\beta}_{[J,M]^{\rm dc}}}
 \nn \\
  &=&
 \sum_{n\in\mZ+\alpha-\alpha'}
 \sum_{p=M-J}^{M+J}
 y^{\alpha'-\alpha-\frac{2}{k}(M+n)}
 \frac{\chi_{J,M+n,n+p}(\tau,\beta)(1-q_o)}
      {(1+yq_o^{\frac{1}{2}+p+n})(1+y^{-1}q_o^{\frac{1}{2}-p-n})}.
\end{eqnarray}
 The wave function thus becomes
\begin{equation}
  U_{[J,M]^{\rm dc}}(j,m,\beta)
 =
 e^{\frac{4\pi iM}{k}(m+\beta)}
  (\tfrac{k\pi}{2})^{\frac{1}{2}}\nu^{j+\frac{1}{2}}
  \frac{\Gamma(1+j-m)\Gamma(1+j+m)\sin\{\tfrac{\pi}{k}(2J+1)(2j+1)\}}
       {\Gamma(2j+2)\Gamma(\frac{2j+1}{k})\sin\frac{(2j+1)\pi}{k}}.
\end{equation}
These A-branes $\ket{A_{[J,M]^{\rm dc}}}$ are also independent of $M$,
and labelled by a single positive integer $n=2J+1$.
So we also denote them by $\ket{A_{[n]}}$.
The case $J=M=0$ corresponds to the identity A-brane $\ket{A_{[1]}}$
analyzed previously.

%%%%%%%%%%%%%%%%%%%%%%%%%%%%%%%%%%%%%%%%%%%
\subsection{Modular Bootstrap for B-branes}
%%%%%%%%%%%%%%%%%%%%%%%%%%%%%%%%%%%%%%%%%%%

One might expect that the wave functions for B-branes are obtained
through a similar analysis of annulus amplitudes, but it turns out not
the case.

Based on the free field picture, we found that the boundary
operator $[B^{l(s)}_m]$ between B-branes satisfy the momentum
quantization law $m+s\in k\mZ+{\rm const}$.
So the annulus amplitudes as seen from the open string channel
should be sums over characters labelled by $(l,m,s)$ with the above
constraint.
However, the simple shift of $m$ by $k\mZ$ is not an isomorphism
of representations of $N=2$ superconformal algebra, especially
for irrational $k$ and $(l,m,s)$ belonging to chiral representations.
The simple mixtures of chiral and non-chiral representations will
not lead to annulus amplitudes with nice modular transformation property,
i.e. they will not have a sensible closed string channel interpretation.
In particular, for irrational $k$, it is difficult to think of
spectrum of open string states between identity B-branes.
If there is no identity brane, then the modular bootstrap analysis
for B-branes will not be as powerful as it was for A-branes.

For A-branes, the periodicity under R-rotation $\alpha\to\alpha+1$
was the key in finding the correct open string spectrum.
However, for B-branes this does not seem to yield any useful
information on which representations to sum over.
Consider an open string state labelled by $(l,m,s)$ and
stretched between two B-branes, and what happens to it when one of
the B-branes is R-rotated once in an adiabatic way so that it
returns to itself.
$s$ will increase by one as before, but this time $m$ will decrease by
one as well in order to meet with the momentum quantization law.
The new state is related to the original state by the action
of supercurrent (\ref{GVope}), so the two states are within the same
representation space of boundary superconformal algebra.

When $k$ is an integer, there is a candidate for open string spectrum
between identity B-branes, because then the sum over $k\mZ$ shifts of
the quantum number $m+s$ can be interpreted as the sum over $k\mZ$ spectral
flows.
This is expressed in terms of characters as follows:
\begin{eqnarray}
\lefteqn{
  \sum_{m\in\frac12\mZ+\beta}\int_{{\cal C}_0}\frac{dj}{ik}
  e^{\frac{i\pi}{k}(2j+1)(2J+1)-\frac{4\pi iMm}{k}}
  \chi_{j,m,\beta}(\tau_c,-\alpha)
}
 \nn\\ &=&
  \sum_{n\in k\mZ}e^{-2\pi i\chat\alpha\beta-\frac{4\pi i\beta M}{k}}
  \chi_{J,M+n+\alpha,\alpha}(\tau_o,\beta)
 \nn\\ &\stackrel{k\in\mZ}{=}&
  \sum_{n\in k\mZ+\alpha}e^{-2\pi i\chat\alpha\beta-\frac{4\pi i\beta M}{k}}
  \chi_{J,M+n,n}(\tau_o,\beta).
\end{eqnarray}
The sum over $k\mZ$ spectral flows of identity character has a nice
modular transformation property.
Although this might be extended to the cases with rational $k$ by
a suitable orbifolding, we will continue to focus on integer $k$.

Let us go on and see whether we can re-write the annulus amplitude
and obtain an analytic expression for wave function in consistency with
the reflection relation of bulk operators.
Denoting by $T_{[1]}(j,m,\beta)$ the wave function for the identity
B-brane, one finds
\begin{eqnarray}
  T_{[1]}(j,m,\beta) &=& R(j,m,-m)T_{[1]}(-j-1,m,\beta),
 \nn \\
  T_{[1]}(j,m,\beta)T_{[1]}(-j-1,-m,-\beta)
  &=& -\frac{\pi\sin\pi(2j+1)\sin\tfrac{\pi(2j+1)}{k}}
            {2k \sin\pi(j+m)\sin\pi(j-m)}.
\end{eqnarray}
Note that $R(j,m,-m)=e^{2\pi im}R(j,m,m)$ under the quantization
condition $m\in\frac12\mZ$.
The first equation is solved by
\begin{equation}
 T_{[1]}(j,m,\beta) = (\tfrac{\pi}{2})\nu^{j+\frac12}
 \frac{\Gamma(1+j+m)\Gamma(1+j-m)}{\Gamma(2j+2)\Gamma(\frac{2j+1}{k})}
 \left\{\hat T(j)+e^{2\pi im}\hat T(-j-1)\right\}
\end{equation}
and the second one yields
\begin{equation}
  2\hat T(j)\hat T(-j-1)=1,~~~\hat T(j)^2 + \hat T(-j-1)^2=0,
\end{equation}
which has no solution at $j=-1/2$.
This shows that there is no analytic wave function for
identity B-brane even for integer $k$.

This result seems in contradiction with the known classification of
B-branes in minimal model, where we do have identity B-brane.
Naively, Liouville theory would have to have the same set
of B-branes as in minimal model when $k$ is sent to a
negative integer.
This apparent contradiction is due to the consistency
with reflection relation we imposed on B-branes.
Minimal models are theories without continuous spectrum of
representations, and we only consider bulk operators
$V^{j(s,\bar s)}_{m,\bar m}$ with $2j\in\mZ_{\ge0}$ and do not care
about reflection relations.
Classification of B-branes in such models therefore needs a different
treatment, and the modular bootstrap analysis should work.

We will not go into any more detail on these special models
since irrational models with continuous spectrum are of our main interest.

%%%%%%%%%%%%%%%%%%%%%%%%%%%%%%%%%%%%%%%%%%%%%%%
\section{One-Point Functions on a Disc}%%%%%%%%
%%%%%%%%%%%%%%%%%%%%%%%%%%%%%%%%%%%%%%%%%%%%%%%

Here we derive the wave functions for boundary states using Ward
identity of disc correlators containing degenerate fields.
We will see that all the wave functions for A-branes obtained in
previous section satisfy the constraint arising from Ward identity.
For B-branes, this is the only way available for obtaining
wave functions.

The main idea of this analysis is the application of the techniques
invented in \cite{Teschner} to disc correlators.
The analysis along this path has been done in Liouville theory in
\cite{Fateev-ZZ} and $N=1$ super-Liouville theory in
\cite{Fukuda-H2, Ahn-RS}.
For $N=2$ Liouville theory, relevant disc correlators have been
partially analyzed in \cite{Ahn-SY1, Ahn-SY2}.

%%%%%%%%%%%%%%%%%%%%%%%%%%%%%%%%%%%%%%%%%%%%%%%
\subsection{A-branes}%%%%%%%%%%%%%%%%%%%%%%%%%%
%%%%%%%%%%%%%%%%%%%%%%%%%%%%%%%%%%%%%%%%%%%%%%%

The wave functions $U$ for various A-branes were defined
so as to agree with disc one-point structure constants.
Namely, the one point function of bulk operators on the upper
half plane is given by
\begin{equation}
  \vev{V^{j(s,\bar{s})}_{m,\bar{m}}(z,\bar z)}_{A}=
  |z-\bar{z}|^{-2h}U_A(j,m,s)\delta_{m,\bar{m}}\delta_{s,\bar s}.
\end{equation}
A powerful constraint on $U$ can be derived from the conformal
bootstrap of disc two-point function involving degenerate operators.
In the following we study those containing $j=1/2$ or $j=k/2$
degenerate fields.
We will use the OPE formulae of bulk operators involving
$j=1/2$ and $j=k/2$ operators (\ref{OPE1}), (\ref{OPE2}),
as well as the expressions for reflection coefficients for
bulk operators (\ref{refcoef}).

%%%%%%%%%%%%%%%%%%%%%%%%%%%%%%%%%%%%%%%%%%%%%%%%%%%%%%%%%
\subsubsection{$\bf \vev{\,V^{1/2}V^j\,}$ for A-branes}%%
%%%%%%%%%%%%%%%%%%%%%%%%%%%%%%%%%%%%%%%%%%%%%%%%%%%%%%%%%
We start with the following correlator
\begin{equation}
\vev{V^{1/2}_{n,n}(z_0)V^{j(s,s)}_{m,m}(z_1)}
 = |z_{0\bar{1}}|^{-4h_0}
   |z_{1\bar{1}}|^{2h_0-2h_1}F(z),~~~
  (n=\pm\tfrac{1}{2},~~~
   z\equiv\left|\tfrac{z_{01}}{z_{0\bar{1}}}\right|^2)
\label{p12pj}
\end{equation}
where $h_0=\frac{4n^2-3}{4k}, h_1=\frac{(m+s)^2-j(j+1)}{k}+\frac{s^2}{2}$.
$V^{1/2}_{n,n}$ does not satisfy the quantization law for $\theta$-momentum
and winding number but is perturbatively well-defined.
$F(z)$ is a solution of a certain differential equation that
arises from superconformal Ward identity, and is expressed
as the following integral
\begin{equation}
  F(z) = z^{\frac{2mn-j}{k}}(1-z)^{-\frac{1}{k}}
         \int dt |t|^{\frac{2j}{k}}|t-z|^{\frac{1}{k}}|t-1|^{\frac{1}{k}}
         \left\{\frac{m}{t}+\frac{n}{t-z}-\frac{n}{t-1}\right\}.
\label{2PT-F1}
\end{equation}
This expression is easily obtained from the free field realization
as a correlator with one screening operator (denoted as
$\tilde\mu\tilde S$ previously) inserted.
The cross-ratio $z$ takes values in $0\le z\le 1$,
so we divide the real line into four segments
\begin{equation}
 (0)~[-\infty,0]~~~
 (1)~[0,z]~~~
 (2)~[z,1]~~~
 (3)~[1,\infty],
\end{equation}
and define $F_i(z)$ by the $t$-integration over the $i$-th segment.
Then the s-channel basis diagonalizing the monodromy
around $z=0$ is given by $F_1$ and $F_3$:
\begin{eqnarray}
    \frac{k\Gamma(1-\frac{2j+1}{k})}
         {\{m+2n(j+1)\}\Gamma(\frac{1}{k})\Gamma(-\frac{2j+2}{k})}
  F_3(z) &=&
  F_+^s(z)   ~\sim~ z^{\frac{2mn-j}{k}},
 \nn \\
    \frac{k\Gamma(1+\frac{2j+1}{k})}
         {(m-2nj)\Gamma(\frac{2j}{k})\Gamma(\frac{1}{k})}
  F_1(z) &=&
  F_-^s(z)   ~\sim~ z^{\frac{2mn+j+1}{k}}.
\label{2PT-F1S}
\end{eqnarray}
$F(z)$ in (\ref{p12pj}) should therefore be written in terms of them as
\begin{equation}
  F(z) = \sum_\pm C_\pm(n,n;j,m,m)U(j\pm\tfrac{1}{2},m+n,s)F_\pm^s(z).
\end{equation}
The t-channel basis diagonalizing the monodromy around $z=1$ is given by
\begin{eqnarray}
 -\frac{\Gamma(-\frac{2}{k})}
       {m\Gamma(\frac{2j}{k})\Gamma(-\frac{2j+2}{k})}
  F_0(z) &=&  F_+^t(z) ~\sim~ (1-z)^{-\frac{1}{k}},
 \nn \\
  \frac{\Gamma(\frac{2}{k})}
       {n\Gamma(\frac{1}{k})\Gamma(\frac{1}{k})}
  F_2(z) &=&  F_-^t(z) ~\sim~ (1-z)^{\frac{1}{k}},
\end{eqnarray}
and the two bases are related via
\begin{equation}
\begin{array}{rcl}
  F_+^s &=& x_{++}F_+^t + x_{+-}F_-^t, \\
  F_-^s &=& x_{-+}F_+^t + x_{--}F_-^t,
\end{array}
\end{equation}
\begin{equation}
\begin{array}{l}
 x_{++} = \displaystyle
   \frac{2m\Gamma(1-\frac{2j+1}{k})\Gamma(\frac{2}{k})}
        {\{m+2n(j+1)\}\Gamma(1-\frac{2j}{k})\Gamma(\frac{1}{k})}, \\
 x_{-+} = \displaystyle
  \frac{2m\Gamma(1+\frac{2j+1}{k})\Gamma(\frac{2}{k})}
       {(m-2nj)\Gamma(1+\frac{2j+2}{k})\Gamma(\frac{1}{k})},
\end{array}
\begin{array}{l}
 x_{+-} = \displaystyle
  -\frac{2nk\Gamma(1-\frac{2j+1}{k})\Gamma(-\tfrac{2}{k})}
        {\{m+2n(j+1)\}\Gamma(-\frac{2j+2}{k})\Gamma(-\tfrac{1}{k})}, \\
 x_{--} = \displaystyle
  -\frac{2nk\Gamma(1+\frac{2j+1}{k})\Gamma(-\tfrac{2}{k})}
        {(m-2nj)\Gamma(\frac{2j}{k})\Gamma(-\tfrac{1}{k})}.
\end{array}
\label{xs01}
\end{equation}
The term proportional to $F_-^t$ in the t-channel represents
the operator $V^{1/2}_{n,n}$ approaching the boundary
and turning into identity operator:
\begin{equation}
 V^{1/2}_{n,n}(z) \to u(n)|z-\bar{z}|^{\frac{1}{k}} + \cdots.
\end{equation}
We thus obtain the following recursion relation for $U$:
\begin{equation}
 u(n)
 U(j,m,s) = \sum_\pm x_{\pm-}C_\pm(nn,jmm)U(j\pm\tfrac{1}{2},m+n,s).
\end{equation}
or more explicitly
\begin{eqnarray}
\lefteqn{
 \frac{u(n)\Gamma(-\tfrac{1}{k})}{\nu^{\frac12}\Gamma(-\frac{2}{k})}
 \frac{U(j,m,s)}{U_{[1]}(j,m,s)}
 \sin\tfrac{(2j+1)\pi}{k}
} \nn\\ &=&
 \frac{U(j+\tfrac{1}{2},m+n,s)}{U_{[1]}(j+\tfrac{1}{2},m+n,s)}
 \sin\tfrac{(2j+2)\pi}{k}
+\frac{U(j-\tfrac{1}{2},m+n,s)}{U_{[1]}(j-\tfrac{1}{2},m+n,s)}
 \sin\tfrac{2j\pi}{k}.
\end{eqnarray}
The wave functions obtained in the previous subsection all satisfy
this constraint with 
\begin{equation}
 u_{[J,M]}(n) = \frac{2\nu^{\frac12}\Gamma(-\frac2k)}{\Gamma(-\frac1k)}
 e^{\frac{4\pi iMn}{k}}\cos\tfrac{(2J+1)\pi}{k}
\end{equation}
for all branes labelled by $[J,M]$ (non-chiral non-degenerate branes,
degenerate branes $[J_{r,s},M]$, chiral branes $[J,M]^\pm$ and
degenerate chiral branes $[J,M]^{\rm dc}$).

%%%%%%%%%%%%%%%%%%%%%%%%%%%%%%%%%%%%%%%%%%%%%%%%%%%%%%%%%
\subsubsection{$\bf \vev{\,V^{k/2}V^j\,}$ for A-branes}%%
%%%%%%%%%%%%%%%%%%%%%%%%%%%%%%%%%%%%%%%%%%%%%%%%%%%%%%%%%

We can derive another recursion relation from the two-point
function involving $j=\frac{k}{2}$ degenerate representation.
Consider the following correlator on a disc:
\begin{equation}
\vev{V^{k/2}_{n\bar{n}}(z_0)V^{j(s,s)}_{m\bar{m}}(z_1)}
 = z_{0\bar1}^{-2h_0}
   z_{1\bar0}^{-h_0-\bar h_0-h_1+\bar h_1}
   z_{1\bar1}^{h_0+\bar h_0-h_1-\bar h_1}
   z_{\bar0\bar1}^{h_0-\bar h_0+h_1-\bar h_1}
   F(z).~~~~
\label{pk2pj}
\end{equation}
where $z=\left|\tfrac{z_{01}}{z_{0\bar{1}}}\right|^2$ and
\begin{equation}
  h_0 = \tfrac{n^2}{k}-\tfrac{k+2}{4},~~~
  \bar h_0 = \tfrac{\bar n^2}{k}-\tfrac{k+2}{4},~~~
  h_1 = \tfrac{(m+s)^2-j(j+1)}{k}+\tfrac{s^2}{2},~~~
  \bar h_1 = \tfrac{(\bar m+s)^2-j(j+1)}{k}+\tfrac{s^2}{2}.
\end{equation}
The conservation of R-charge requires
\begin{equation}
 n+m=\bar{n}+\bar{m}.
\end{equation}
The function $F$ is expressed as a contour integral of the form
\begin{eqnarray}
F(z) &=&
   z^{\frac{2nm}{k}-j}(1-z)^{-\frac{2n\bar n}{k}-\frac{k}{2}}
   \int_C dwd\hat{w}
       |w|^{j+m}|w-z|^{\frac{k}{2}+n}|w-1|^{\frac{k}{2}-\bar{n}}
 \nn \\ && \hskip35mm\times
      |\hat{w}|^{j-m}
      |\hat{w}-z|^{\frac{k}{2}-n}
      |\hat{w}-1|^{\frac{k}{2}+\bar{n}}
      |w-\hat{w}|^{-k-1}.
\label{Fqrt}
\end{eqnarray}
This can easily be derived using free fields and screening operators,
and is shown to satisfy the Ward identity.
Note that the solution is unique except for the choice of contours:
at first sight it would appear that by flipping $j$ to $-j-1$
we would obtain a new solution, but it is actually not the case.
As in the previous paragraph, we assume $z$ to take values in
$0\le z\le 1$ and divide the real line into four segments.
Different contours give different functions, and we denote various
functions as follows:
\begin{equation}
 F_{1\hat{1}}(z) \leftrightarrow \{0<w<\hat{w}<z\},~~~~
 F_{\hat{1}2}(z) \leftrightarrow \{0<\hat{w}<z<w<1\},~~{\rm etc.}
\end{equation}
The basis of contour integrals in the s-channel that diagonalizes
the monodromy around $z=0$ is given by the following six:
\begin{equation}
    F_{1\hat{1}},~F_{\hat{1}1},~
    F_{3\hat{3}},~F_{\hat{3}3},~
    F_{1\hat{3}},~F_{\hat{1}3},
\end{equation}
but only four linear combinations out of them are indeed the
solutions of differential equation.
The reason for this is that, since the integrals are along segments,
one must always worry about the boundary term when checking that
these integrals indeed satisfy a differential equation.
A simple way to analyze this is to see whether one can replace the
contours ending on points $0,1,z,\infty$ by those encircling them.
For example, $F_{1\hat3}$ might fail to satisfy a differential
equation due to the boundary $w=0,w=z$ and $\hat w=1,\hat w=\infty$, but
this is not the case since one can replace the contours by those
not ending on those points.
Such replacements of contours are not possible for $F_{1\hat1}$ or
$F_{\hat 11}$, but a certain linear combination of them does have
a closed contour integral expression.
In this way one finds that there are only four solutions as listed below:
(in the following we denote
 $\bs(x)\equiv\sin(\pi x), \bc(x)\equiv \cos(\pi x)$)
\begin{eqnarray}
 F_+^s &=&
 -\frac{\Gamma(-j-\frac{k}{2}+m+n)
        \Gamma(-j-\frac{k}{2}-m-n)\Gamma(-2j)}
       {\pi\Gamma(-j+\bar{m})\Gamma(-j-\bar{m})\Gamma(-2j-k-1)}
  \left\{  \bs(\tfrac{k}{2}-\bar{n})F_{3\hat{3}}
          +\bs(\tfrac{k}{2}+\bar{n})F_{\hat{3}3}
  \right\},
 \nn \\
 F_-^s &=&
 -\frac{\Gamma(1+j-\frac{k}{2}+m+n)\Gamma(1+j-\frac{k}{2}-m-n)\Gamma(2j+2)}
       {\pi\Gamma(1+j+m)\Gamma(1+j-m)\Gamma(2j-k+1)}
  \left\{  \bs(\tfrac{k}{2}-n)F_{1\hat{1}}
          +\bs(\tfrac{k}{2}+n)F_{\hat{1}1}
  \right\},
 \nn \\
 F_\uparrow^s
 &=&
  \frac{\Gamma(2+j+\frac{k}{2}-m-n)
        \Gamma(1+\frac{k}{2}-j-m-n)}
       {\Gamma(1+j-m)\Gamma(-j-\bar{m})
        \Gamma(1+\frac{k}{2}-n)\Gamma(1+\frac{k}{2}-\bar{n})}
  F_{\hat{1}3},
 \nn \\
 F_\downarrow^s
 &=&
  \frac{\Gamma(2+j+\frac{k}{2}+m+n)
        \Gamma(1+\frac{k}{2}-j+m+n)}
       {\Gamma(1+j+m)\Gamma(-j+\bar{m})
        \Gamma(1+\frac{k}{2}+n)\Gamma(1+\frac{k}{2}+\bar{n})}
  F_{1\hat{3}}.
\label{sbasis}
\end{eqnarray}
They form the s-channel basis of solutions with the asymptotics
\begin{equation}
 F_+^s \sim  z^{\frac{2nm}{k}-j}, ~~
 F_-^s \sim  z^{\frac{2nm}{k}+j+1},~~
 F_\uparrow^s \sim   z^{\frac{2nm}{k}+\frac{k}{2}-m-n+1},~~
 F_\downarrow^s \sim z^{\frac{2nm}{k}+\frac{k}{2}+m+n+1}.
\end{equation}
These asymptotics are easily derived by using the function $G_k$
and its properties summarized in the appendix.
$F(z)$ in (\ref{pk2pj}) should therefore be expressed as
\begin{eqnarray}
 e^{-i\pi(h_0+h_1)-\frac{i\pi}{k}(n\bar m-\bar nm)}F(z)
  &=&
  \sum_\pm
  \tilde{C}_\pm(n\bar n;jm\bar m)U(j\pm\tfrac{k}{2},m+n,s)F_\pm^s(z)
 \nn \\ &&
 +\sum_{\updownarrow}\tilde{C}_\updownarrow(n\bar n;jm\bar m)
  U(j,m+n\mp\tfrac{k}{2}\mp1,s\pm1)F_\updownarrow^s(z)
\label{fx2}
\end{eqnarray}
On the other hand, the basis in the t-channel diagonalizing
the monodromy around $z=1$ is given by
\begin{equation}
    F_{0\hat{0}},~F_{\hat{0}0},~
    F_{2\hat{2}},~F_{\hat{2}2},~
    F_{0\hat{2}},~F_{\hat{0}2}.
\end{equation}
 The two bases are related as follows:
\begin{eqnarray}
  F_{1\hat{1}} &=&
  F_{0\hat{0}}\frac{\bs(j-m)\bs(j+\bar{m})}{\bs(k)\bs(k+m-\bar{m})}
 -F_{\hat{0}0}\frac{\bs(j-\bar{m})\bs(k+j+m)}{\bs(k)\bs(k+m-\bar{m})}
 \nn \\ &&
 +\left\{
   \bs(\tfrac{k}{2}+\bar{n})F_{2\hat{2}}
  +\bs(\tfrac{k}{2}-\bar{n})F_{\hat{2}2}
  \right\}
  \frac{\bs(\frac{k}{2}+n)}{\bs(k)\bs(m-\bar{m})}
 -F_{0\hat{2}}\frac{\bs(\tfrac{k}{2}+\bar{n})\bs(j+\bar{m})}
                   {\bs(m-\bar{m})\bs(k+m-\bar{m})}, \nn \\
%%%%
 F_{\hat{1}1}
 &=&
      -F_{0\hat{0}}
   \frac{\bs(j+\bar{m})\bs(k+j-m)}{\bs(k)\bs(k-m+\bar{m})}
      +F_{\hat{0}0}
   \frac{\bs(j-\bar{m})\bs(j+m)}{\bs(k)\bs(k-m+\bar{m})}
 \nn \\ &&
      -\left\{\bs(\tfrac{k}{2}-\bar{n})F_{\hat{2}2}
             +\bs(\tfrac{k}{2}+\bar{n})F_{2\hat{2}}\right\}
   \frac{\bs(\frac{k}{2}-n)}{\bs(k)\bs(m-\bar{m})}
      +F_{\hat{0}2}
   \frac{\bs(\tfrac{k}{2}-\bar{n})\bs(j-\bar{m})}
        {\bs(k-m+\bar{m})\bs(m-\bar{m})},
 \nn \\
%%%%
  F_{3\hat{3}} &=&
  F_{0\hat{0}}\frac{\bs(j-m)\bs(j+\bar{m})}{\bs(k)\bs(k-m+\bar{m})}
 +F_{\hat{0}0}\frac{\bs(j+m)\bs(k-j+\bar{m})}{\bs(k)\bs(k-m+\bar{m})}
 \nn \\ &&
 -\left\{
  \bs(\tfrac{k}{2}+n)F_{2\hat{2}}
 +\bs(\tfrac{k}{2}-n)F_{\hat{2}2}
  \right\}
 \frac{\bs(\frac{k}{2}+\bar{n})}{\bs(k)\bs(m-\bar{m})}
 +F_{\hat{0}2}
 \frac{\bs(\tfrac{k}{2}+n)\bs(j-m)}{\bs(m-\bar{m})\bs(k-m+\bar{m})},
 \nn \\
%%%%
  F_{\hat{3}3} &=&
  F_{0\hat{0}}\frac{\bs(j-m)\bs(k-j-\bar{m})}{\bs(k)\bs(k+m-\bar{m})}
 +F_{\hat{0}0}\frac{\bs(j+m)\bs(j-\bar{m})}{\bs(k)\bs(k+m-\bar{m})}
 \nn \\ &&
 +\left\{
  \bs(\tfrac{k}{2}+n)F_{2\hat{2}}
 +\bs(\tfrac{k}{2}-n)F_{\hat{2}2}
  \right\}
 \frac{\bs(\frac{k}{2}-\bar{n})}{\bs(k)\bs(m-\bar{m})}
 -F_{0\hat{2}}
 \frac{\bs(\tfrac{k}{2}-n)\bs(j+m)}{\bs(m-\bar{m})\bs(k+m-\bar{m})},
 \nn \\
%%%%
 F_{1\hat{3}}
 &=&
  F_{\hat{0}0}
   \frac{\bs(j+m)\bs(j-\bar{m})-\bs(k-j+m)\bs(k+j+\bar{m})}
        {\bs(k-m+\bar{m})\bs(k+m-\bar{m})}
 -F_{0\hat{0}}
   \frac{2\bc(k)\bs(j-m)\bs(j+\bar{m})}
        {\bs(k-m+\bar{m})\bs(k+m-\bar{m})}
 \nn \\ &&
 -F_{0\hat{2}}
  \frac{\bs(j+\bar{m})\bs(\frac{k}{2}-n)}
       {\bs(m-\bar{m})\bs(k+m-\bar{m})}
 +F_{\hat{0}2}
   \frac{\bs(j-m)\bs(\tfrac{k}{2}-\bar{n})}
        {\bs(k-m+\bar{m})\bs(m-\bar{m})},
 \nn \\
%%%%
 F_{\hat{1}3} &=&
  F_{0\hat{0}}
  \frac{\bs(j+\bar{m})\bs(j-m)-\bs(k-j-m)\bs(k+j-\bar{m})}
       {\bs(k-m+\bar{m})\bs(k+m-\bar{m})}
 -F_{\hat{0}0}
  \frac{2\bc(k)\bs(j-\bar{m})\bs(j+m)}
       {\bs(k-m+\bar{m})\bs(k+m-\bar{m})}
 \nn \\ &&
 +F_{\hat{0}2}
  \frac{\bs(\tfrac{k}{2}+n)\bs(j-\bar{m})}
       {\bs(m-\bar{m})\bs(k-m+\bar{m})}
 -F_{0\hat{2}}
  \frac{\bs(\tfrac{k}{2}+\bar{n})\bs(j+m)}
       {\bs(m-\bar{m})\bs(k+m-\bar{m})}.
\end{eqnarray}
The t-channel describes the degenerate operator
$V^{k/2}_{n\bar{n}}$ approaching the boundary and decomposing
into a sum of boundary operators.
To derive a recursion relation for the one-point structure constants,
we would first be interested in the terms proportional to the
boundary $j=0$ operator.
They should behave like $\sim(1-z)^{-\frac{2n\bar n}{k}+\frac{k}{2}+1}$
and are proportional to $F_{2\hat{2}}$ or $F_{\hat{2}2}$.
Very surprisingly, those functions do not appear when we express
the functions $F^s_\pm,F^s_\updownarrow$ in terms of t-channel basis.
Therefore we have to focus on the terms proportional to
\begin{eqnarray}
 F_\uparrow^t &=&
  \frac{\Gamma(1+m-\bar{m})\Gamma(k+m-\bar{m}+2)F_{0\hat{2}}}
       {\Gamma(j+m+1)\Gamma(-j-\bar{m})
        \Gamma(\frac{k}{2}-n+1)\Gamma(\frac{k}{2}+\bar{n}+1)}
  ~\sim~
  (1-z)^{-\frac{2n\bar n}{k}+\frac{k}{2}-n+\bar{n}+1},
 \nn \\
 F_\downarrow^t &=&
  \frac{\Gamma(1-m+\bar{m})\Gamma(k-m+\bar{m}+2)F_{\hat{0}2}}
       {\Gamma(j-m+1)\Gamma(-j+\bar{m})
        \Gamma(\frac{k}{2}+n+1)\Gamma(\frac{k}{2}-\bar{n}+1)}
  ~\sim~
  (1-z)^{-\frac{2n\bar n}{k}+\frac{k}{2}+n-\bar{n}+1},
\end{eqnarray}
 which correspond to the terms in the self-OPE
\begin{eqnarray}
 V^{k/2}_{n\bar{n}}(z) &\to&
 |z-\bar{z}|^{-\frac{2n\bar n}{k}+\frac{k}{2}-n+\bar{n}+1}
 \tilde{u}^\uparrow(n,\bar{n})
 B^{\frac k2(1)}_{n-\bar{n}-\frac{k}{2}-1}(z)
 \nn \\ &&
+|z-\bar{z}|^{-\frac{2n\bar n}{k}+\frac{k}{2}+n-\bar{n}+1}
 \tilde{u}^\downarrow(n,\bar{n})
 B^{\frac k2(-1)}_{n-\bar{n}+\frac{k}{2}+1}(z)
 + \cdots
\label{pk2so}
\end{eqnarray}
We will also have to consider the bulk-to-boundary propagators
 for some special case:
\begin{eqnarray}
 \vev{B^{k/2 (\pm1)}_{n}(x)V^{j(s,s)}_{m\bar{m}}(z)}
 &=&
   |z-\bar{z}|^{h_2-h_1-\bar h_1}
   (x-z)^{\bar h_1-h_1-h_2}(x-\bar z)^{h_1-\bar h_1-h_2}
 \nn \\ && \times 
   \delta(m-\bar{m}+n\pm\tfrac{k}{2}\pm1)
   U^\pm(j,m,\bar{m}).
\label{bbp}
\end{eqnarray}
where $h_1,\bar h_1$ are the same as before and
$h_2=\frac{(n\pm1)^2}{k}-\frac k4$.
The s/t-channel bases are related as follows:
\begin{equation}
\begin{array}{rcl}
 F_+^s
  &=& x_{+\uparrow  }\Gamma(-m+\bar{m})F_\uparrow^t
    + x_{+\downarrow}\Gamma(m-\bar{m})F_\downarrow^t +\cdots, \\
 F_-^s
  &=& x_{-\uparrow  }\Gamma(-m+\bar{m})F_\uparrow^t
    + x_{-\downarrow}\Gamma(m-\bar{m})F_\downarrow^t +\cdots,\\
 F_\uparrow^s
  &=& x_{\uparrow\uparrow  }\Gamma(-m+\bar{m})F_\uparrow^t
    + x_{\uparrow\downarrow}\Gamma(m-\bar{m})F_\downarrow^t +\cdots,  \\
 F_\downarrow^s
  &=& x_{\downarrow\uparrow  }\Gamma(-m+\bar{m})F_\uparrow^t
    + x_{\downarrow\downarrow}\Gamma(m-\bar{m})F_\downarrow^t +\cdots,
\end{array}
\end{equation}
\begin{eqnarray}
 x_{+\uparrow}
  &=&
  \frac{\Gamma(-1-k-m+\bar{m})
        \Gamma(-j-\frac{k}{2}+m+n)\Gamma(-j-\frac{k}{2}-m-n)\Gamma(-2j)}
       {\Gamma(-j-m)\Gamma(-j+\bar{m})
        \Gamma(-\frac{k}{2}+n)\Gamma(-\frac{k}{2}-\bar{n})\Gamma(-2j-k-1)},
 \nn \\
 x_{-\uparrow}
 &=&
  \frac{\Gamma(-1-k-m+\bar{m})
        \Gamma(1+j-\frac{k}{2}+m+n)\Gamma(1+j-\frac{k}{2}-m-n)\Gamma(2j+2)}
       {\Gamma(1+j-m)\Gamma(1+j+\bar{m})
        \Gamma(-\frac{k}{2}+n)\Gamma(-\frac{k}{2}-\bar{n})
        \Gamma(2j-k+1)},
 \nn \\
 x_{\uparrow\uparrow}
 &=&
  \frac{\Gamma(-1-k-m+\bar{m})
        \Gamma(2+j+\frac{k}{2}-m-n)\Gamma(1+\frac{k}{2}-j-m-n)}
       {\Gamma(1+j-m)\Gamma(-j-m)
        \Gamma(1+\frac{k}{2}-\bar{n})\Gamma(-\frac{k}{2}-\bar{n})},
 \nn \\
 x_{\downarrow\uparrow}
 &=&
  \frac{\Gamma(-1-k-m+\bar{m})
        \Gamma(2+j+\frac{k}{2}+m+n)\Gamma(1+\frac{k}{2}-j+m+n)}
       {\Gamma(1+j+\bar{m})\Gamma(-j+\bar{m})
        \Gamma(1+\frac{k}{2}+n)\Gamma(-\frac{k}{2}+n)}.
\label{st1}
\end{eqnarray}
The coefficients $x_{\pm\downarrow},x_{\updownarrow\downarrow}$
are obtained from $x_{\pm\uparrow},x_{\updownarrow\uparrow}$ by the exchange
$m\leftrightarrow\bar{m}$, $n\leftrightarrow\bar{n}$.

The basis change law  becomes singular when $m-\bar{m}$
is an integer, which is actually the case for all the perturbatively
well-defined vertex operators $V^{j(s,s)}_{m\bar{m}}$.
As a consequence, the solutions of differential equation develop
a logarithm at $z\sim 1$ and signal the emergence of a logarithmic
operator on the boundary.
This logarithm can be understood in the following way.
Let us focus on the case $m=\bar{m}$, and recall that the boundary
operators $B^{k/2(\pm1)}_m$ with $m=\mp(\frac{k}{2}+1)$ are
expected from the representation theory to behave like identity.
Looking at the basis change law above, it is expected
that $B^{k/2(\pm1)}_{m\mp\frac{k}{2}\mp1}$ approaches identity
as $m\to 0$ with divergent coefficient:
\begin{equation}
  B^{k/2 (\pm1)}_{m\mp\frac{k}{2}\mp1}
 ~\stackrel{m\to 0}{\to}~ c^\updownarrow\Gamma(\pm m)\times{\bf 1}.
\end{equation}
Since we have two sets of operators (both parametrized by $m$)
approaching the identity as $m\to 0$, one can define the logarithmic
operator by their difference.
This is analogous to the case of the free boson theory of $\phi$,
where we have continuously many primary operators $e^{ia\phi}$.
$e^{\pm ia\phi}$ have the same conformal weight $h=\frac{a^2}{2}$,
except at $h=0$ we have two operators $1$ and $\phi$, the latter of
which is logarithmic and is obtained by $a$-derivative.

The above argument also shows that the bulk-boundary propagators
(\ref{bbp}) become proportional to $U(j,m)$ when $m=\bar{m}$:
\begin{equation}
  U^\pm(j,m,\bar{m})\stackrel{m\to\bar{m}}{\to}
  c^\updownarrow\Gamma(\mp m\pm\bar{m})U(j,m).
\end{equation}
Using this together with (\ref{fx2}), (\ref{st1}) we obtain another
relation between one-point structure constants:
\begin{eqnarray}
  c^\updownarrow\tilde{u}^\updownarrow(n,n)U(j,m,s)
 &=&
  ~~x_{+\updownarrow}\tilde{C}_+(nn;jmm)
  U(j+\tfrac{k}{2},m+n,s)
 \nn \\ &&
   +x_{-\updownarrow}\tilde{C}_-(nn;jmm)
  U(j-\tfrac{k}{2},m+n,s)
 \nn \\ &&
   +x_{\uparrow\updownarrow}\tilde{C}_\uparrow(nn;jmm)
  U(j,m+n-\tfrac{k}{2}-1,s+1)
 \nn \\ &&
   +x_{\downarrow\updownarrow}\tilde{C}_\downarrow(nn;jmm)
  U(j,m+n+\tfrac{k}{2}+1,s-1).
\label{Urec2a}
\end{eqnarray}
or more explicitly,
\begin{eqnarray}
\lefteqn{
  c^\updownarrow\tilde{u}^\updownarrow(n,n)
  \frac{\Gamma(-\frac{k}{2}+n)\Gamma(-\frac{k}{2}-n)}
       {k\nu^{\frac k2}\Gamma(-1-k)}
 \times \hat U(j,m,s)
} \nn\\
 &=&
  -\hat U(j+\tfrac k2,m+n,s)+\hat U(j,m+n-\tfrac k2-1,s+1)
 \nn\\ &&
  -\hat U(j-\tfrac k2,m+n,s)+\hat U(j,m+n+\tfrac k2+1,s-1),
 \nn\\ &&
 \hat U(j,m,s) ~=~
  \frac{U(j,m,s)\bs(2j)}
       {U_{[1]}(j,m,s)\bs(j+m)\bs(j-m)}.
\end{eqnarray}
The wave functions obtained in the previous section
all satisfy this equation.
The structure constants $c^\updownarrow, \tilde{u}^\updownarrow(n,n)$ satisfy
\begin{equation}
  c^\updownarrow_{[J,M]}\tilde u^\updownarrow_{[J,M]}(n,n) =
  4\sin\{(J+M)\pi\}\sin\{(J-M)\pi\}
  \frac{e^{\frac{4\pi iMn}{k}}k\nu^{\frac k2}\Gamma(-1-k)}
       {\Gamma(-\frac{k}{2}+n)\Gamma(-\frac{k}{2}-n)}
\end{equation}
for all A-branes.
Note that it vanishes for the A-branes corresponding to chiral
representations $[J,M]^\pm$ and $[J,M]^{\rm dc}$.

%%%%%%%%%%%%%%%%%%%%%%%%%%%%%%%%%%%%%%%%%%%%%%%
\subsection{B-branes}%%%%%%%%%%%%%%%%%%%%%%%%%%
%%%%%%%%%%%%%%%%%%%%%%%%%%%%%%%%%%%%%%%%%%%%%%%

The disc one-point function for B-branes takes the form
\begin{equation}
  \vev{V^{j(s,\bar s)}_{m,\bar m}(z)}_B = |z-\bar z|^{-2h}
  T(j,m,s)\delta_{m+\bar m,0}\delta_{s+\bar s,0}
\end{equation}
where $m+\bar m=s+\bar s=0$ follows from $L_0-\bar L_0=J_0+\bar J_0=0$.
The functional form of $T$ is largely determined from symmetry
argument.
First, since $s$ is the momentum of bosonization of conserved
$U(1)$ current, it is reasonable to assume its $s$-dependence
to be simply $e^{ias}$ for some constant $a$.
Then the boundary condition on supercurrents $T_F^\pm=\bar T_F^\pm$ require
\begin{equation}
 T(j,m,s) = e^{ia(m+s)}\nu^{j+\frac12}
          \frac{\Gamma(-2j)\Gamma(-\frac{2j+1}{k})}
               {\Gamma(-j+m)\Gamma(-j-m)}T_0(j,m)
\end{equation}
where some functions of $j$ were put for later convenience,
and $T_0$ is periodic in $m$ with unit period.
Noticing that $m\in\frac12\mZ$ for perturbatively well-defined
operators, one finds such periodic functions are proportional
either to $1$ or $e^{2\pi im}$.
Writing $T_0(j,m)=\hat T(j)+e^{2\pi im}\check T(j)$, one finds
$\check T(j)=\hat T(-j-1)$ from the reflection relation.
Thus all we are left with is to determine an unknown function $\hat T(j)$ in
the {\it ansatz}
\begin{equation}
 T(j,m,s) = e^{ia(m+s)}\nu^{j+\frac12}
          \frac{\Gamma(-2j)\Gamma(-\frac{2j+1}{k})}
               {\Gamma(-j+m)\Gamma(-j-m)}
  \{\hat T(j)+e^{2\pi im}\hat T(-j-1)\}.
\end{equation}
One might think of other ansaetze, but they all
reduce to the above one under the condition $m\in\frac12\mZ$.
For example, the ansatz
\begin{equation}
 T(j,m,s) = e^{ia(m+s)}\nu^{j+\frac12}
          \frac{\Gamma(1+j+m)\Gamma(1+j-m)}
               {\Gamma(2j+2)\Gamma(\frac{2j+1}{k})}
  \{\hat f(j)+e^{2\pi im}\hat f(-j-1)\}
\end{equation}
is related to the previous ansatz by
\begin{equation}
  \hat T(j)\pm \hat T(-j-1) = -\frac{2}{k}
  \frac{\bs(2j)\bs(\frac{2j+1}{k})}{\bc(2j)\mp1}\{f(j)\pm f(-j-1)\}.
\end{equation}
The analysis of the disc correlators for B-branes proceeds
in a similar way as for A-branes.
First, the disc two-point function containing $j=1/2$ degenerate
operator takes the form
\begin{equation}
\vev{V^{1/2}_{n,-n}(z_0)V^{j(s,-s)}_{m,-m}(z_1)}
 = |z_{0\bar{1}}|^{-4h_0}
   |z_{1\bar{1}}|^{2h_0-2h_1}F(z),~~~
  (n=\pm\tfrac{1}{2},~~~
   z\equiv\left|\tfrac{z_{01}}{z_{0\bar{1}}}\right|^2)
\end{equation}
where $F(z)$ is the same contour integral expression as for A-branes,
\begin{equation}
  F(z) = z^{\frac{2mn-j}{k}}(1-z)^{-\frac{1}{k}}
         \int dt |t|^{\frac{2j}{k}}|t-z|^{\frac{1}{k}}|t-1|^{\frac{1}{k}}
         \left\{\frac{m}{t}+\frac{n}{t-z}-\frac{n}{t-1}\right\}.
\end{equation}
Using the same bases $F_\pm^s$ as before, one finds
\begin{eqnarray}
  F(z) &=&  \sum_\pm C_\pm(n,-n;j,m,-m)T(j\pm\tfrac{1}{2},m+n,s)F_\pm^s(z)
 \nn\\
       &=& t(n)T(j,m,s)F_-^t(z)+\cdots,
\end{eqnarray}
where $t(n)$ is the self OPE coefficient of $V^{1/2}_{n,-n}$
turning into boundary identity operator,
\begin{equation}
  V^{1/2}_{n,-n}(z) \to t(n)|z-\bar{z}|^{\frac{1}{k}} + \cdots.
\end{equation}
We thus obtain a recursion relation for $T$:
\begin{equation}
 t(n)T(j,m,s) =
 ~ \sum_\pm x_{\pm-}C_\pm(n,-n;j,m,-m)T(j\pm\tfrac{1}{2},m+n,s),
\end{equation}
where $x_{\pm-}$ are given in (\ref{xs01}).
In terms of $\hat T$ it becomes simple.
Introducing
\begin{equation}
 \hat T_\pm = \hat T(j)\pm\hat T(-j-1)
\end{equation}
one finds
\begin{equation}
  q\hat T_\pm(j) = \hat T_\mp(j+\tfrac12) - \hat T_\mp(j-\tfrac12),~~~
  q=\frac{t(n)\Gamma(-\frac1k)}{e^{ian}\nu^{\frac12}\Gamma(-\frac2k)}.
\label{hatT1}
\end{equation}

Next, the two-point function containing $j=k/2$ operator is
\begin{equation}
\vev{V^{k/2}_{n,-\bar{n}}(z_0)V^{j(s,-s)}_{m,-\bar{m}}(z_1)}
 = z_{0\bar1    }^{-2h_0}
   z_{1\bar0    }^{-h_0-\bar h_0-h_1+\bar h_1}
   z_{1\bar1    }^{ h_0+\bar h_0-h_1-\bar h_1}
   z_{\bar0\bar1}^{ h_0-\bar h_0+h_1-\bar h_1}
   F(z),
\end{equation}
where $z=\left|\tfrac{z_{01}}{z_{0\bar{1}}}\right|^2$ and $F(z)$
is the same contour integral (\ref{Fqrt}) as was given for A-branes.
$F(z)$ should be expressed in terms of s-channel basis as
\begin{equation}
 e^{-i\pi(h_0+h_1)+\frac{i\pi}{k}(n\bar m-\bar nm)}
 F(z) ~=~
  \sum_\pm\tilde{C}_\pm(n,-\bar{n};j,m,-\bar{m})
  T(j\pm\tfrac{k}{2},m+n,s)F_\pm^s(z),
\end{equation}
because the one-point function vanishes for operators with
$s+\bar s\neq 0$ and therefore we cannot have terms proportional to
$\tilde C_\updownarrow$ in the right hand side.
As before, after rewriting $F(z)$ in t-channel basis
we focus on the terms proportional to $F_\updownarrow^t$
which correspond to the following terms in the self-OPE
\begin{eqnarray}
 V^{k/2}_{n,-\bar n}(z) &\to&
 |z-\bar{z}|^{-\frac{2n\bar n}{k}+\frac{k}{2}-n+\bar n+1}
 \tilde{t}^\uparrow(n,-\bar n)
 B^{k/2(1)}_{n-\bar n-\frac{k}{2}-1}(z)
 \nonumber \\ &&
+|z-\bar{z}|^{-\frac{2n\bar n}{k}+\frac{k}{2}+n-\bar n+1}
 \tilde{t}^\downarrow(n,-\bar n)
 B^{k/2(-1)}_{n-\bar n+\frac{k}{2}+1}(z)
 + \cdots
\end{eqnarray}
The two-point function exhibits a logarithmic behavior at $z=1$
when $n-\bar n=0$, and we interpret it as the degeneracy of
the following boundary operators:
\begin{equation}
  B^{k/2(\pm1)}_{m\mp\frac{k}{2}\mp1}
 ~\stackrel{m\to 0}{\to}~ c^\updownarrow\Gamma(\mp m)\times {\bf 1}.
\end{equation}
Thus we obtain a recursion relation:
\begin{eqnarray}
  c^\updownarrow\tilde{t}^\updownarrow(n,-n)T(j,m,s)
 &=&
  ~~x_{+\updownarrow}\tilde{C}_+(n,-n;j,m,-m)T(j+\tfrac{k}{2},m+n,s)
 \nonumber \\ &&
   +x_{-\updownarrow}\tilde{C}_-(n,-n;j,m,-m)T(j-\tfrac{k}{2},m+n,s)
\end{eqnarray}
where $x_{\pm\updownarrow}$ are the ones given in (\ref{st1}).
In terms of $\hat T(j)$ this can be rewritten as
\begin{equation}
 p_0\hat T_\pm(j) = \hat T_\pm(j+\tfrac k2)+\hat T_\pm(j-\tfrac k2),~~~
 p_1\hat T_\pm(j) = \hat T_\mp(j+\tfrac k2)-\hat T_\mp(j-\tfrac k2),
\label{hatT2}
\end{equation}
where
\begin{eqnarray}
 p_{0,1} = \left.
  \frac{c^\updownarrow\tilde t^\updownarrow(n,-n)
        \Gamma(-\frac k2+n)\Gamma(-\frac k2-n)}
       {e^{ian}k\nu^{\frac k2}\Gamma(-k-1)}
 \right|_{2n={\rm even,odd}}
\end{eqnarray}

A one-parameter family of solutions to (\ref{hatT1}), (\ref{hatT2})
can be found easily:
\begin{equation}
  \hat T(j) =   \exp(2j+1)u,~~~
  q   = 2\sinh u,~~~
  p_0 = 2\cosh ku,~~~
  p_1 = 2\sinh ku.
\end{equation}
Using labels $[J,M]$ instead of $(a,u)$, we summarize the result
for B-branes below.
\begin{eqnarray}
 T_{[J,M]}(j,m,s) &=& T_0\nu^{j+\frac12}e^{\frac{4\pi iM}{k}(m+s)}
 \frac{\Gamma(-2j)\Gamma(-\frac{2j+1}{k})}{\Gamma(-j+m)\Gamma(-j+m)}
 \nn\\ && \times
 \left\{e^{\frac{i\pi}{k}(2j+1)(2J+1)}
       +e^{2\pi im}e^{-\frac{i\pi}{k}(2j+1)(2J+1)} \right\}, \nn\\
 t(n) &=& 2ie^{\frac{4\pi iMn}{k}}\nu^{\frac12}
  \frac{\Gamma(-\frac2k)}{\Gamma(-\frac1k)}\sin\{\tfrac{\pi}{k}(2J+1)\},\nn\\
 c^\updownarrow\tilde t^\updownarrow(n,-n) &=&
  \frac{e^{\frac{4\pi iMn}{k}}k\nu^{\frac k2}\Gamma(-k-1)}
       {\Gamma(-\frac k2+n)\Gamma(-\frac k2-n)}
  (e^{i\pi(2J+1)}+e^{2\pi in}e^{-i\pi(2J+1)}).
\label{Bwave}
\end{eqnarray}

%%%%%%%%%%%%%%%%%%%%%%%%%%%%%%%%%%%%%%%%%%%%%%%%%
\section{Boundary Interactions}%%%%%%%%%%%%%%%%%%
%%%%%%%%%%%%%%%%%%%%%%%%%%%%%%%%%%%%%%%%%%%%%%%%%

Here we discuss the Lagrangian description of various boundary
states and possible boundary interactions in $N=2$ Liouville theory.
Some aspects of this issue have been studied in
\cite{Ahn-Y,Ahn-SY1,Ahn-SY2}.

In general $N=(2,2)$ Landau-Ginzburg models defined by the action
\begin{equation}
  S = \int d^2z d^4\theta K(\Phi^i,\bar\Phi^i)
    + \int d^2z d\theta^+d\bar\theta^+ W(\Phi^i)
    + \int d^2z d\theta^-d\bar\theta^- \bar W(\bar\Phi^i),
\end{equation}
there are several ways to preserve supersymmetry on worldsheets with
boundary\cite{Govindarajan-JS,Hori-IV,Hori,Hellerman-KLM,
              Kapustin-L,Brunner-HLS}.
One way is to put boundary conditions on fields $\Phi^j$;
the boundary states are then naturally associated to the submanifolds
of $\Phi^j$-space defined by the boundary conditions.
A-branes are Lagrangian submanifolds which should also be
pre-images of a straight line in complex $W$-plane,
whereas B-branes are holomorphic submanifolds which
are level-sets of $W$ \cite{Govindarajan-JS, Hori-IV}.
More recently it has been found that the {\it matrix factorization}
enables one to describe B-branes in LG models in terms of
certain boundary interactions which involve a Chan Paton
degree of freedom \cite{Kapustin-L,Brunner-HLS}.
In this section we first propose the form of boundary interaction
for B-branes using this approach, and reproduce a few disc structure
constants obtained in the previous section from perturbative
computation.
Then we make a similar proposal for A-branes.

\subsection{B-branes}

Consider as a LG theory of a single chiral field $\Phi$,
and assume the superpotential $W$ factorizes as
$W(\Phi)=\frac12f(\Phi)g(\Phi)$.
Then on the B-boundary\cite{Hori} defined by
\begin{equation}
  z=\bar z,~~~ \theta^+=\bar\theta^+,~~~\theta^-=\bar\theta^-,
\end{equation}
one introduces the boundary supercovariant derivative
\begin{equation}
  D^B_\pm ~=~ \frac{\partial}{\partial\theta^\pm} -i\theta^\mp\partial_x,
  ~~~~(x\equiv {\rm Re}(z))
\end{equation}
and the fermionic superfields $\Gamma,\bar\Gamma$ satisfying
\begin{equation}
  D^B_-\Gamma = g(\Phi),~~~
  D^B_+\bar\Gamma = \bar g(\bar\Phi),
\end{equation}
in terms of which the boundary interaction is expressed in the
following way:
\begin{equation}
  S_{\rm boundary} ~=~ -\frac12\oint dx\left[
  \int d\theta^+d\theta^- \bar\Gamma\Gamma
 +\int d\theta^+\Gamma f(\Phi)
 +\int d\theta^-\bar\Gamma\bar f(\bar\Phi)
  \right].
\end{equation}
Using the $\theta$-expansion
\begin{equation}
\begin{array}{rcl}
 \Gamma &=& \lambda + \theta^-g(\phi)+\theta^+ G
           -i\theta^+\theta^-\left\{
           \partial_x\lambda+\sqrt2g'(\phi)(\psi_++\bar\psi_+)\right\},\\
\bar\Gamma &=& \bar\lambda +\theta^+\bar g(\bar\phi)+\theta^-\bar G
           +i\theta^+\theta^-\left\{
             \partial_x\bar\lambda
            +\sqrt2\bar g'(\bar\phi)(\psi_-+\bar\psi_-)\right\}
\end{array}
\end{equation}
the boundary interaction can be rewritten as follows
\begin{eqnarray}
  {\cal L}_{\rm boundary} &=&
 -i\bar\lambda\dot\lambda
  +\tfrac12|g|^2 -\tfrac12|G|^2-\tfrac12Gf-\tfrac12\bar G\bar f
 \nn\\ &&
  +\tfrac{i}{\sqrt{2}}\{\lambda f'-\bar\lambda g'\}(\psi_++\bar\psi_+)
  +\tfrac{i}{\sqrt{2}}\{\bar\lambda\bar f'-\lambda\bar g'\}(\psi_-+\bar\psi_-).
\end{eqnarray}
The B-type supersymmetry variation of $S_{\rm boundary}$ precisely
cancels the surface term arising from the variation of the bulk action.
It is easy to rotate the boundary condition by R-symmetry,
although we have not taken it into account explicitly.

We apply this prescription to the B-type boundary states in $N=2$
Liouville theory.
At first sight, non-trivial factorizations break the invariance
under the unit period shift of $\theta$, but the theory remains
invariant if we let the boundary fermions $\lambda,\bar\lambda$
(or more precisely all the components in the superfields
$\Gamma,\bar\Gamma$) transform as $g(\Phi), \bar g(\bar\Phi)$.
We will only consider the cases with
\[f\sim g\sim  W^{1/2},\]
because a copy of $N=2$ superconformal symmetry is unbroken
only for this choice \cite{Ahn-Y}.
Note also that the boundary interactions then become precisely the
{\it holomorphic square roots} of bulk interactions,
and the theory can still be regarded as a perturbed free CFT
after the irrelevant terms are discarded.
After suitably normalizing boundary fields and incorporating the effects
of nonzero worldsheet curvature, the boundary action can be
written as follows:
\begin{eqnarray}
\lefteqn{
\oint dx\left[
     \bar\lambda\partial_x\lambda -\sqrt{\tfrac2k}\tfrac{K\rho}{4\pi}\right]
    +\mu_BS_B+\mu_{\bar B}S_{\bar B}
    +\bar\mu_B\bar S_B+\bar\mu_{\bar B}\bar S_{\bar B}
}\nn\\
&=&  \oint dx\left[
     \bar\lambda\partial_x\lambda -\sqrt{\tfrac2k}\tfrac{K\rho}{4\pi}
    -(\mu_B\lambda+\mu_{\bar B}\bar\lambda)e^{-\sqrt{\frac k2}\phi_L+iH_L}
    -e^{-\sqrt{\frac k2}\bar\phi_L-iH_L}
     (\bar\mu_B\lambda+\bar\mu_{\bar B}\bar\lambda)
  \right],
\label{bintsB}
\end{eqnarray}
where $K$ denotes the curvature of the boundary appearing in
the Euler number formula
\begin{equation}
 \chi = 2-2\sharp(\mbox{handles})-\sharp(\mbox{holes})
      =\int_\Sigma\frac{\sqrt{g}R}{4\pi} +\int_{\partial\Sigma}\frac{K}{2\pi}.
\end{equation}
The terms proportional to $\mu_B,\bar\mu_B$, etc will be called
the boundary screening operators.
From the condition $fg=2W$ one finds
\begin{equation}
 \mu_B\mu_{\bar B} ~=~ \bar\mu_B\bar\mu_{\bar B}~=~ \frac{\mu k}{2\pi}.
\label{factor}
\end{equation}
Note that the boundary fermions were renormalized to have the standard
propagator,
\begin{equation}
 \vev{\lambda(x)\bar\lambda(x')}~=~
 \vev{\bar\lambda(x)\lambda(x')}~=~\frac12{\rm sign}(x-x').
\end{equation}
It also follows from this that any non-vanishing correlator of
$\lambda,\bar\lambda$ is taking values $\pm1/2$, e.g.,
\begin{equation}
 \vev{\lambda(x_1)\bar\lambda(x_{\bar 1})\cdots
      \lambda(x_n)\bar\lambda(x_{\bar n})} ~=~
 \vev{\bar\lambda(x_1)\lambda(x_{\bar 1})\cdots
      \bar\lambda(x_n)\lambda(x_{\bar n})} ~=~ \frac12~~~
 (x_1>x_{\bar 1}>\cdots>x_n>x_{\bar n}).
\end{equation}

The boundary fermions introduce the Chan-Paton degree of
freedom on each boundary.
For example, the Hamiltonian quantization of the theory on the strip
$(0\le \sigma\le \pi,~\tau\in\mR)$ has two sets of fermions
$\lambda_0(\tau),\bar\lambda_0(\tau)$ and
$\lambda_\pi(\tau),\bar\lambda_\pi(\tau)$, which under the free field
approximation satisfy the standard anti-commutation relation.
So the Chan-Paton space is two-dimensional for each boundary,
and is spanned by $\ket{0}$ and $\ket{1}=\bar\lambda\ket{0}$
where $\ket{0}$ is annihilated by $\lambda$.

From the viewpoint of perturbed free CFT, one can also consider
the following interaction:
\begin{equation}
\tilde\mu_B\tilde S_B
~=~  -\tilde\mu_B\oint dx
      (\lambda\bar\lambda-\bar\lambda\lambda)
      (\psi_+\psi_--i\sqrt{2k}\partial\theta)e^{-\sqrt{\frac2k}\rho_L}
\label{tilmub}
\end{equation}
which is also a holomorphic square root of a bulk screening operator.
The above operator depends on boundary fermions in a strange manner,
but the reason will be explained shortly.

\subsubsection{Computation of disc correlators}

The relations between the labels of boundary states and the boundary
couplings can be obtained by computing some disc structure constants
from free field approach.
Let us begin by setting up the consistent rules for computing
correlators on the upper half plane.
Namely, we need to be able to calculate correlators so that they
are either invariant or flipping sign under re-orderings of
operators appearing in a correlator $\vev{\cdots}$, when all
the operators are in the physical spectrum and their Grassmann
parity is suitably defined.

The Neumann boundary conditions on the fields $\rho,\theta,H$
correlate the left- and right-moving sectors of the theory.
We evaluate it using the propagators,
\begin{equation}
 \vev{\rho_L(z)\rho_R(\bar z')}~=~
 \vev{\theta_L(z)\theta_R(\bar z')}~=~
 \vev{H_L(z)H_R(\bar z')}~=~ -\ln(z-\bar z')+\frac{i\pi}2.
\end{equation}
The Wick contraction of free fields gives, after taking the
factor (\ref{cocycle}) into account, the following rule
for correlators of bulk operators:
\begin{equation}
\vev{\tprod_iV^{j_i(s_i,\bar s_i)}_{m_i,\bar m_i}(z_i,\bar z_i)}~\sim~
\prod_i|z_i-\bar z_i|^{\gamma_{i\bar i}}
\prod_{i<j}
 (z_i     -     z_j)^{\gamma_{ij}}
 (z_i     -\bar z_j)^{\gamma_{i\bar j}}
 (\bar z_i-     z_j)^{\gamma_{\bar ij}}
 (\bar z_i-\bar z_j)^{\gamma_{\bar i\bar j}},
\end{equation}
where $\gamma_{ij}\equiv\frac{2}{k}\{(m_i+s_i)(m_j+s_j)-j_ij_j\}+s_is_j$, etc.
The Wick contraction involving boundary operators is defined simply by
\begin{equation}
\begin{array}{rcl}
 V^{j_i(s_i,\bar s_i)}_{m_i,\bar m_i}(z)
 B^{j_a(s_a)}_{m_a}(x)
 &\sim& (z-x)^{\gamma_{ia}}(\bar z-x)^{\gamma_{\bar ia}}, \\
 B^{j_a(s_a)}_{m_a}(x)
 V^{j_i(s_i,\bar s_i)}_{m_i,\bar m_i}(z)
 &\sim& (x-z)^{\gamma_{ia}}(x-\bar z)^{\gamma_{\bar ia}}, \\
 B^{j_a(s_a)}_{m_a}(x)
 B^{j_b(s_b)}_{m_b}(x')
 &\sim& (x-x')^{\gamma_{ab}}.
\end{array}
\end{equation}
The first and the second lines generically differ by phase.
To ensure $V(z)B(x)=\pm B(x)V(z)$, we therefore require that
the operator $V^{j(s,\bar s)}_{m,\bar m}$ appearing in the
correlator $\vev{\cdots}$ between two boundary operators connected
by the boundary state $B^\alpha_{[J,M]}$ yields a phase factor\footnote{
If the correlator contain no boundary operators, one takes the unique
boundary state appearing on the boundary and consider
the similar phase factor.},
\begin{equation}
 \exp\left(\frac{2\pi i(M+\alpha)}{k}(m+s-\bar m-\bar s)
            +i\pi\alpha(s-\bar s)\right).
\end{equation}
In addition, we require that the boundary operator $B^{j(s)}_m$
connecting the boundary states $B^\alpha_{[J,M]}$ and
$B^{\alpha'}_{[J',M']}$ satisfy
\begin{equation}
  m+s\in M-M'+\alpha-\alpha'+\frac{k}{2}N,~~~~
    s\in      \alpha-\alpha'+S.~~~~(N,S\in\mZ)
\label{ospecB}
\end{equation}
We require that the physical operators with even (odd) $N$ are
accompanied by even (odd) numbers of boundary fermions, and define
the Grassmann parity of the boundary operator to be $N+S$ mod 2.
Note that this rule makes all the boundary interaction terms in
(\ref{bintsB}) Grassmann-even.
On the other hand, the Grassmann parity of the bulk operator
$V^{j(s,\bar s)}_{m,\bar m}$ is $s-\bar s$ mod 2.
In order to get the (anti-)commutativity in accordance with this
assignment of Grassmann parity, we have to require that the bulk
operator $V^{j(s,\bar s)}_{m,\bar m}$ in disc correlators should
behave like
\begin{equation}
 e^{ \frac{i\pi}{2}(m-\bar m)}\lambda\bar\lambda
+e^{-\frac{i\pi}{2}(m-\bar m)}\bar\lambda\lambda.
\end{equation}
Finally, the boundary interaction $\tilde\mu_B\tilde S_B$ of (\ref{tilmub})
has to be proportional to $\lambda\bar\lambda-\bar\lambda\lambda$
in order to commute with other boundary interactions.

Let us compute some disc structure constants using the free field
prescription.
To begin with, we compute the coefficient $t(n)$ appearing in
\begin{equation}
 V^{1/2}_{n,-n}(z,\bar z) \to t(n) |z-\bar z|^{\frac1k}+\cdots.
\end{equation}
It is given by a free field correlator with one insertion of
the boundary screening operator $\tilde\mu_B\tilde S_B$,
\begin{equation}
  t(n) ~=~
  \vev{B^{-1}_0\cdot V^{1/2}_{n,-n}\cdot
      (-\tilde\mu_B\tilde S_B)}_{\rm free}
  ~=~
   -\frac{4k\pi\tilde\mu_B\Gamma(-\frac2k)}{\Gamma(-\frac1k)^2}
    e^{\frac{4\pi iMn}{k}},
\end{equation}
and the comparison of this with the analysis of disc two-point function
yields,
\begin{eqnarray}
 \tilde\mu_B &=&
 -i\sin\{\tfrac\pi k(2J+1)\}
  \frac{\nu^{\frac12}\Gamma(-\tfrac1k)}{2\pi k}.
\label{tmub}
\end{eqnarray}
Let us next evaluate the product of OPE coefficients
$c^\updownarrow\tilde t^\updownarrow(n,\bar n)$
They were defined in the previous section as follows,
\begin{eqnarray}
  V^{k/2}_{n,\bar n} &\longrightarrow&
  \sum_{(+,\uparrow),(-,\downarrow)}
  \tilde t^\updownarrow(n,\bar n)
  |z-\bar z|^{\frac{2n\bar n}{k}+\frac k2+1\mp(n+\bar n)}
  B^{\frac k2(\pm1)}_{n+\bar n\mp\frac k2\mp1}+\cdots, \nn\\
  \tilde t^\updownarrow(n,\bar n)
  B^{k/2(\pm 1)}_{(n+\bar n)\mp\frac k2\mp1}
  &\stackrel{n+\bar n\to 0}{\longrightarrow}&
  c^\updownarrow\tilde t^\updownarrow(n,-n)\Gamma(\pm (n+\bar n))\cdot{\bf 1}.
\end{eqnarray}
Although we were not aware in the previous section, $t^\updownarrow$
are linear in the boundary fermions and the product
$c^\updownarrow\tilde t^\updownarrow(n,-n)$ involves the algebra
of boundary fermions.
Calculating them as follows,
\begin{eqnarray}
  \tilde t^\uparrow(n,\bar n)
 &\sim& \vev{B^{-k/2-1(-1)}_{k/2+1-n-\bar n}V^{k/2}_{n,\bar n}
          (-\mu_BS_B-\mu_{\bar B}S_{\bar B})}_{\rm free}
 \nn\\ &\sim&
     \{\mu_B\lambda+(-)^{n-\bar n}\mu_{\bar B}\bar\lambda\}
     \frac{2\pi\Gamma(-k-1-n-\bar n)}
          {\Gamma(-\frac k2+n)\Gamma(-\frac k2+\bar n)}
     e^{\frac{2\pi iM(n-\bar n)}{k}}, \nn\\
 (a\lambda+b\bar\lambda)B^{\frac k2(1)}_{m-\frac k2-1}(0)
 &\sim & -(\bar\mu_B\bar S_B+\bar\mu_{\bar B}\bar S_{\bar B})
 (a\lambda+b\bar\lambda)B^{\frac k2(1)}_{m-\frac k2-1}(0) \nn\\
 &\sim & 
 (a\bar\mu_{\bar B}+b\bar\mu_B)\int_0^\Lambda x^{m-1}
 ~\sim~ (a\bar\mu_{\bar B}+b\bar\mu_B)\Gamma(m),
\end{eqnarray}
we find
\begin{equation}
 c^\uparrow t^\uparrow(n,-n) ~=~
 (\mu_B\bar\mu_{\bar B}+(-)^{2n}\mu_{\bar B}\bar\mu_B)
     \frac{2\pi\Gamma(-k-1)}
          {\Gamma(-\frac k2+n)\Gamma(-\frac k2-n)}e^{\frac{4\pi iMn}{k}}.
\end{equation}
By comparing this with (\ref{Bwave}) we obtain
\begin{equation}
  \mu_B\bar\mu_{\bar B} ~=~ -\frac{k\mu}{2\pi}e^{2 \pi iJ},~~~
  \mu_{\bar B}\bar\mu_B ~=~ -\frac{k\mu}{2\pi}e^{-2\pi iJ}.
\end{equation}
Combining this with (\ref{factor}) we can determine the boundary
couplings up to a single phase.
In the next section we set the couplings as follows,
\begin{equation}
 (\mu_B,\mu_{\bar B},\bar\mu_B,\bar\mu_{\bar B}) ~=~
 i\sqrt{\frac{k\mu}{2\pi}}
 (e^{i\pi(J-M)},e^{-i\pi(J-M)},e^{-i\pi(J+M)},e^{i\pi(J+M)})
\end{equation}
and compute the reflection coefficients of boundary operators.
We will read off the open string spectrum from it and find a
precise agreement with the result of modular bootstrap of
annulus amplitudes.

\subsection{A-branes}

For A-branes in LG models, it is not known how to construct
boundary interactions.
However, in the framework of perturbed free CFT, nothing seems to
prevent us from incorporating the boundary screening operators
of the same form.
In this and the following sections we will try to reproduce
some disc structure constants involving A-branes
using the following boundary action,
\begin{eqnarray}
\lefteqn{
\oint dx\left[
     \bar\lambda\partial_x\lambda -\sqrt{\tfrac2k}\tfrac{K\rho}{4\pi}\right]
    +\mu_AS_A+\bar\mu_A\bar S_A+\tilde\mu_A\tilde S_A
}\nn\\
&=&  \oint dx\left[
     \bar\lambda\partial_x\lambda -\sqrt{\tfrac2k}\tfrac{K\rho}{4\pi}
    -\mu_A\lambda e^{-\sqrt{\frac k2}\phi_L+iH_L}
    -\bar\mu_Ae^{-\sqrt{\frac k2}\bar\phi_L-iH_L}\bar\lambda
 \right.\nn\\ && \left. \hskip18mm
    -\tilde\mu_A
      (\lambda\bar\lambda-\bar\lambda\lambda)
      (\psi_+\psi_--i\sqrt{2k}\partial\theta)e^{-\sqrt{\frac2k}\rho_L}
  \right].
\label{bintsA}
\end{eqnarray}
An important difference between A- and B-type boundaries
in performing free CFT computation is that the boundary condition
on free fields $\theta, H$ is Dirichlet for A-type and Neumann for B-type.
In the same sense, we should regard the boundary interactions as
carrying nonzero winding number or momentum for A- or B-type
boundaries, respectively.

As we did in the case of B-branes, we regard the system as that of
free fields perturbed by bulk and boundary screening operators and
compute various disc correlators perturbatively.
The correlation between left and right-moving sectors is given by
the following set of propagators,
\begin{equation}
 \vev{\rho  _L(z)\rho  _R(\bar z')} ~=~
-\vev{\theta_L(z)\theta_R(\bar z')} ~=~
-\vev{H     _L(z)H     _R(\bar z')} ~=~
 -\ln(z-\bar z')+i\pi.
\end{equation}
The bulk operator $V^{j(s,\bar s)}_{m,\bar m}$ appearing in a
correlator $\vev{\cdots}$ between two boundary operators connected
by A-type boundary state $A^\alpha_{[J,M]}$ yields a factor
\begin{equation}
 \exp\left(
  \frac{2\pi i}{k}(M+\alpha)(m+\bar m+s+\bar s)+{i\pi}\alpha(s+\bar s)
     \right)
 \{e^{ \frac{i\pi}{2}(m+\bar m)}\lambda\bar\lambda
  +e^{-\frac{i\pi}{2}(m+\bar m)}\bar\lambda\lambda \}.
\end{equation}
In order for the bulk and boundary operators to (anti-)commute
in accordance with their Grassmann parity, we require the physical
boundary operator $B^{j(s)}_m$ between A-branes
$A^\alpha_{[JM]}$ and $A^{\alpha'}_{[J'M']}$ to satisfy
\begin{equation}
\begin{array}{rcll}
  \lambda\bar\lambda B^{j(s)}_m,
  \bar\lambda\lambda B^{j(s)}_m &\cdots&
  m\in M-M'+\mZ,&
  s\in \alpha-\alpha'+\mZ, \\
  \lambda B^{j(s)}_m            &\cdots&
  m\in M-M'-\frac k2+\mZ,&
  s\in \alpha-\alpha'+\mZ, \\
  \bar\lambda B^{j(s)}_m        &\cdots&
  m\in M-M'+\frac k2+\mZ,&
  s\in \alpha-\alpha'+\mZ.
\end{array}
\label{Aphys}
\end{equation}
The fermion number of boundary operator is given by
$s-\alpha+\alpha'-\sharp(\lambda)+\sharp(\bar\lambda)$ and is always
an integer.
In the next section we will discuss a little more about the
above condition for the physical spectrum.

One can relate the boundary couplings $(\mu_A,\bar\mu_A,\tilde\mu_A)$
with the labels $[J,M]$ of A-branes by computing the coefficients
$u(n)$, $\tilde u^\updownarrow(n,\bar n)$ and $c^\updownarrow$
perturbatively and comparing the results with those in the
previous section.
One finds
\begin{eqnarray}
 \tilde\mu_A &=&
 -\cos\{\tfrac\pi k(2J+1)\}
  \frac{\nu^{\frac12}\Gamma(-\tfrac1k)}{2\pi k}, \nn\\
  \mu_A\bar\mu_A &=&
  \frac{2k\mu}{\pi}\sin\pi(J+M)\sin\pi(J-M).
\end{eqnarray}
In the next section we compute the reflection coefficients of
boundary operators using the values of $(\mu_A,\bar\mu_A)$
\begin{equation}
 \mu_A     ~=~ \sqrt{\frac{2k\mu}{\pi}}\sin\pi(J-M),~~~~~
 \bar\mu_A ~=~ \sqrt{\frac{2k\mu}{\pi}}\sin\pi(J+M).
\label{muA}
\end{equation}

~

Let us finally point out that the bulk-boundary propagator
$\vev{V^{k/2}_{n,\pm n}B^{-1}_0}$ ($\pm$ signs correspond
to A- and B-branes respectively) exactly vanishes when evaluated
as a screening integral.
This is in consistency with the observation of the previous
section that the self-OPE of degenerate operator $V^{k/2}_{n,\pm n}$
does not yield boundary identity operator in a simple manner.

%%%%%%%%%%%%%%%%%%%%%%%%%%%%%%%%%%%%%%%%%%%%%%%%%%%%%%%%
\section{Boundary Reflection Coefficients}%%%%%%%%%%%%%%
%%%%%%%%%%%%%%%%%%%%%%%%%%%%%%%%%%%%%%%%%%%%%%%%%%%%%%%%

Now that the wave functions for various boundary states are available,
one can obtain the spectrum of open strings between any branes
from the modular transformation property of annulus amplitudes.
For example, the annulus amplitude between two A-branes both
corresponding to non-chiral non-degenerate representations
is calculated as
\begin{eqnarray}
 Z &=& \bra{A^{\alpha,\beta}_{[J,M]}}
       e^{i\pi\tau_c(L_0+\bar{L}_0-\frac{c}{12})}
       \ket{A^{\alpha',-\beta}_{[J',M']}}
 \nn \\
   &=& \sum_{m+\beta\in\frac{k}{2}\mZ}\int\frac{dj}{i\pi}
       U_{[J,M]}(-j-1,-m,-\beta)U_{[J',M']}(j,m,\beta)
       \chi_{j,m+\beta,\beta}(\tau_c,\alpha'-\alpha).
\end{eqnarray} 
Rewriting this as a sum of characters in the open string channel
we obtain
\begin{eqnarray}
 Z &=& \sum_{n\in\mZ+\alpha-\alpha'}
  e^{2\pi i\beta(\alpha'-\alpha-\frac{2}{k}(n+M-M'))}
  \int_{-\infty}^\infty ds\left\{
   \rho_0^A(s|J,J')\chi_{-\frac{1}{2}+is,n+M-M',n}(\tau_o,\beta)
 \right.  \nn \\ && \hskip50mm
  +\sum_\pm \left. 
   \rho_1^A(s|J,J')\chi_{-\frac{1}{2}+is,n+M-M'\pm\frac{k}{2},n}(\tau_o,\beta)
  \right\},
\label{aaA}
\end{eqnarray}
with
\begin{eqnarray}
 \rho_0^A(s|J,J') &=& \int_{-\infty}^\infty dp e^{2\pi ips}
   \frac{\cosh\{\pi p(2J+1)\}\cosh\{\pi p(2J'+1)\}\cosh\{k\pi p\}}
        {\sinh(\pi p)\sinh(\pi kp)},
 \nn \\
 \rho_1^A(s|J,J') &=& \int_{-\infty}^\infty dp e^{2\pi ips}
   \frac{\cosh\{\pi p(2J+1)\}\cosh\{\pi p(2J'+1)\}}
        {2\sinh(\pi p)\sinh(\pi kp)}.
\end{eqnarray}
As a non-trivial consistency check, here we try to read off
the spectrum from a different approach using the reflection
coefficients of boundary operators.
This also enables one to find and check the correspondences between
various form of boundary interactions and the boundary states.
We will heavily apply the techniques developed in $N=0$ and $N=1$
Liouville theories\cite{Fateev-ZZ, Fukuda-H2} that use boundary
degenerate operators.

Let us take the upper half plane with A- or B-type boundary conditions.
The boundary operators are labelled by the two D-branes they
are ending on, as well as their Chan-Paton indices.
Note that each single D-brane is defined with two-dimensional
Chan-Paton space.
So the boundary operators are $2\times2$ matrices.
We denote the four matrix elements of boundary operators as
\begin{equation}
[\lambda\bar\lambda B^{l(s)}_m]^X_{~X'},~~~
[\bar\lambda\lambda B^{l(s)}_m]^X_{~X'},~~~
[\lambda            B^{l(s)}_m]^X_{~X'},~~~
[\bar\lambda        B^{l(s)}_m]^X_{~X'}
\end{equation}
where $X$ and $X'$ are the sets of parameters specifying the boundary
states appearing on its left and right:
\begin{equation}
  X = \{[J,M];\alpha\},~~~ X'= \{[J',M'];\alpha'\}.
\end{equation}
$\alpha$ labels the rotation by R-symmetry.

In this section we order the boundary operators in OPE
formulae or correlators as they appear on the real axis.
In computing correlators using the prescription of previous section,
we need to re-order the operators first and then Wick contract.
Note that this also involves the re-ordering of composite
operators like $\lambda\bar\lambda B^{l(s)}_m$.

%%%%%%%%%%%%%%%%%%%%%%%%%%%%%%%%%%%%%%%%%%%%%%%
\subsection{A-branes}%%%%%%%%%%%%%%%%%%%%%%%%%%
%%%%%%%%%%%%%%%%%%%%%%%%%%%%%%%%%%%%%%%%%%%%%%%

In the previous section we have argued that the boundary operator
$[B^{l(s)}_m]^X_{~X'}$ can connect the two boundary states
$X=\{[J,M],\alpha\}$ and $X'=\{[J',M'],\alpha'\}$ only when it
satisfies a certain quantization condition.
From the annulus amplitude (\ref{aaA}) one finds
\begin{equation}
  s\in \alpha-\alpha'+\mZ,~~~
  m\in M-M'+\mZ~~{\rm or}~ M-M'\pm \tfrac k2+\mZ.
\end{equation}
Recall that the A-branes in $N=2$ Liouville theory are extending
orthogonally to the periodic direction and $M$ can be understood
as their position.
Boundary operators therefore carry winding number $m$, and the above
condition says it can differ from naive values $M-M'+\mZ$ by
$\pm \frac k2$.
This mild breaking of winding number quantization law
should be understood as due to boundary interaction terms.
Moreover, the mild-ness of the winding number violation should be
due to the boundary fermions.
Naively, if there were boundary operators with $m=M-M'+\frac{k}{2}$,
one would get operators with $m=M-M'+\frac{kN}{2}~(N\ge2)$ by fusing them.
However, the operators with $N\ge 2$ never appear if we require
\begin{equation}
\begin{array}{lcl}
  m\in M-M'+\mZ &{\rm for}&
  [\lambda\bar\lambda B^{l(s)}_m]^X_{~X'},~
  [\bar\lambda\lambda B^{l(s)}_m]^X_{~X'}, \\
  m\in M-M'+\tfrac{k}{2}+\mZ &{\rm for}&
  [\bar\lambda B^{l(s)}_m]^X_{~X'}, \\
  m\in M-M'-\tfrac{k}{2}+\mZ &{\rm for}&
  [\lambda B^{l(s)}_m]^X_{~X'},
\end{array}
\end{equation}
simply because $(\bar\lambda)^N$ vanishes if $N\ge 2$.
This is consistent with the physical condition (\ref{Aphys})
we found from locality in the free field picture.

As a reflection coefficient of boundary operators, we first consider
\begin{equation}
  [B^{l(s)}_m]^{X}_{~X'} = d(l,m,s;X;X')[B^{-l-1(s)}_m]^{X}_{~X'}.
\end{equation}
Let us first find out what kind of matrix the coefficient $d$ is.
Recall that the reflection coefficients are related to the two-point
functions:
\begin{equation}
  d(l,m,s;X;X') ~\sim~ \vev{[B^{l(s)}_m]^X_{~X'}[B^{l(-s)}_{-m}]^{X'}_{~X}}.
\end{equation}
From the quantization law of $m$ it follows that the boundary
operator $\lambda\bar\lambda B$ has nonzero two-point functions
with $\lambda\bar\lambda B$ or $\bar\lambda\lambda B$, and $\lambda B$
has nonzero two-point functions only with $\bar\lambda B$.
For the operators $\lambda\bar\lambda B,\bar\lambda\lambda B$
the reflection coefficient therefore becomes a $2\times2$ matrix:
\begin{equation}
  \left(\begin{array}{c}
  \lambda\bar\lambda B^{l(s)}_m \\
  \bar\lambda\lambda B^{l(s)}_m
  \end{array}\right)
~=~
  \left(\begin{array}{cc}
    d^{\lambda\bar\lambda}_{~\lambda\bar\lambda} &
    d^{\lambda\bar\lambda}_{~\bar\lambda\lambda} \\
    d^{\bar\lambda\lambda}_{~\lambda\bar\lambda} &
    d^{\bar\lambda\lambda}_{~\bar\lambda\lambda}
  \end{array}\right)
  \left(\begin{array}{c}
  \lambda\bar\lambda B^{-l-1(s)}_m \\
  \bar\lambda\lambda B^{-l-1(s)}_m
  \end{array}\right).
\end{equation}
For the operators $\lambda B,\bar\lambda B$ the reflection coefficients
will be ordinary numbers.
Finding these coefficients is our primary goal in the
following arguments.

To begin with, let us consider the $2\times2$ matrix-valued
reflection coefficient $d(l,m,s;X;X')$ for boundary operators
$\lambda\bar\lambda B, \bar\lambda\lambda B$.
First of all, it follows from the quantization condition on $m$ and $s$
that
\begin{equation}
 d(l,m,s;X;X') \sim
 \delta^{(1)}_{m,M-M'}\delta^{(1)}_{s,\alpha-\alpha'}
\end{equation}
To derive further constraints on $d$, we have to analyze the
boundary OPEs involving degenerate operators.
Consider the following OPEs
\begin{eqnarray}
  \left(\begin{array}{c}
  ~[\lambda\bar\lambda B^{l(s)}_m]^{X}_{~X'}\\
  ~[\bar\lambda\lambda B^{l(s)}_m]^{X}_{~X'}
  \end{array}\right)
  \times~[B^{k/2}_{m_0}]^{X'}_{~X''}
 &\longrightarrow&
  \tilde c^R_+
  \left(\begin{array}{c}
 ~[\lambda\bar\lambda B^{l+k/2(s)}_{m+m_0}]^{X}_{~X''}\\
 ~[\bar\lambda\lambda B^{l+k/2(s)}_{m+m_0}]^{X}_{~X''}
  \end{array}\right)
 +\tilde c^R_-
  \left(\begin{array}{c}
 ~[\lambda\bar\lambda B^{l-k/2(s)}_{m+m_0}]^{X}_{~X''}\\
 ~[\bar\lambda\lambda B^{l-k/2(s)}_{m+m_0}]^{X}_{~X''}
  \end{array}\right)
 \nn\\~
  [B^{k/2}_{m_0}]^{X''}_{~X}\times
  \left(\begin{array}{c}
  ~[\lambda\bar\lambda B^{l(s)}_{m}]^{X}_{~X'}\\
  ~[\bar\lambda\lambda B^{l(s)}_{m}]^{X}_{~X'}
  \end{array}\right)
 &\longrightarrow&
  \tilde c^L_+
  \left(\begin{array}{c}
 ~[\lambda\bar\lambda B^{l+k/2(s)}_{m+m_0}]^{X''}_{~X'}\\
 ~[\bar\lambda\lambda B^{l+k/2(s)}_{m+m_0}]^{X''}_{~X'}
  \end{array}\right)
 +\tilde c^L_-
  \left(\begin{array}{c}
 ~[\lambda\bar\lambda B^{l-k/2(s)}_{m+m_0}]^{X''}_{~X'}\\
 ~[\bar\lambda\lambda B^{l-k/2(s)}_{m+m_0}]^{X''}_{~X'}
  \end{array}\right)
\end{eqnarray}
where $X$'s are abbreviations for the labels of branes,
\begin{equation}
X=[JM\alpha],~~X'=[J'M'\alpha'],~~X''=[J''M''\alpha'']
\end{equation}
and $m\in M-M'+\mZ,~m_0=M'-M'',~\alpha'-\alpha''=0$
in the first line and similarly for the second line.
The coefficients $\tilde c^{R,L}_\pm$ are $2\times2$ matrices
in the same way as $d$.
One finds as usual $\tilde c^{R,L}_+\equiv 1$, and $\tilde c^{R,L}_-$
give a set of recursion relations for the reflection coefficient.
For example, by considering the term proportional to
$B^{-l-1+\frac k2(s)}_{m+m_0}$ in the product of
$B^{l(s)}_m$ and $B^{k/2}_{m_0}$ one obtains
\begin{eqnarray}
 d(l,m,s;X;X') &=&
  \tilde c^R_-(l,m,s,m_0;X;X';X'')
  d(l-\tfrac k2,m+m_0,s;X;X''),\nn\\
 d(l,m,s;X;X') &=&
  \tilde c^L_-(l,m,s,m_0;X'';X;X')
  d(l-\tfrac k2,m+m_0,s;X'';X').
\label{brec1}
\end{eqnarray}
Another set of recursion relations follows by considering
the term proportional to $B^{-l-1-\frac k2(s)}_{m+m_0}$,
\begin{eqnarray}
 d(l+\tfrac k2,m+m_0,s;X;X'') &=&
 d(l,m,s;X;X') \tilde c^R_-(-l-1,m,s,m_0;X;X';X''),\nn\\
 d(l+\tfrac k2,m+m_0,s;X'';X') &=&
 d(l,m,s;X;X') \tilde c^L_-(-l-1,m,s,m_0;X'';X;X'),
\label{brec2}
\end{eqnarray}
but the former two are related to the latter two due to
\begin{equation}
  d(l,m,s;X;X')d(-l-1,m,s;X;X')=1.
\label{dcons1}
\end{equation}
Also, the two equations in (\ref{brec1}) are not independent
once we notice
\begin{equation}
\begin{array}{rcl}
  d(l,m,s;X;X')^t &=& d(l,-m,-s;X';X),\\
  \tilde c^L_-(l,m,s,m_0;X'';X;X')^t &=&
  \tilde c^R_-(-l-1+\tfrac k2,-m-m_0,-s,m_0;X';X'';X),
\end{array}
\label{dcons2}
\end{equation}
from their relation to disc correlators.

The matrix coefficients $\tilde c^{L,R}_-$ are calculated as screening
integrals:
\begin{eqnarray}
  \tilde c^L_-(l,m,s,m_0;X;X';X'')
 &=& \vev{B^{l(s)}_mB^{k/2}_{m_0}B^{-l-1+k/2(-s)}_{-m-m_0}}
 \nn\\
 &=& \vev{B^{l(s)}_mB^{k/2}_{m_0}B^{-l-1+k/2(-s)}_{-m-m_0}
          (-\mu S-\bar\mu\bar S+\mu_AS_A\bar\mu_A\bar S_A)}_{\rm free}.
\end{eqnarray}
There are two kinds of contributions to the coefficient $\tilde c_-$,
one proportional to $\mu_A\bar\mu_A$ and the other proportional to
$\mu$ or $\bar\mu$.
The first ones are expressed as the following integral
\begin{equation}
  I = \int dsd\bar s |s|^{l-m}|\bar s|^{l+m}
 |1-s|^{\frac k2-m_0}|1-\bar s|^{\frac k2+m_0}|s-\bar s|^{-k-1},
\end{equation}
with different integration domains.
As before, we consider the three segments of the real line
\[
(1)~ [-\infty,0],~~~(2)~[0,1],~~~(3)~[1,\infty]
\]
and denote the integrals with suffices indicating the integration
domain of $s,\bar s$.
For example,
\[
 I_{1\bar 1}\Leftrightarrow \{s<\bar s<0\},~~
 I_{\bar 12}\Leftrightarrow \{\bar s<0< s<1\},~~
 I_{2\bar 3}\Leftrightarrow \{0< s<1 <\bar s\},~~~{\rm etc.}
\]
These integrals can be expressed in terms of the functions $G_k$
defined in the appendix, where some useful formulae are also
presented.
The integrals $I_{1\bar 1}$ and $I_{\bar 11}$ have simple expressions
\begin{eqnarray}
 I_{1\bar 1} &=&
 \frac{\Gamma(1+l+m)\Gamma( -l+\frac k2+m+m_0)}
      {\Gamma( -l+m)\Gamma(1+l-\frac k2+m+m_0)}\Gamma(-2l-1)\Gamma(2l-k+1),
 \nn\\
 I_{\bar 11} &=&
 \frac{\Gamma(1+l-m)\Gamma( -l+\frac k2-m-m_0)}
      {\Gamma( -l-m)\Gamma(1+l-\frac k2-m-m_0)}\Gamma(-2l-1)\Gamma(2l-k+1)
\label{I11}
\end{eqnarray}
but others do not.
Using them, the matrix elements are computed as
\begin{eqnarray}
 \tilde c^R_-(l,m,s,m_0;X;X';X'')^{\lambda\bar\lambda}_{~\lambda\bar\lambda}
 &=&
  \mu_A  \bar\mu_A   I_{1\bar 1}
 +\mu'_A \bar\mu'_A  I_{2\bar 2}
 +\mu''_A\bar\mu''_A I_{3\bar 3}
 +\mu'_A \bar\mu''_A I_{2\bar 3}
 +I_0(\mu,\bar\mu)^{\lambda\bar\lambda}_{~\lambda\bar\lambda},
 \nn\\
 \tilde c^R_-(l,m,s,m_0;X;X';X'')^{\bar\lambda\lambda}_{~\bar\lambda\lambda}
 &=&
  \mu_A  \bar\mu_A  I_{\bar 11}
 +\mu'_A \bar\mu'_A I_{\bar 22}
 +\mu''_A\bar\mu''_AI_{\bar 33}
 +\mu''_A\bar\mu'_A  I_{\bar 23}
 +I_0(\mu,\bar\mu)^{\bar\lambda\lambda}_{~\bar\lambda\lambda},
 \nn\\
  \tilde c^R_-(l,m,s,m_0;X;X';X'')^{\lambda\bar\lambda}_{~\bar\lambda\lambda}
 &=&
 \{ \mu'_A \bar\mu_A   I_{\bar 12}
  +\mu''_A\bar\mu_A   I_{\bar 13}\} (-)^{s-\alpha+\alpha'},
 \nn\\
 \tilde c^R_-(l,m,s,m_0;X;X';X'')^{\bar\lambda\lambda}_{~\lambda\bar\lambda}
 &=&
 \{ \mu_A  \bar\mu'_A  I_{1\bar 2}
  +\mu_A  \bar\mu''_A I_{1\bar 3}\} (-)^{s-\alpha+\alpha'}.
\end{eqnarray}
and similarly for $\tilde c_-^L$.
In this expression, the coupling constants $\mu_A,\bar\mu_A$ are
functions of $(J,M)$ as given in (\ref{muA}) and similarly for
those with primes.
The diagonal elements of $\tilde c^{R}_-$ also have
${\cal O}(\mu,\bar\mu)$ contribution $I_0$ which are given by
\begin{eqnarray}
 (I_0)^{\lambda\bar\lambda}_{~\lambda\bar\lambda} &=&
 -\frac{k\mu}{\pi}\left\{
  \bc(2M-k)I_{\bar 11} +\bc(2M' )I_{2\bar 2} +\bc(2M'')I_{3\bar 3}
 +\bc(M'+M''-\tfrac k2)I_{2\bar 3}
 \right\}
 \nn \\
 (I_0)^{\bar\lambda\lambda}_{~\bar\lambda\lambda} &=&
 -\frac{k\mu}{\pi}\left\{
  \bc(2M+k)I_{1\bar 1} +\bc(2M' )I_{\bar 22} +\bc(2M'')I_{\bar 33}
 +\bc(M'+M''+\tfrac k2)I_{\bar 23}
 \right\}.
\end{eqnarray}
After the $\mu_A,\bar\mu_A$ are substituted with the functions (\ref{muA}),
the expression for $\tilde c^{L,R}_-$ simplifies under the following
assumption
\begin{quote}
{\it the degenerate operators $[B^{k/2}_m]^X_{~X'}$ only appear between
     branes $X=[J,M,\alpha]$ and $X'=[J',M',\alpha']$
     satisfying $J-J'=\tfrac k2$.}
\end{quote}
The OPE coefficients $\tilde c^{L,R}_-$ are then given by
\begin{eqnarray}
\lefteqn{
  \left.\tilde c^R_-(l,m,s,m_0;X;X';X'')\right|
  _{m_0=M'-M'',~J'=J''+\frac k2,~ \alpha'=\alpha''}
}\nn\\ &=&
   \frac{2k\mu}{\pi}
   \left(\begin{array}{cc}\gamma &0\\0& \bar\gamma\end{array}\right)
   \left(\begin{array}{cc}
   \bs(J+M) & -\bs(J'+l-M) \\ -\bs(J'+l+M) & \bs(J-M) \end{array}\right)
 \nn\\ && \times
   \left(\begin{array}{cc} I_{1\bar 1} &0\\0& I_{\bar 11} \end{array}\right)
   \left(\begin{array}{cc}
   \bs(J-M) & \bs(J'+l-k-M) \\ \bs(J'+l-k+M) & \bs(J+M) \end{array}\right)
   \left(\begin{array}{cc}\bar\gamma &0\\0& \gamma\end{array}\right),\nn\\
\lefteqn{
  \left.\tilde c^L_-(l,m,s,m_0;X'';X;X')\right|
  _{m_0=M''-M,~J''=J+\frac k2,~ \alpha=\alpha''}
}\nn\\ &=&
   \frac{2k\mu}{\pi}
   \left(\begin{array}{cc}\gamma &0\\0& \bar\gamma\end{array}\right)
   \left(\begin{array}{cc}
   \bs(J'-M') & -\bs(J-l+M') \\ -\bs(J-l-M') & \bs(J'+M') \end{array}\right)
 \nn\\ && \times
   \left(\begin{array}{cc} I_{\bar 11} &0\\0& I_{1\bar 1}\end{array}\right)
   \left(\begin{array}{cc}
   \bs(J'+M') & \bs(J-l+k+M') \\ \bs(J-l+k-M') & \bs(J'-M') \end{array}\right)
   \left(\begin{array}{cc}\bar\gamma &0\\0& \gamma\end{array}\right),
\nn\\ &&
 \gamma ~=~ \exp\left\{\tfrac{i\pi}{2}(M-M'-m+\alpha-\alpha'-s)\right\}
\end{eqnarray}
where $I_{1\bar 1}, I_{\bar 11}$ are as given in (\ref{I11}).

It turns out very non-trivial to check that the set of
recursion relations are consistent (solvable) and has a solution
with the appropriate symmetry properties (\ref{dcons1}),(\ref{dcons2}).
We find that the solution can be written in terms of the special
functions $\bG$ and $\bS$ introduced in \cite{Fateev-ZZ}
(see the appendix for their definitions) as follows:
\begin{eqnarray}
 d(l,m,s;X;X') &=&
 (\nu b^{2-2b^2})^{l+\frac12}
 \frac{\bG(b(2l+1))}{\bG(-b(2l+1))}
 \bS(b(l+J+J'+2))\bS(b(l+J-J'+1))
 \nn\\ && \times
 \bS(b(l-J+J'+1))\bS(b(l-J-J'))
 \times \hat d(l,m,s;X;X'),
\end{eqnarray}
where $b\equiv k^{-1/2}$ as before.
The matrix part $\hat d(l,m,s;X;X')$ is given by
\begin{eqnarray}
 \hat d(l,m,s;X;X') &=& 4
 \left(\begin{array}{cc}\gamma&0\\0&\bar\gamma \end{array}\right)
 \left(\begin{array}{cc}
 \bs(J'-M') & -\bs(J-l+M') \\ -\bs(J-l-M') & \bs(J'+M') \end{array}\right)
 \left(\begin{array}{cc}
 \frac{\Gamma(1+l-m)}{\Gamma(-l-m)} &0 \\
 0& \frac{\Gamma(1+l+m)}{\Gamma(-l+m)} \end{array}\right)
 \nn\\ && \hskip18mm\times
 \left(\begin{array}{cc}
 \bs(J'+M') & -\bs(J+l+M') \\ -\bs(J+l-M') & \bs(J'-M') \end{array}\right)
 \left(\begin{array}{cc}\bar\gamma&0\\0&\gamma \end{array}\right)
 \nn\\  &=& 4
 \left(\begin{array}{cc}\gamma&0\\0&\bar\gamma \end{array}\right)
 \left(\begin{array}{cc}
 \bs(J+M) & -\bs(J'+l-M) \\ -\bs(J'+l+M) & \bs(J-M) \end{array}\right)
 \left(\begin{array}{cc}
 \frac{\Gamma(1+l+m)}{\Gamma(-l+m)} &0 \\
 0& \frac{\Gamma(1+l-m)}{\Gamma(-l-m)} \end{array}\right)
 \nn\\ && \hskip18mm\times
 \left(\begin{array}{cc}
 \bs(J-M) & -\bs(J'-l-M) \\ -\bs(J'-l+M) & \bs(J+M) \end{array}\right)
 \left(\begin{array}{cc}\bar\gamma&0\\0&\gamma \end{array}\right),
 \nn\\&&
 \gamma ~=~ \exp\left\{\tfrac{i\pi}{2}(M-M'-m+\alpha-\alpha'-s)\right\}
\end{eqnarray}

We can solve similar equations for reflection
coefficients for operators $[\lambda B^{l(s)}_m]$ or
$[\bar\lambda B^{l(s)}_m]$.
The reflection coefficients $d_\lambda,d_{\bar\lambda}$ for these
operators are ordinary numbers, so one expects them to be
proportional to $\frac{\Gamma(1+l\pm m)}{\Gamma(-l\pm m)}$.
Then $\tilde c^{L,R}_-$ should be proportional to either
$I_{1\bar 1}$ or $I_{\bar 11}$ and not their linear combination.
Let us consider the case of $[\lambda B^{l(s)}_m]$ and calculate
the OPE coefficients $\tilde c^{L,R}_-$ as screening integrals.
We find that the sum of bulk and boundary screening integrals
takes the simple form:
\begin{eqnarray}
  \tilde c^R_-(l,m,s,m_0;X;X',X'') &=&
 -\frac{2k\mu}{\pi}\bs(l+J+J'-\tfrac k2)\bs(l-J+J'-\tfrac k2)I_{1\bar 1},
  \nn\\
  \tilde c^L_-(l,m,s,m_0;X;X',X'') &=&
 -\frac{2k\mu}{\pi}\bs(l-J-J'-\tfrac k2)\bs(l-J+J'-\tfrac k2)I_{1\bar
 1}.
\end{eqnarray}
where we imposed again $m_0=M'-M'',~ J'=J''+\frac k2,~\alpha'=\alpha''$
in the first line and similarly in the second line, too.
The reflection coefficient $d_\lambda(l,m;X;X')$ is obtained as
before by solving a set of recursion relations.
One finds
\begin{eqnarray}
  d_\lambda(l,m;X;X') &=& (\nu b^{2-2b^2})^{l+\frac12}
 \frac{\Gamma(1+l+m)}{\Gamma(-l+m)} \frac{\bG(b(2l+1))}{\bG(-b(2l+1))}
 \nn\\ && \times
 \bS(b(l+J+J'+2+\tfrac k2))\bS(b(l+J-J'+1+\tfrac k2))
 \nn\\ && \times
 \bS(b(l-J+J'+1+\tfrac k2))\bS(b(l-J-J'+\tfrac k2)).
\end{eqnarray}

Boundary chiral operators are expected to satisfy a different
kind of reflection relation, which should be of the form
\begin{equation}
 [\bar\lambda B^{l(s)}_{\mp l}]^X_{~X'} ~=~
  d^\pm(l,s;X;X')\left(\begin{array}{cc}
  ~[\lambda\bar\lambda B^{\tilde l(s+1)}_{\pm\tilde l}]^X_{~X'} \\
  ~[\bar\lambda\lambda B^{\tilde l(s+1)}_{\pm\tilde l}]^X_{~X'}
  \end{array}\right),
\end{equation}
where $\tilde l=-l-1-\frac k2$.
Here $d^\pm$ is a two-component row vector with components
$(d^\pm_{\lambda\bar\lambda},d^\pm_{\bar\lambda\lambda})$.
Once we know the OPE coefficient
$\tilde c^R_\downarrow=((\tilde c^R_\downarrow)_{\lambda\bar\lambda},
                        (\tilde c^R_\downarrow)_{\bar\lambda\lambda})$
defined by
\begin{equation}
  [\lambda B^{l(s)}_l]^X_{~X'}[B^{k/2}_m]^{X'}_{~X''}
  \longrightarrow 
 ((\tilde c^R_\downarrow)_{\lambda\bar\lambda},~
  (\tilde c^R_\downarrow)_{\bar\lambda\lambda})
  \left(\begin{array}{c}
  ~[\lambda\bar\lambda B^{l(s-1)}_{l+m+1+k/2}]^X_{~X''} \\
  ~[\bar\lambda\lambda B^{l(s-1)}_{l+m+1+k/2}]^X_{~X''}
  \end{array}\right)
  +\cdots,
\end{equation}
the reflection coefficient $d^-$ can be calculated as
\begin{eqnarray}
  d^-(l,s;X;X') &=& \tilde c^R_\downarrow(l,m,s;X;X';X'')
  d(l,l+m+\tfrac k2+1,s-1;X;X''),\nn\\
&&  ( m~=~M'-M'',~~~~
      J'-J'' ~=~\tfrac k2,~~~~
      \alpha'=\alpha''),
\end{eqnarray}
and similarly for the other one.
After some computation one obtains
\begin{eqnarray}
(c^R_\downarrow)(l,m,s;X;X';X'')
 &=&
-\sqrt{\frac{2k\mu}{\pi}}
 \frac{\Gamma(2l+1)\Gamma(-2l-1-m-\frac k2)}{\Gamma(-m-\frac k2)}
 \nn\\&&
 \times\left(\bs(J'+M'+2l)(-)^{\alpha-\alpha'-s},~\bs(J+M)\right),
\end{eqnarray}
and
\begin{eqnarray}
 d^\pm(l,s;X;X') &=& \pm2(\nu b^{-2b^2})^{l+\frac 12+\frac k4}
 \frac{\bG(b(2l+k+1))}{\bG(-b(2l+1))}
\nn\\ && \times
 \bS(b(l+J+J'+2+\tfrac k2))
 \bS(b(l+J-J'+1+\tfrac k2))
\nn\\ && \times
 \bS(b(l-J+J'+1+\tfrac k2))
 \bS(b(l-J-J'  +\tfrac k2)) \times \hat d^\pm(l,s;X;X'),
\nn\\
\hat d^+(l,s;X;X') &=&
 \left(\bs(J-M),~(-)^{\alpha-\alpha'-s-1}\bs(J'-M')\right), \nn\\
\hat d^-(l,s;X;X') &=&
 \left((-)^{\alpha-\alpha'-s-1}\bs(J'+M'),~\bs(J+M)\right).
\end{eqnarray}

There are some consistency checks we can do.
As an example, one can consider another important boundary OPE
involving $l=1/2$ degenerate operators
\begin{eqnarray}
 ~[B^l_m]^X_{~X'}[B^{1/2}_{m_0}]^{X'}_{~X''}
  &\longrightarrow&
  c^R_+[B^{l+1/2}_{m+m_0}]^X_{~X''} +
  c^R_-[B^{l-1/2}_{m+m_0}]^X_{~X''}, \nn\\
 ~[B^{1/2}_{m_0}]^{X''}_{~X}[B^l_m]^X_{~X'}
  &\longrightarrow&
  c^L_+[B^{l+1/2}_{m+m_0}]^{X''}_{~X'} +
  c^L_-[B^{l-1/2}_{m+m_0}]^{X''}_{~X'},
\end{eqnarray}
and calculate $c^{R,L}_-$ as screening integrals
which are proportional to $\tilde\mu_A$.
On the other hand, they are also calculated as certain ratios
of the reflection coefficients obtained above.
Comparing the two results we obtain
\begin{equation}
  \tilde\mu_A ~=~
  -\nu^{\frac12}\frac{\Gamma(-\frac1k)}{2k\pi}
            \cos\tfrac\pi k(2J+1),
\end{equation}
in consistency with the result of $\tilde\mu_A$ of the previous
section (\ref{tmub}).

Finally, let us try reading off the open string spectrum from
the reflection coefficients and matching with the result of
modular bootstrap analysis (\ref{aaA}).
The reflection coefficients are essentially the phase shifts of
wave functions that are scattered off the Liouville wall, so by
taking its log derivative with respect to the Liouville momentum
($l$ quantum number) one should be able to read off the spectrum density.
For $d(l,m,s;X;X')$ which is a matrix-valued quantity, it is natural
to define the phase shift by the logarithm of its determinant.
Discarding the factors which are independent of $J,J'$ and irrelevant
one obtains
\begin{eqnarray}
\log\det d(l,m,s;X;X')
 &\sim&
  \log[ \bS(b(l+J+J'+2))\bS(b(l+J+J'+2+k))
 \nn\\ && \hskip4mm \times
        \bS(b(l+J-J'+1))\bS(b(l+J-J'+1+k))
 \nn\\ && \hskip4mm \times
        \bS(b(l-J+J'+1))\bS(b(l-J+J'+1+k))
 \nn\\ && \hskip4mm \times
        \bS(b(l-J-J'))\bS(b(l-J-J'+k)) ]
 \nn\\ && \hskip-30mm ~=~
 -2\int_{-\infty}^\infty\frac{dp}{p}
  \frac{e^{(2l+1)\pi p}
        \cosh\{(2J+1)\pi p\}\cosh\{(2J'+1)\pi p\}\cosh\{k\pi p\}}
       {\sinh(\pi p)\sinh(k\pi p)}.
\end{eqnarray}
The $l$-derivative of this agrees with the spectral function $\rho^A_0$
in (\ref{aaA}) up to numerical factors.
Similarly, the logarithm of $d_\lambda(l,m,s;X;X')$ is given by
\begin{eqnarray}
\log d_\lambda(l,m,s;X;X')
 &\sim&
 -\int_{-\infty}^\infty\frac{dp}{p}
  \frac{e^{(2l+1)\pi p}\cosh\{(2J+1)\pi p\}\cosh\{(2J'+1)\pi p\}}
       {\sinh(\pi p)\sinh(k\pi p)},
\end{eqnarray}
and its $l$-derivative agrees with $\rho^A_1$ in (\ref{aaA}) up to
numerical factors.

%%%%%%%%%%%%%%%%%%%%%%%%%%%%%%%%%%%%%%%%%%%
\subsection{B-brane}%%%%%%%%%%%%%%%%%%%%%%%
%%%%%%%%%%%%%%%%%%%%%%%%%%%%%%%%%%%%%%%%%%%

Using the wave functions for B-branes one can compute the
open string spectrum between two B-branes $X=\{[J,M],\alpha\}$
and $X'=\{[J',M'],\alpha'\}$:
\begin{eqnarray}
 Z &=& \bra{B^{\alpha,\beta}_{[J,M]}}e^{i\pi\tau_c(L_0+\bar L_0-\frac c{12})}
       \ket{B^{\alpha',-\beta}_{[J',M']}}
 \nn\\ &=&
   \sum_{m\in\frac12\mZ}\int_{{\cal C}_0}\frac{dj}{i\pi}
   T_{[J,M]}(-j-1,-m,-\beta)T_{[J',M']}(j,m,\beta)
   \chi_{j,m+\beta,\beta}(\tau_c,\alpha'-\alpha)
 \nn\\ &=&
 \frac{2kT_0^2}{\pi}
 e^{-2\pi i\beta(\alpha-\alpha'+\frac2k(\alpha-\alpha'+M-M'))}
 \sum_{m\in k\mZ+M-M'+\alpha-\alpha'}
 \nn\\ && 
 \int ds
\left\{
   \chi_{-\frac12+is,m,\alpha-\alpha'}(\tau_o,\beta)\rho^B_0(s|J,J')
  +\chi_{-\frac12+is,m+\frac k2,\alpha-\alpha'}(\tau_o,\beta)\rho^B_1(s|J,J')
 \right\},
\end{eqnarray}
where $T_{[J,M]}(j,m,\beta)$ is the wave function of (\ref{Bwave})
and $T_0$ is its normalization which is so far undetermined.
The spectral functions $\rho^B_{0,1}$ are given by
\begin{eqnarray}
  \rho^B_0(s|J,J') &=&
  \int dp\frac{e^{2\pi ips}
  \left[\cosh\{2\pi p(J-J')\}\cosh(k\pi p)+\cosh\{2\pi p(J+J'+1)\}\right]}
              {\sinh(\pi p)\sinh(k\pi p)},\nn\\
  \rho^B_1(s|J,J') &=&
  \int dp\frac{e^{2\pi ips}
  \left[\cosh\{2\pi p(J+J'+1)\}\cosh(k\pi p)+\cosh\{2\pi p(J-J')\}\right]}
              {\sinh(\pi p)\sinh(k\pi p)}.
\label{specB}
\end{eqnarray}

We would like to reproduce this from the boundary reflection coefficients.
Notice first of all that the condition on physical open string operators
(\ref{ospecB}) is in accordance with the spectrum that can be read
off from the annulus amplitude.
Let us recapitulate it here:
\begin{equation}
\begin{array}{lcl}
  s\in \alpha-\alpha'+\mZ,~~~
  m+s \in \alpha-\alpha'-M+M'+k\mZ &\mbox{for}&
  [\lambda\bar\lambda B^{l(s)}_m]^X_{~X'},~
  [\bar\lambda\lambda B^{l(s)}_m]^X_{~X'},\\
  s\in \alpha-\alpha'+\mZ,~~~
  m+s \in \alpha-\alpha'-M+M'+k\mZ+\frac k2 &\mbox{for}&
  [    \lambda B^{l(s)}_m]^X_{~X'},~
  [\bar\lambda B^{l(s)}_m]^X_{~X'}.
\end{array}
\end{equation}

We first consider the $2\times2$ matrix-valued reflection coefficient
$d(l,m,s;X;X')$ for the operators
$ (\lambda\bar\lambda B^{l(s)}_m,~\bar\lambda\lambda B^{l(s)}_m)$,
which is defined in the same way as for A-branes.
To obtain it, we analyze the recursion relations arising from
the boundary OPEs involving $l=k/2$ degenerate operators.
Calculation of the OPE coefficients $\tilde c^{L,R}_-$ goes in
a similar way as before, except that the bulk screening operators
do not show up.
Under the assumption that the degenerate operator $[B^{k/2}_m]^X_{~X'}$
connects two boundary states only when $J=J'+\frac k2$,
we obtain
\begin{eqnarray}
\lefteqn{
  \tilde c^R_-(l,m,s,m_0;X;X';X'')
  |_{m_0=M'-M'',~J'=J''+\frac k2,~\alpha'=\alpha''}
} \nn\\ &=&
 -\frac{ik\mu}{\pi}\bs(l+J+J'-\tfrac k2)
 \left(\begin{array}{rr}\gamma&0\\0&\bar\gamma\end{array}\right) 
 \left(\begin{array}{rr}\xi&-\bar\xi\\
                        -\bar\xi&\xi\end{array}\right) 
 \left(\begin{array}{cc} I_{1\bar 1} & 0 \\ 0 & I_{\bar 11} \end{array}\right) 
 \left(\begin{array}{cc}\xi'&\bar\xi'\\
                        \bar\xi'&\xi'\end{array}\right)
 \left(\begin{array}{rr}\bar\gamma&0\\0&\gamma\end{array}\right) ,
\nn\\
\lefteqn{
  \tilde c^L_-(l,m,s,m_0;X'';X;X')
  |_{m_0=M''-M,~J''=J+\frac k2,~\alpha''=\alpha}
} \nn\\ &=&
 -\frac{ik\mu}{\pi}\bs(J+J'-l+\tfrac k2)
 \left(\begin{array}{rr}\gamma&0\\0&\bar\gamma\end{array}\right) 
 \left(\begin{array}{rr}\bar\xi&-\xi\\
                        -\xi&\bar\xi\end{array}\right) 
 \left(\begin{array}{cc} I_{\bar 11} & 0 \\ 0 & I_{1\bar1}\end{array}\right) 
 \left(\begin{array}{cc}\bar\xi'&\xi'\\
                        \xi'&\bar\xi'\end{array}\right)
 \left(\begin{array}{rr}\bar\gamma&0\\0&\gamma\end{array}\right) ,
 \nn\\
 &&
  \gamma ~=~ e^{\frac{i\pi}{2}(M-M'-m+\alpha-\alpha'-s)},~~~~
  \xi  ~=~ e^{\frac{i\pi}{2}(J-J'-l)},~~~~
  \xi' ~=~ e^{\frac{i\pi}{2}(J-J'-l+k)}.
\end{eqnarray}
A number of recursion relations for $d$ are obtained easily,
and by solving them we find
\begin{eqnarray}
 d(l,m,s;X;X') &=& -i(\nu b^{2-2b^2})^{l+\frac12}
  \frac{\bG(b(2l+1))}{\bG(-b(2l+1))}
  \bS(b(l+J+J'+2+\tfrac k2))\bS(b(l+J-J'+1))
 \nn\\ && \times
  \bS(b(l-J+J'+1))\bS(b(l-J-J'+\tfrac k2))
 \nn\\ && \times
 \left(\begin{array}{cc}\gamma&0\\0&\bar\gamma\end{array}\right) 
\left(\begin{array}{rr}\xi&-\bar\xi\\
                       -\bar\xi&\xi\end{array}\right) 
 \left(\begin{array}{cc} \frac{\Gamma(1+l+m)}{\Gamma(-l+m)} & 0 \\
                      0 &\frac{\Gamma(1+l-m)}{\Gamma(-l-m)}
       \end{array}\right) 
 \left(\begin{array}{cc} \tilde\xi&\bar{\tilde\xi}\\
                         \bar{\tilde\xi}&\tilde\xi\end{array}\right)
 \left(\begin{array}{cc}\bar\gamma&0\\0&\gamma\end{array}\right) 
 \nn\\ &&
 \xi       ~=~ e^{\frac{i\pi}{2}(J-J'-l)},~~~~
 \tilde\xi ~=~ e^{\frac{i\pi}{2}(J-J'+l+1)},~~~~
 \gamma    ~=~ e^{\frac{i\pi}{2}(M-M'-m+\alpha-\alpha'-s)}.
\end{eqnarray}

For boundary operators between B-branes, the reflection coefficient
for those proportional to $\lambda,\bar\lambda$
also becomes $2\times 2$ matrix,
\begin{equation}
 \left(\begin{array}{c}
 \lambda     B^{l(s)}_m \\
 \bar\lambda B^{l(s)}_m
 \end{array}\right)
 ~=~
 \left(\begin{array}{cc}
 d'^\lambda_{~\lambda}       & d'^\lambda_{~\bar\lambda} \\
 d'^{\bar\lambda}_{~\lambda} & d'^{\bar\lambda}_{~\bar\lambda}
 \end{array}\right)
 \left(\begin{array}{c}
 \lambda     B^{-l-1(s)}_m \\
 \bar\lambda B^{-l-1(s)}_m
 \end{array}\right).
\end{equation}
The calculation of the reflection coefficient $d'(l,m,s;X,X')$
proceeds in the same way as before.
We only present the final result,
\begin{eqnarray}
 d'(l,m,s;X;X') &=&
-i(\nu b^{2-2b^2})^{l+\frac12}
  \frac{\bG(b(2l+1))}{\bG(-b(2l+1))}
  \bS(b(l+J+J'+2)\bS(b(l+J-J'+1+\tfrac k2))
 \nn\\ &&\times
  \bS(b(l-J+J'+1+\tfrac k2))\bS(b(l-J-J'))
 \nn\\ &&\times
 \left(\begin{array}{cc}\gamma&0\\0&\bar\gamma\end{array}\right) 
 \left(\begin{array}{rr}\eta&\bar\eta\\
                        \bar\eta&\eta\end{array}\right) 
 \left(\begin{array}{cc} \frac{\Gamma(1+l+m)}{\Gamma(-l+m)} & 0 \\
                    0 &  \frac{\Gamma(1+l-m)}{\Gamma(-l-m)}
       \end{array}\right) 
 \left(\begin{array}{rr} \tilde\eta&-\bar{\tilde\eta}\\
                        -\bar{\tilde\eta}&\tilde\eta\end{array}\right)
 \left(\begin{array}{cc}\bar\gamma&0\\0&\gamma\end{array}\right) 
 \nn\\ &&
 \eta        ~=~ e^{\frac{i\pi}{2}(J+J'+l)},~~~~
 \tilde\eta  ~=~ e^{\frac{i\pi}{2}(J+J'-l-1)},~~~~
 \gamma      ~=~ e^{\frac{i\pi}{2}(M+M'+m-\alpha+\alpha'+s)}.
\end{eqnarray}

The reflection relation for chiral operators can also
be obtained in the same way as for A-brane case.
It can be put in the following form,
\begin{eqnarray}
 \left(\begin{array}{cc}
 \xi & \bar\xi \\ \bar\xi & \xi \end{array}\right)
 \left(\begin{array}{cc}
 \gamma_- & 0 \\ 0 & \bar{\gamma}_- \end{array}\right)
 \left(\begin{array}{c}
 \lambda\bar\lambda B^{l(s)}_{l}\\
 \bar\lambda\lambda B^{l(s)}_{l}
 \end{array}\right)
 &=& d^-\cdot
 \left(\begin{array}{rr}
 \eta & -\bar\eta \\ -\bar\eta & \eta \end{array}\right)
 \left(\begin{array}{cc}
 \tilde\gamma_- & 0 \\ 0 & \bar{\tilde\gamma}_- \end{array}\right)
 \left(\begin{array}{c}
 \lambda     B^{\tilde l(s-1)}_{-\tilde l}\\
 \bar\lambda B^{\tilde l(s-1)}_{-\tilde l}
 \end{array}\right)
 ~=~ \left(\begin{array}{c}\ast \\ 0 \end{array}\right)
 , \nn\\
 \left(\begin{array}{cc}
 \xi & \bar\xi \\ \bar\xi & \xi \end{array}\right)
 \left(\begin{array}{cc}
 \gamma_+ & 0 \\ 0 & \bar{\gamma}_+ \end{array}\right)
 \left(\begin{array}{c}
 \lambda\bar\lambda B^{l(s)}_{-l} \\
 \bar\lambda\lambda B^{l(s)}_{-l}
 \end{array}\right)
 &=& d^+\cdot
 \left(\begin{array}{rr}
 \eta & -\bar\eta \\ -\bar\eta & \eta \end{array}\right)
 \left(\begin{array}{cc}
 \tilde\gamma_+ & 0 \\ 0 & \bar{\tilde\gamma}_+ \end{array}\right)
 \left(\begin{array}{c}
 \lambda     B^{\tilde l(s+1)}_{\tilde l} \\
 \bar\lambda B^{\tilde l(s+1)}_{\tilde l}
 \end{array}\right)
 ~=~ \left(\begin{array}{c}0 \\ \ast \end{array}\right),
 \nn\\
\begin{array}{rcl}
 \xi &=& e^{\frac{i\pi}{2}(J-J'-l)} \\
 \gamma_\mp &=& e^{\frac{i\pi}{2}(M-M'\mp l+\alpha-\alpha'-s)}
\end{array}
 && 
\begin{array}{rcl}
 \eta &=& e^{\frac{-i\pi}{2}(J+J'+l)} \\
 \tilde\gamma_\mp &=& e^{\frac{i\pi}{2}(M+M'\mp\tilde l-\alpha+\alpha'+s\mp1)}
\end{array}
\end{eqnarray}
The right hand sides of these equations mean that a suitable linear
combination of $\lambda\bar\lambda B^{l(s)}_{\pm l}$ and
$\bar\lambda\lambda B^{l(s)}_{\pm l}$ has a partner
under the reflection $l\leftrightarrow \tilde l$, while another
suitable linear combination should be regarded as zero.
The coefficients $d^{\mp}$ are given by
\begin{eqnarray}
 d^\mp(l,s;X;X') &=&
 \mp(\nu b^{-2b^2})^{l+\frac 12+\frac k4}
 \frac{\bG(b(2l+k+1))}{\bG(-b(2l+1))}
 \nn\\ && \times
 \bS(b(l+J+J'+2+\tfrac k2))
 \bS(b(l-J-J'+1+k))
 \nn\\ && \times
 \bS(b(l+J-J'+1+k))
 \bS(b(l-J-J'+\tfrac k2)).
\end{eqnarray}

Finally, let us calculate the open string spectrum between two B-branes
using boundary reflection coefficients.
The spectral densities of boundary operators proportional to
$(\lambda\bar\lambda,~\bar\lambda\lambda)$ or $(\lambda,~\bar\lambda)$
are obtained as suitable derivatives of
$(\log \det d)$ or $(\log\det d')$.
\begin{eqnarray}
 \log\det d &\sim& 
 -\int_{-\infty}^\infty \frac{dp}{p}
  \frac{e^{(2l+1)\pi p}
        \left\{ \cosh 2\pi p(J+J'+1)
               +\cosh k\pi p\cosh 2\pi p(J-J') \right\}}
       {\sinh(\pi p)\sinh(k\pi p)}       ,\nn\\
 \log\det d' &\sim& 
 -\int_{-\infty}^\infty \frac{dp}{p}
  \frac{e^{(2l+1)\pi p}
        \left\{ \cosh 2\pi p(J-J')
               +\cosh k\pi p\cosh 2\pi p(J+J'+1) \right\}}
       {\sinh(\pi p)\sinh(k\pi p)}.
\end{eqnarray}
These are in precise agreement with the spectral densities
$\rho^B_0,\rho^B_1$ of (\ref{specB}).
This result also suggests that the correct normalization of
the wave functions for the B-branes (\ref{Bwave}) is to set
\begin{equation}
 T_0 ~=~ \sqrt{\frac{\pi}{4k}}.
\end{equation}

%%%%%%%%%%%%%%%%%%%%%%%%%%%%%%%%%%%%%%%%%%%%
\section{Concluding Remarks}%%%%%%%%%%%%%%%%
%%%%%%%%%%%%%%%%%%%%%%%%%%%%%%%%%%%%%%%%%%%%

We now understand the branes in $N=2$ Liouville theory as boundary
states, which are algebraic objects satisfying boundary conditions
on $N=2$ supercurrents, and also in terms of worldsheet actions
containing boundary interactions.
We have obtained an explicit correspondence between two descriptions,
and various structure constants of the theory on the disc have been
analyzed to the same extent as for the $N=0$ and $N=1$
Liouville theories.

The boundary interactions for B-branes proposed in this paper
can be understood within the framework of Landau-Ginzburg theory,
but the ones for A-branes are new.
It is therefore necessary to understand the properties of these
interactions from various viewpoints, such as mirror coset
model.

For A-branes, the description in terms of boundary interactions
is expected to apply only to those corresponding to non-degenerate
representations.
Some degenerate A-branes might be described by the theory on a pseudosphere
(a recent work \cite{Ahn-SY2} has analyzed this issue).
For B-branes, the relation between the labels of branes and
the representations of $N=2$ superconformal algebra is less clear.

We have not paid much attention to the open or closed
string states belonging to discrete representations.
Although they will not invalidate the analysis of the present
paper, they will play a significant role in certain problems
in string theory.
It is also important to understand the modular transformation
property of characters for these representations.

As a perturbed linear dilaton CFT, $N=2$ Liouville theory has
a structure very similar to the sine-Liouville theory, which
is believed to be dual to the bosonic $SL(2,\mR)/U(1)$
coset model describing two-dimensional black hole.
The boundary states in the sine-Liouville theory are expected
to be described by a similar set of boundary interactions including
boundary fermions.
It would be interesting to study the D-branes in these related
models and their Wick-rotated cousins along the same path.

~

\subsubsection{Acknowledgements}

It is a pleasure to thank T. Eguchi, J. Hashiba, K. Hori, T. Kimura,
A. Pakman, D. Page and Y. Sugawara for useful discussions and comments.
The early part of this work was done in collaboration with J. Hashiba
and then with K. Hori.
A part of the work was carried out during the author's stay at
the RIMS and YITP at Kyoto university.

The work for the second version started during the collaboration with
S. Ribault on related models which lead to the discovery of some
fatal errors in the previous version.
A part of the work was carried out during the visit to the
ETH Zurich, and the author thanks the string theory group at
the Institute for Theoretical Physics for hospitality and
discussions.

~

~

\appendix

%%%%%%%%%%%%%%%%%%%%%%%%%%%%%%%%%%%%%%%%
\section{Some Useful Formulae}%%%%%%%%%%
%%%%%%%%%%%%%%%%%%%%%%%%%%%%%%%%%%%%%%%%

In the main text we frequently used
\begin{equation}
  \gamma(x)\equiv\frac{\Gamma(x)}{\Gamma(1-x)},~~~
  \bs(x)\equiv\sin(\pi x), ~~~
  \bc(x)\equiv\cos(\pi x).
\end{equation}
The functions $\Upsilon, \bG$ and $\bS$ are defined by
\begin{eqnarray}
  \log\bG(x) &=& \int_0^\infty \frac{dt}{t}
  \left[ \frac{e^{-Qt/2}-e^{-xt}}{(1-e^{-bt})(1-e^{-t/b})}
        +\frac{e^{-t}}{2}\left(\tfrac{Q}{2}-x\right)^2
        +\frac{1}{t}\left(\tfrac{Q}{2}-x\right)\right],
 \nonumber \\
 \log\Upsilon(x) &=& \int_0^\infty \frac{dt}{t}
 \left[e^{-2t}\left(\tfrac{Q}{2}-x\right)^2
      -\frac{\sinh^2\{(\frac{Q}{2}-x)t\}}
            {\sinh(bt)\sinh(t/b)}\right],~~~~~
 \nonumber \\
  \log\bS(x) &=& \int_0^\infty\frac{dt}{t}
  \left[ \frac{2x-Q}{t}
        -\frac{\sinh\{(x-\frac{Q}{2})t\}}
              {2\sinh(\frac{bt}{2})\sinh(\frac{t}{2b})}
  \right],
\end{eqnarray}
where $Q=b+b^{-1}$, and are characterized by the shift relations
\begin{eqnarray}
  \bG(x+b)=\bG(x)\frac{b^{\frac{1}{2}-bx}\Gamma(bx)}{\sqrt{2\pi}},
& \Upsilon(x+b) = \Upsilon(x)b^{1-2bx}\gamma(bx),
& \bS(x+b) = \bS(x)2\sin(b\pi x),
 \nonumber \\
  \bG(x+\tfrac{1}{b})=
  \bG(x)\frac{b^{\frac{x}{b}-\frac{1}{2}}\Gamma(\tfrac{x}{b})}{\sqrt{2\pi}},
& \Upsilon(x+\tfrac{1}{b}) = \Upsilon(x)b^{\frac{2x}{b}-1}\gamma(\tfrac{x}{b}),
& \bS(x+\tfrac{1}{b}) = \bS(x)2\sin(\tfrac{\pi x}{b})
\end{eqnarray}
Note also that
\begin{equation}
  \Upsilon(x)=\bG(x)\bG(Q-x),~~~
  \bS(x)=\frac{\bG(Q-x)}{\bG(x)}.
\end{equation}
$\bG(x)$ has poles at $x=-mb-nb^{-1}~(m,n\in\mZ_{\ge 0})$
and no poles.

The functions $\eta(\tau)$ and $\vartheta(\nu,\tau)$
are defined by ($q\equiv e^{2\pi i\tau}, z\equiv e^{2\pi i\alpha}$)
\begin{eqnarray}
  \eta(\tau) &=& q^{\frac{1}{24}}\prod_{n\ge 1}(1-q^n),\nonumber \\
  \vartheta(\alpha,\tau)  &=&
  \prod_{n\ge1}(1-q^n)(1+zq^{n-\frac{1}{2}})(1+z^{-1}q^{n-\frac{1}{2}})
  ~=~
  \sum_{n\in\mZ}q^{\frac{n^2}{2}}z^n,
\end{eqnarray}
and obey the modular S transformation law:
\begin{equation}
  \vartheta(\alpha,\tau_o) =
 (-i\tau_c)^{\frac{1}{2}}q_c^{\frac{\alpha^2}{2}}
  \vartheta(-\alpha\tau_c,\tau_c),~~~
  \eta(\tau_o) =
 (-i\tau_c)^{\frac{1}{2}}\eta(\tau_c)
 ~~~~(\tau_o\tau_c=-1).
\label{Str}
\end{equation}

\vskip4mm

In the main text we often encountered the contour integrals of the
following form:
\begin{eqnarray}
\lefteqn{
 \int_{0<s<t<1} dsdt
 s^a(1-s)^b t^{\bar{a}}(1-t)^{\bar{b}}(t-s)^{-k-1}
} \nn \\
 &=&
 \frac{\Gamma(1+a+\bar{a}-k)\Gamma(1+b+\bar{b}-k)\Gamma(a+1)\Gamma(\bar{b}+1)}
      {\Gamma(a-\bar{c}+1)\Gamma(\bar{b}-c+1)}
 \nn \\ && \times
 {}_3F_2(a+1,\bar{b}+1,k-c-\bar{c};a-\bar{c}+1,\bar{b}-c+1;1)
 ~~~\equiv~
 G_k\gsix{a}{b}{c}{\bar{a}}{\bar{b}}{\bar{c}},
 \nn \\ &&
 ~~~(c=k-1-a-b,~~\bar{c}=k-1-\bar{a}-\bar{b})
\end{eqnarray}
 The function $G_k$ satisfies the equalities
\begin{equation}
 c+\bar{c}=k ~~\Rightarrow~~ 
  G_k\gsix{a}{b}{c}{\bar{a}}{\bar{b}}{\bar{c}} =
 \frac{\Gamma(-a-\bar{a}-1)\Gamma(-b-\bar{b}-1)\Gamma(a+1)\Gamma(\bar{b}+1)}
      {\Gamma(-\bar{a})\Gamma(-b)},
\end{equation}
\begin{equation}
  \bs(a)\bs(\bar{a})G_k\gsix{\bar{a}}{\bar{b}}{\bar{c}}{a}{b}{c}
 +\bs(a)\bs(k-\bar{a})G_k\gsix{a}{b}{c}{\bar{a}}{\bar{b}}{\bar{c}}
 =
  \bs(c)\bs(\bar{c})G_k\gsix{c}{b}{a}{\bar{c}}{\bar{b}}{\bar{a}}
 +\bs(k-c)\bs(\bar{c})G_k\gsix{\bar{c}}{\bar{b}}{\bar{a}}{c}{b}{a}.
\end{equation}

\newpage

\begin{center}{\large\sc References}\end{center}\par

\end{document}